\documentclass[aps, prd, reprint, twocolumn, nofootinbib, superscriptaddress]{revtex4-2}
\usepackage{adjustbox}
\usepackage{amsmath}
\usepackage{amssymb}
\usepackage{amsfonts}
\usepackage{amsthm}
\usepackage{bm}
\usepackage{booktabs, array}
\usepackage{caption}
\usepackage{dcolumn}
\usepackage{mathrsfs}
\usepackage{soul}
\usepackage[T1]{fontenc}
\usepackage[caption=false]{subfig}
\usepackage{graphicx} 
\usepackage{multirow}
\usepackage{makecell}
\usepackage[dvipsnames]{xcolor}
\usepackage{tikz}
\usepackage[breaklinks=true]{hyperref}
\usetikzlibrary{shapes.geometric, arrows}
\usepackage{orcidlink}
\usepackage[T1]{fontenc}
\usetikzlibrary{positioning, arrows.meta, fit, shapes.geometric}

\newcommand{\bGR}{{\mbox{\tiny bGR}}}
\newcommand{\dCS}{{\mbox{\tiny dCS}}}
\newcommand{\cloud}{{\mbox{\tiny cloud}}}

\newcommand{\geo}{{\mbox{\tiny geo}}}
\newcommand{\eff}{{\mbox{\tiny eff}}}

\hypersetup{
    colorlinks=true,
    linkcolor=NavyBlue,
    filecolor=Magenta,
    urlcolor=NavyBlue,
    citecolor=NavyBlue
}

\DeclareSymbolFont{toneitalic}{T1}{\familydefault}{m}{it}
\DeclareMathSymbol{\cpartial}{\mathord}{toneitalic}{"F0}

\newcommand{\UIUC}{Illinois  Center  for  Advanced  Studies  of  the  Universe \&
Department of Physics, University of Illinois at Urbana-Champaign, Urbana, Illinois 61801, USA}
\newcommand{\Caltech}{Theoretical Astrophysics 350-17, California Institute of Technology, Pasadena, CA 91125, USA}
\newcommand{\Guelph}{Department of Physics, University of Guelph, Guelph, Ontario, N1G 2W1, Canada}

\allowdisplaybreaks
\begin{document}

\title{Evolving extreme mass-ratio inspirals in a perturbed Schwarzschild spacetime}

\author{Michael LaHaye}
\thanks{These authors contributed equally.}
\affiliation{\Guelph}

\author{Colin Weller\,\orcidlink{0000-0001-5173-5638}}
\thanks{These authors contributed equally.}
\affiliation{\Caltech}

\author{Dongjun Li\,\orcidlink{0000-0002-1962-680X}}
\thanks{Contact author: \href{dongjun@illinois.edu}{dongjun@illinois.edu}}
\affiliation{\Caltech}
\affiliation{\UIUC}

\author{Patrick Bourg\,\orcidlink{0000-0003-0015-0861}}
\affiliation{Institute for Mathematics, Astrophysics and Particle Physics, Radboud University, Heyendaalseweg 135, 6525 AJ Nijmegen, The Netherlands}

\author{Yanbei Chen}
\affiliation{\Caltech}

\author{Huan Yang \orcidlink{0000-0002-9965-3030}}
\thanks{Contact author: \href{hyangdoa@tsinghua.edu.cn}{hyangdoa@tsinghua.edu.cn}} 
\affiliation{Department of Astronomy, Tsinghua University, Beijing 100084, China}

\date{\today}

\begin{abstract}
In this work, we develop the modified Teukolsky formalism that describes the GW radiation from a point mass orbiting around a perturbed Schwarzschild BH. This perturbation of the background spacetime induces a secular change in the orbital phase of the point mass. In turn, this causes a modification in the GW flux, which can be used to probe the background spacetime. We explicitly apply this formalism to a bumpy Schwarzschild spacetime as a proof of principle. The results pave the way for the description of EMRIs in generic perturbed Kerr spacetime in future developments.
\end{abstract}

\maketitle
\section{Introduction}
\label{sec:introduction}

Precisely measuring black hole (BH) spacetimes using extreme mass-ratio inspirals (EMRIs), systems composed of a stellar-mass compact object orbiting a supermassive BH, is a primary science objective for space-based gravitational wave (GW) detectors \cite{LISA:2022kgy}, such as LISA \cite{LISA:2017pwj, LISA:2024hlh}, Taiji \cite{Hu:2017mde, Ruan:2018tsw}, and TianQin \cite{TianQin:2015yph, TianQin:2020hid}. Among the main extragalactic sources for these detectors \cite{LISA:2022yao, LISA:2024hlh}, EMRIs are one of the most sensitive probes of BH spacetimes, as their long-duration signals, accumulating hundreds of thousands of orbital cycles \cite{Barack:2018yly}, can encode subtle spacetime deformations around these supermassive BHs \cite{LISA:2022kgy}. For example, LISA is expected to measure the spacetime quadrupole moment with an accuracy of $\sim 10^{-4}M^3$ \cite{Tahura:2023qqt}, where $M$ is the supermassive BH mass. Deformations of BH spacetimes can come from theories beyond general relativity (bGR), such as dynamical Chern-Simons (dCS) gravity \cite{Alexander:2009tp, Yunes:2009hc, Cardoso:2009pk, Molina:2010fb, Pani:2011xj, Yagi:2012ya, Owen:2021eez, Wagle:2021tam, Srivastava:2021imr, Wagle:2023fwl, Chung:2025gyg, Li:2025fci, Lam:2025elw}, Einstein dilaton Gauss-Bonnet (EdGB) gravity \cite{Pani:2009wy, Yunes:2011we, Blazquez-Salcedo:2016enn, Blazquez-Salcedo:2017txk, Kleihaus:2015aje, Pierini:2021jxd, Pierini:2022eim, Chung:2024ira, Chung:2024vaf, Blazquez-Salcedo:2024oek}, and higher-derivative gravity \cite{Cano:2019ore, Cano:2022wwo, Cano:2023tmv, Cano:2023jbk, Cano:2024ezp, Maenaut:2024oci}. Furthermore, astrophysical environments around supermassive BHs, such as ultralight bosonic clouds \cite{Arvanitaki:2009fg, Arvanitaki:2010sy, Herdeiro:2014goa, Zhang:2018kib, Berti:2019wnn, Zhang:2019eid, Maselli:2020zgv, Baumann:2021fkf, Baumann:2022pkl, Duque:2023seg, Brito:2023pyl, Dyson:2025dlj, Li:2025ffh}, dark matter halos \cite{Hernquist:1990be, Gondolo:1999ef, Ullio:2001fb, Cardoso:2022whc, Destounis:2022obl, Gliorio:2025cbh, Mitra:2025tag, Fernandes:2025osu}, and accretion disks \cite{Shakura:1972te, Petrich:1988zz, Barausse:2007dy, Yunes:2011ws, Kocsis:2011dr, Babichev:2012sg, Kimura:2021dsa, Campos:2025zag}, can also backreact on the BH geometry. More drastic departures from BHs in general relativity (GR) include horizonless BH mimickers, or exotic compact objects \cite{Cardoso:2019rvt, LISA:2022kgy}, such as bosonic stars \cite{Kaup:1968zz, Siemonsen:2020hcg, Jetzer:1991jr, Herdeiro:2014goa, Brito:2015pxa, Herdeiro:2020jzx}, wormholes \cite{Einstein:1935tc, Morris:1988cz, Morris:1988tu, Hawking:1988ae, Visser:1989kh, Visser:1995cc, Damour:2007ap}, gravastars \cite{Mazur:2001fv, Mazur:2004fk, Visser:2003ge}, and fuzzballs \cite{Lunin:2001jy, Lunin:2002qf, Mathur:2005zp, Skenderis:2008qn, Mathur:2008nj}, which can have different multipole moments from BHs in GR \cite{Tahura:2023qqt} and result in observable GW signals \cite{Ryan:1995wh, Ryan:1997hg, Barack:2006pq, Vigeland:2009pr, Vigeland:2011ji}.

Spherically and axially symmetric BHs in vacuum GR are well described by the Schwarzschild \cite{Schwarzschild:1916uq} and Kerr \cite{Kerr:1963ud} metrics, respectively. Measuring possible deviations from Schwarzschild or Kerr spacetimes via EMRIs requires accurate modeling of the EMRI orbital evolution. Within GR, the long-term evolution of EMRIs can be formulated in a two-timescale framework using gravitational self-force theory \cite{Mino:1996nk, Hughes:1999bq, Hinderer:2008dm, Hughes:2016xwf, VanDeMeent:2018cgn, Barack:2018yvs, Miller:2020bft, Pound:2021qin}, where the conserved quantities such as energy, angular momentum, and Carter constant are evolved within the radiation reaction timescale, using the GW fluxes towards space infinity and the BH horizon. Accurate calculations of GW fluxes for rotating (Kerr) BHs rely on
the Newman-Penrose (NP) formalism \cite{Newman:1961qr} to describe curvature perturbations induced by the companion stellar-mass object on the central supermassive BH. These curvature perturbations are encoded in the perturbations of the Weyl scalars $\Psi_0$ and $\Psi_4$, which satisfy decoupled and separable second-order partial differential equations, referred to as the \textit{Teukolsky equations} \cite{Teukolsky:1973ha, Press:1973zz, Teukolsky:1974yv}. The fluxes of conserved quantities toward the horizon and space infinity can then be determined from the solutions of $\Psi_0$ and $\Psi_4$, respectively.

Our goal is to use BH perturbation theory to model the EMRI evolution in a deformed Kerr background, arising from bGR corrections, environmental effects, or BH mimickers. One challenge of extending the EMRI modeling procedures for a Kerr BH to a deformed Kerr background is that the Teukolsky formalism requires the background spacetime to be Ricci flat and Petrov type D, which are conditions not necessarily satisfied by deformed BHs or BH mimickers \cite{Yunes:2009hc, Yunes:2011we, Yagi:2012ya, Herdeiro:2014goa, Kleihaus:2015aje, Bamber:2021knr, DeLuca:2021ite, Owen:2021eez, Fernandes:2022gde, Yagi:2023eap}. A resolution to this challenge was formulated in \cite{Li:2022pcy, Hussain:2022ins} by developing a \textit{modified Teukolsky formalism} (MTF) for non-Ricci-flat and algebraically general BHs. The key idea of the MTF is to perform a two-parameter expansion of the NP equations using $\epsilon$ and $\zeta$, where $\epsilon$ is the mass ratio between the secondary and the supermassive BH, and $\zeta$ characterizes the strength of bGR or environmental effects. Under such an expansion, one can get a set of decoupled and separable Teukolsky-like equations for $\Psi_0$ and $\Psi_4$, where deformations of background BH spacetimes are converted to effective sources coupled to GWs in vacuum GR \cite{Li:2022pcy, Hussain:2022ins}. 

The MTF has already been applied to study ringdown in higher-derivative gravity \cite{Cano:2023tmv, Cano:2023jbk, Cano:2024ezp, Maenaut:2024oci} and dynamical Chern-Simons gravity \cite{Wagle:2023fwl, Li:2025fci}. More recently, \cite{Li:2025ffh} extended the MTF to model EMRIs under bGR or environmental effects and applied it to EMRIs with circular equatorial orbits embedded in ultralight complex scalar clouds formed via superradiance. A similar effort by \cite{Polcar:2025yto} applied the MTF to study EMRIs surrounded by ring-like matter, which serves as a toy model of accretion disks or nuclear stellar distributions. More broadly, this two-parameter expansion approach has been widely applied in studying environmental effects on EMRIs around nonrotating \cite{Brito:2023pyl, Rahman:2025mip, Datta:2025ruh} and rotating BHs \cite{Dyson:2025dlj}, where the MTF is applicable in principle. Despite all the progress, none of the previous literature has explicitly computed the source terms within the modified Teukolsky equations for an EMRI system, which is crucial for calculating modified gravitational flux, although some of the procedures for studying ringdown in \cite{Cano:2023tmv, Wagle:2023fwl} might still apply here.

To resolve this issue, we are applying the MTF to study EMRIs around deformed supermassive BHs by explicitly calculating the source terms in the modified Teukolsky equations for a generic deformed background spacetime. As a proof of principle, this work focuses on nonrotating BHs with generic axisymmetric deformations, but the procedures developed here can be directly generalized to deformed rotating BHs. Since the flux balance law for the Carter constant in vacuum GR may not necessarily hold in more general background spacetimes, we focus on circular and equatorial orbits to avoid complications arising from inclined or eccentric trajectories. Such orbits are typical of wet EMRIs formed through disk-assisted capture and migration in active galactic nuclei, as interactions with the disk largely reduce the eccentricity and inclination of the secondary before it enters the LISA frequency band \cite{Pan:2021oob, Pan:2021ksp}. 

Moreover, deviations of the background metric often arise from corrections to the Einstein-Hilbert action or backreactions of matter and other fields, leading to additional gravitational self-interactions (e.g., in higher-derivative gravity) or interactions with other nonmetric fields (e.g., for ultralight scalar clouds). These additional interactions are usually captured in some effective stress-energy tensors, but the details of which rely on the underlying bGR theory or matter model. For this reason, we ignore these model-dependent contributions to gravitational flux and focus on the more universal contribution due to background geometry corrections in this work. This work serves as a necessary middle step in using the MTF to model the EMRI evolution in physically motivated scenarios, such as ultralight scalar clouds considered in \cite{Li:2025ffh}. Following \cite{Ryan:1995wh, Ryan:1997hg, Collins:2004ex, Barack:2006pq, Vigeland:2009pr, Vigeland:2011ji, Johannsen:2011dh}, this work also paves the road for conducting parametrized tests of BH spacetimes via EMRIs. 

The rest of the paper is organized as follows. In Sec.~\ref{sec:MTF}, we review the MTF first developed in \cite{Li:2022pcy} and its extension to EMRIs in \cite{Li:2025ffh}. In Sec.~\ref{sec:bacground_NP}, we review the tetrad and associated Geroch–Held–Penrose (GHP) operators \cite{Geroch:1973am} defined on the background BH spacetime in vacuum GR (i.e., Schwarzschild metric for nonrotating BHs). In Sec.~\ref{sec:perturbed_NP}, we calculate the $\mathcal{O}(\zeta^1,\epsilon^0)$ corrections to the NP quantities due to generic deviations of the Schwarzschild metric, the latter of which can be replaced by theory-dependent corrections due to bGR \cite{Yunes:2009hc, Yunes:2011we, Yagi:2012ya, Yunes:2011we, Kleihaus:2015aje, Cano:2019ore, Lam:2025elw} and environmental effects \cite{Herdeiro:2014goa, Brito:2023pyl, Cardoso:2022whc, Duque:2023seg, Destounis:2022obl, Gliorio:2025cbh, Mitra:2025tag, Fernandes:2025osu, Petrich:1988zz, Babichev:2012sg, Kimura:2021dsa, Campos:2025zag} or parametrized deviations of the background spacetime \cite{Ryan:1995wh, Ryan:1997hg, Collins:2004ex, Barack:2006pq, Vigeland:2009pr, Vigeland:2011ji, Johannsen:2011dh, Rezzolla:2014mua, Konoplya:2016jvv, Carson:2020dez, Yagi:2023eap}. We then evaluate the NP quantities at $\mathcal{O}(\zeta^0,\epsilon^1)$ due to the GW radiation driven by the secondary in vacuum GR, where the perturbed metric is obtained by solving the Regge-Wheeler (RW) \cite{Regge:1957td} and Zerilli-Moncrief (ZM) \cite{Zerilli:1970se, Moncrief:1974am} equations directly. In Sec.~\ref{sec:construct_source}, we show how to construct the source terms of the modified Teukolsky equations from the perturbed NP quantities at $\mathcal{O}(\zeta^1,\epsilon^0)$ and $\mathcal{O}(\zeta^0,\epsilon^1)$ calculated in Sec.~\ref{sec:perturbed_NP}. The complete expressions of the source terms are expressed in terms of the RW and ZM functions in vacuum GR and the generic axisymmetric deformations of the Schwarzschild metric. In Sec.~\ref{sec:flux}, we prescribe how to solve for the correction to the GW radiation at $\mathcal{O}(\zeta^1,\epsilon^1)$ using the source terms constructed in Sec.~\ref{sec:perturbed_NP} and compute the associated flux. We will not solve the modified Teukolsky equation for any specific background metric deviation, so only a general procedure is provided. We then discuss future avenues of this work in Sec.~\ref{sec:discussion}. Throughout this work, we work in four dimensions with a metric signature $(-,+,+,+)$, as in \cite{Misner:1973prb}. We use the notation in \cite{Chandrasekhar_1983} for all the NP quantities except the metric signature. Unless otherwise noted, we always use geometrized units in which $G=c=1$.

\begin{figure*}
\centering
\begin{tikzpicture}[
  font=\normalsize,
  box/.style={draw, thick, rounded corners, fill=white, align=center,
              inner sep=6pt, text width=6cm},
  dashedbox/.style={box, dashed, very thick},
  subbox/.style={draw, thin, rounded corners, align=center,
                 inner sep=4pt, text width=5.0 cm, fill=white},
  arrow/.style={-{Latex[length=3mm,width=2mm]}, very thick},
  node distance=2.0cm and 2.6cm
]

\node (background)[dashedbox, align=center, 
    label={[black, yshift=0.2cm]\textbf{$\mathcal{O}(\zeta^{0},\epsilon^{0})$}}] {
    \textbf{Background GR Quantities} \\
    Metric $g_{\mu\nu}$: Eq.~\eqref{eq:KerrMetric}\\
    Tetrad $e_a^{\mu(0,0)}$: Eq.~\eqref{eq:Carter_tetrad}\\
    Spin Coefficients: Eq.~\eqref{eq:SpinCoeiffsBackground}
};

\node[dashedbox, draw=blue!80!black, 
    align=center, below left=1.0cm and -3.0cm of background,
    label={[blue!80!black, yshift=0.2cm,xshift=-1.5cm]\textbf{$\mathcal{O}(\zeta^{1},\epsilon^{0})$}},
    text width=8.1 cm] (stationary) {
    \textbf{Stationary Quantities} \\
    Metric $h_{\mu\nu}$: generic \\
    Tetrad $e_a^{\mu(1,0)}$: Eq.~\eqref{eq:perturbed_tetrad} \\
    Spin Coefficients: Eq.~\eqref{eq:perturbedSpinCoeffs} \\
    Weyl Scalars $\Psi_i^{(1,0)}$: Eq.~\eqref{eq:perturbedWeylScalars}
};

\node[dashedbox, draw=red!80!black, 
    align=center, below right=1.0cm and -3.0cm of background,
    label={[red!80!black,yshift=0.2cm,xshift=1.5cm]\textbf{$\mathcal{O}(\zeta^{0},\epsilon^{1})$}},
    text width=8.1 cm] (radiative) {
    \textbf{Radiative GR Quantities} \\
    Metric $q_{\mu\nu}$: Eqs.~\eqref{eq:qOdd}, \eqref{eq:qEven}, \eqref{eq:PerturbedMetricComponets} \\
    Tetrad $e_a^{\mu(0,1)}$: Eqs.~\eqref{eq:perturbed_tetrad} $(h\to q)$, \eqref{eq:tetrad_rotations_10}, \& \eqref{eq:TetradRotCoeffs} \\
    Spin Coefficients: Eqs.~\eqref{eq:perturbedSpinCoeffs} $(h\to q)$, \eqref{eq:spin_coeffs_rotated}, \& \eqref{eq:TetradRotCoeffs} \\
    Weyl Scalars $\Psi_i^{(0,1)}$: Eqs.~\eqref{eq:perturbedWeylScalars} $(h\to q)$, \eqref{eq:WeylScalarRotation}, \& \eqref{eq:TetradRotCoeffs}
};

\node[dashedbox, draw=orange!85!black,
    align=center, below=4.8cm of background,
    label={[orange!85!black,yshift=0.2cm]\textbf{$\mathcal{O}(\zeta^{1},\epsilon^{1})$}},
    text width=8.5cm] (source) {
    \textbf{Modified Teukolsky Equations} \\
    $H_{0}^{(0,0)}\Psi_0^{(1,1)}
    =\mathcal{S}_{\geo}^{(1,1)}+\mathcal{S}_{\eff}^{(1,1)}
    +\mathcal{S}_{p}^{(1,1)}$~[Eq.~\eqref{eq:master_eqn_non_typeD_Psi0}] \\
    $H_{4}^{(0,0)}\Psi_4^{(1,1)}
    =\mathcal{T}_{\geo}^{(1,1)}+\mathcal{T}_{\eff}^{(1,1)}
    +\mathcal{T}_{p}^{(1,1)}$[Eq.~\eqref{eq:master_eqn_non_typeD_Psi4}] \\
};

\node[subbox, draw=orange!85!black, dashed, very thick,
    below=0.8cm of source.south, anchor=north] (eff) {
    $\boldsymbol{T_{\mu\nu}^{\textbf{\eff}}}$ [Eq.~\eqref{eq:S_bGRExpansion}]\\
    $\mathcal{S}_{\eff}^{(1,1)}$: Tables~\ref{tab:typeAS_bGR} \& \ref{tab:typeBS_bGR}\\
    $\mathcal{T}_{\eff}^{(1,1)}$: Tables~~\ref{tab:typeAT_bGR} \& \ref{tab:typeBT_bGR}
};

\node[subbox, draw=orange!85!black, dashed, very thick,
    left=0.8cm of eff] (geo) {
    \textbf{Geometrical} [Eq.~\eqref{eq:S_geoExpansion}] \\
    $\mathcal{S}_{\geo}^{(1,1)}$: Table~\ref{tab:S_geoTable} \\
    $\mathcal{T}_{\geo}^{(1,1)}$: Table~\ref{tab:T_geoTable}
};

\node[subbox, draw=orange!85!black, dashed, very thick,
    right=0.8cm of eff] (emri) {
    $\boldsymbol{T_{\mu\nu}^{p}}$ [Eq.~\eqref{eq:S_pGRAExpansion}] \\
    $\mathcal{S}_p^{(1,1)}$: Tables~\ref{tab:typeAS_bGR} \& \ref{tab:typeBS_bGR} $(h\to q)$\\
    $\mathcal{T}_p^{(1,1)}$: Tables~\ref{tab:typeAT_bGR} \& \ref{tab:typeBT_bGR} $(h\to q)$
};

\node[dashedbox, draw=orange!85!black, dashed, very thick,
    below=0.8cm of eff] (bottomtext) {
    \textbf{Modified Fluxes} \\
    Future null-infinity: Eq.~\eqref{eq:ModFluxPsi4} \\
    Horizon: Eq.~\eqref{eq:horizon_flux}
};

\draw[-{Latex[length=2mm,width=2mm]}, thick] (background.west) -| (stationary.north);
\draw[-{Latex[length=2mm,width=2mm]}, thick] (background.east) -| (radiative.north);
\draw[-{Latex[length=2mm,width=2mm]}, thick] (stationary.south) -- ([xshift=3.65mm]source.north west);
\draw[-{Latex[length=2mm,width=2mm]}, thick] (radiative.south) -- ([xshift=-3.65mm]source.north east);

\draw[thick] (source.south) -- ([yshift=3mm]eff.north);
\draw[-{Latex[length=2mm,width=2mm]}, thick] ([yshift=5mm]eff.north) -- (eff.north);
\draw[-{Latex[length=2mm,width=2mm]}, thick] ([yshift=5mm]eff.north) -| (geo.north);
\draw[-{Latex[length=2mm,width=2mm]}, thick] ([yshift=5mm]eff.north) -| (emri.north);

\draw[thick] (geo.south) -- ([yshift=-3mm]geo.south);
\draw[thick] (eff.south) -- ([yshift=-3mm]eff.south);
\draw[thick] (emri.south) -- ([yshift=-3mm]emri.south);
\draw[thick] ([yshift=-3mm]emri.south) -- ([yshift=-3mm]eff.south);
\draw[thick] ([yshift=-3mm]geo.south) -- ([yshift=-3mm]eff.south);
\draw[-{Latex[length=2mm,width=2mm]}, thick] ([yshift=-3mm]eff.south) -- (bottomtext.north);

\end{tikzpicture}
\caption{A schematic flowchart listing the relevant quantities and equations for evaluating the modified Teukolsky equation in Eqs.~\eqref{eq:master_eqn_non_typeD_Psi0} and \eqref{eq:master_eqn_non_typeD_Psi4}. The final results of all the source terms are tabulated in Appendix~\ref{sec:SourceCoeffs}. For quantities taking the same form at $\mathcal{O}(\zeta^1,\epsilon^0)$ and $\mathcal{O}(\zeta^0,\epsilon^1)$, one can replace $h_{\mu\nu}$ with $q_{\mu\nu}$ in the former's expression to get the latter, where we denote this procedure as $(h\to q)$. For $\mathcal{S}_{p}^{(1,1)}$ and $\mathcal{T}_{p}^{(1,1)}$, one also needs to set the rotation coefficients $a^{(0,1)}$ and $b^{(0,1)}$ to zero when making the replacement, as detailed in Sec.~\ref{sec:construct_source}.}
\label{fig:flowchart}
\end{figure*}
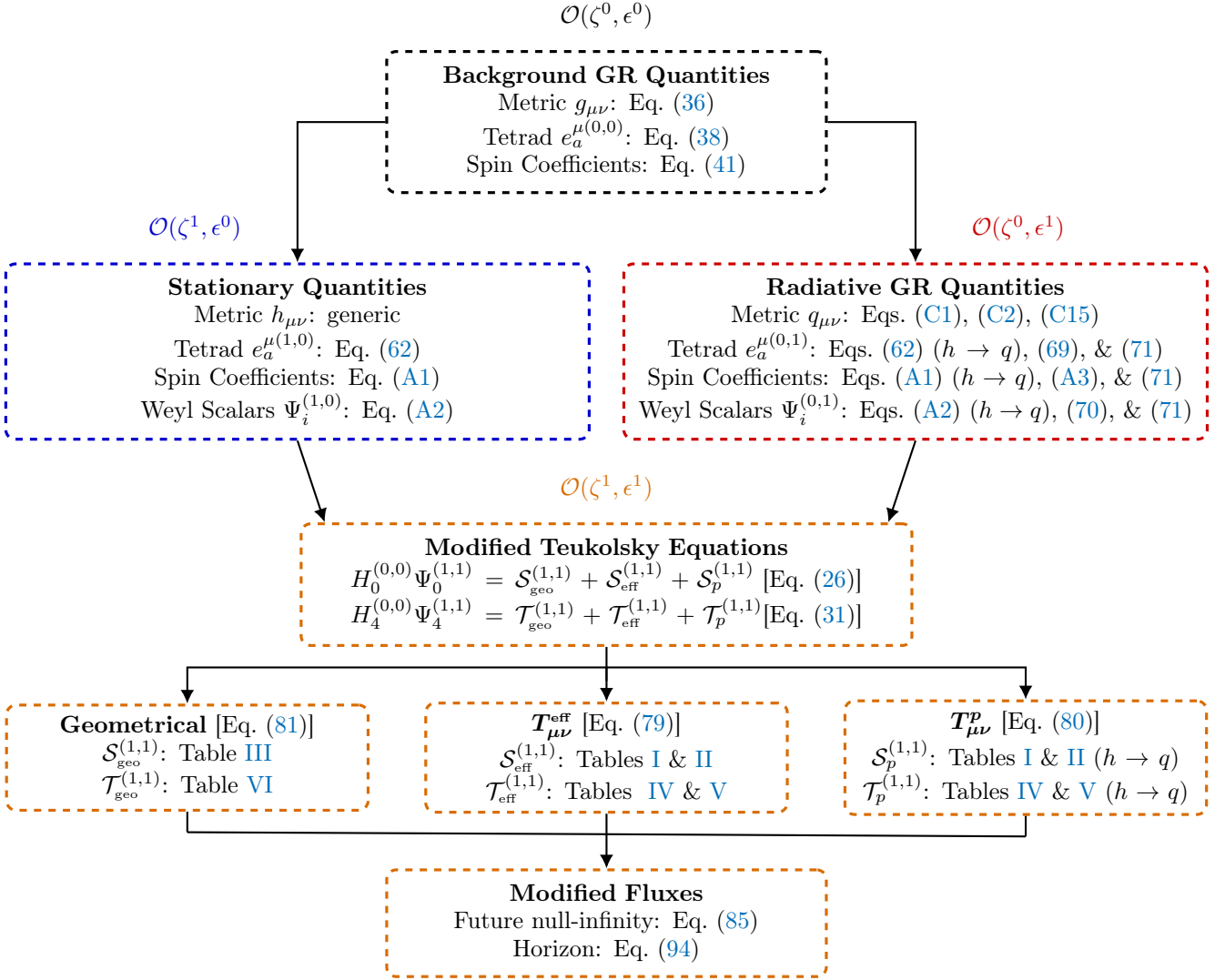


\section{Modified Teukolsky Formalism}
\label{sec:MTF}

In this section, we review the MTF in \cite{Li:2022pcy, Li:2025ffh}. We first discuss the structure of the action for an EMRI system in some bGR theories or embedded in some field or matter distributions. We then introduce an expansion scheme to solve this system perturbatively and derive the corresponding modified Teukolsky equations for the Weyl scalars $\Psi_{0,4}$.

\subsection{The action and expansion scheme}

This work aims to study the EMRI evolution around a perturbatively deformed supermassive BH. Such a deformed BH usually arises from bGR corrections or external matter, so the action describing such a system could be generically written in the following form:
\begin{equation} \label{eq:action}
    S=\int d^4 x\sqrt{-g}\left(\frac{R}{16\pi G}
    +\mathcal{L}_{\bGR}
    +\mathcal{L}_{m}+\mathcal{L}_{p}\right)\,,
\end{equation}
where $\mathcal{L}_{\bGR}$ is the Lagrangian for bGR corrections. Here, we only consider a subset of bGR theories that admits an effective field theory description \cite{Berti:2015itd}, i.e, $\mathcal{L}_{\bGR}$ is proportional to some bGR coupling constant $\alpha_{\bGR}$ such that $\alpha_{\bGR}=0$ in the GR limit. Some examples include dCS gravity \cite{Alexander:2009tp, Yunes:2009hc, Cardoso:2009pk, Molina:2010fb, Pani:2011xj, Yagi:2012ya, Owen:2021eez, Wagle:2021tam, Srivastava:2021imr, Wagle:2023fwl, Chung:2025gyg, Li:2025fci}, EdGB gravity \cite{Pani:2009wy, Yunes:2011we, Blazquez-Salcedo:2016enn, Blazquez-Salcedo:2017txk, Kleihaus:2015aje, Pierini:2021jxd, Pierini:2022eim, Chung:2024ira, Chung:2024vaf, Blazquez-Salcedo:2024oek}, and higher-derivative gravity \cite{Cano:2019ore, Cano:2022wwo, Cano:2023tmv, Cano:2023jbk, Cano:2024ezp, Maenaut:2024oci}. In these theories, $\mathcal{L}_{\bGR}$ usually contains higher-derivative terms of the metric and might couple to other nonmetric fields. For example, in dCS gravity, a pseudoscalar field $\vartheta$ is coupled to the Pontryagin density as follows:
\begin{equation}
    \mathcal{L}_{\bGR}^{\dCS}
    =\frac{\alpha_{\dCS}}{4}\vartheta R_{\nu\mu\rho\sigma}{}^*
    R^{\mu\nu\rho\sigma}\,,
\end{equation}
where ${}^*R^{\mu\nu\rho\sigma}$ is the dual of the Riemann tensor, and $\alpha_{\dCS}$ is the dCS coupling constant. In dCS gravity, besides its coupling to the Pontryagin density, the pseudoscalar field $\vartheta$ also has a Lagrangian for its own dynamics, which we classify as $\mathcal{L}_{m}$, representing corrections to vacuum GR minimally coupled to external fields or matter. For example, in dCS gravity,
\begin{equation}
    \mathcal{L}_{m}^{\dCS}
    =-\frac{1}{2}\partial^{\mu}\vartheta\partial_{\mu}\vartheta
    +V(\vartheta)\,,
\end{equation}
where $V(\vartheta)$ is certain potential in $\vartheta$. This type of correction also appears in non-vacuum environments, such as ultralight bosonic clouds \cite{Arvanitaki:2009fg, Arvanitaki:2010sy, Herdeiro:2014goa, Zhang:2018kib, Berti:2019wnn, Zhang:2019eid, Maselli:2020zgv, Baumann:2021fkf, Baumann:2022pkl, Duque:2023seg, Brito:2023pyl, Dyson:2025dlj, Li:2025ffh}, dark matter halos \cite{Hernquist:1990be, Gondolo:1999ef, Ullio:2001fb, Cardoso:2022whc, Destounis:2022obl, Gliorio:2025cbh, Mitra:2025tag}, and accretion disks \cite{Shakura:1972te, Barausse:2007dy, Yunes:2011ws, Kocsis:2011dr, Babichev:2012sg, Kimura:2021dsa, Campos:2025zag}. For example, for ultralight complex scalar clouds,
\begin{equation}
    \mathcal{L}_{m}^{\cloud}
    =-\partial^\mu\Phi\partial_\mu\bar{\Phi}
    -\mu^2\Phi\bar{\Phi}\,,
\end{equation}
where $\Phi$ is the complex scalar field, and $\mu$ is the scalar mass. Besides $\mathcal{L}_{\bGR}$ and $\mathcal{L}_{m}$, we also have an additional Lagrangian $\mathcal{L}_{p}$ associated with the secondary for an EMRI system. 

Varying the action in Eq.~\eqref{eq:action} with respect to the metric, we then get the Einstein equations with some (effective) stress-energy tensors, i.e.,
\begin{align} \label{eq:EE}
    & G_{\mu\nu}
    =8\pi G\left(T_{\mu\nu}^{\eff}+T^{p}_{\mu\nu}\right)\,, \nonumber\\
    & T_{\mu\nu}^{\eff}\equiv T_{\mu\nu}^{\bGR}+T_{\mu\nu}^{m}
    =-\frac{2}{\sqrt{-g}} \frac{\delta}{\delta g^{\mu\nu}}
    \left[\sqrt{-g}\left(\mathcal{L}_{\bGR}+\mathcal{L}_{m}\right)\right]\,,
\end{align}
where we have combined the stress-energy tensor $T_{\mu\nu}^{\bGR}$ due to bGR corrections with the stress-energy tensor $T_{\mu\nu}^{m}$ due to external fields or matter into $T_{\mu\nu}^{\eff}$. For example, in dCS gravity \cite{Alexander:2009tp, Wagle:2021tam}, 
\begin{align} \label{eq:stress_dCS}
    T_{\mu\nu}^{\eff}
    =& -2\alpha_{\dCS}\left[\nabla_\sigma\vartheta\,
    \epsilon^{\sigma\delta\alpha}{}_{(\mu}\nabla_\alpha R_{\nu)\delta}
    +\nabla_\sigma\nabla_\delta\vartheta\,
    {}^*R^\delta{}_{(\mu\nu)}{}^\sigma\right] \nonumber\\
    & +\partial_\mu\vartheta\partial_\nu\vartheta
    -\frac{1}{2}g_{\mu\nu}\partial^\sigma\vartheta
    \partial_\sigma\vartheta\,,
\end{align}
where the parenthesis in the subscript stands for symmetrization. For an ultralight complex scalar cloud \cite{Brito:2023pyl},
\begin{equation} \label{eq:stress_cloud}
    T_{\mu\nu}^{\eff}
    =2\partial_{(\mu}\Phi\partial_{\nu)}\bar{\Phi}-g_{\mu\nu}
    \left(\partial^\sigma\Phi\partial_\sigma\bar{\Phi}
    +\mu^2\Phi\bar{\Phi}\right)\,.
\end{equation}
The term $T^{p}_{\mu\nu}$ is the stress-energy tensor of a point particle, i.e.,
\begin{equation} \label{eq:stress_particle}
    T_{\mu\nu}^{p}
    =m_p\int u_\mu u_\nu\frac{\delta^{(4)}
    \left(x^\mu-x_p^\mu(\tau)\right)}{\sqrt{-g}}d\tau\,,
\end{equation}
where $m_p$ is the particle's mass, $x_p^{\mu}$ is the particle's worldline, and $u^{\mu}=dx^\mu/d\tau$ is the particle's four-velocity, with $\tau$ being the proper time. For circular equatorial orbits around nonrotating BHs considered in this work, Eq.~\eqref{eq:stress_particle} reduces to
\begin{equation} \label{eq:stress_particle_circular}
   T_{\mu\nu}^{p}
   =u_{\mu}u_{\nu}\frac{m_p}{\left(r^2 u^t\right)}\delta\left[r-r_p\right]
   \delta\left[\varphi-\omega_z t\right]\delta[\theta-\pi/2]\,,
\end{equation}
with
\begin{equation}
    u^a=u^t(1,0,0,\omega_z)\,,\;
    u^t=\frac{1}{\sqrt{1-3M/r_p}}\,,\;
    \omega_z=\sqrt{M/r_p^3}\,.
\end{equation}
The equations of motion for nonmetric fields, such as $\vartheta$ in dCS gravity or $\Phi$ for an ultralight complex scalar cloud, can be found by varying the action in Eq.~\eqref{eq:action} with respect to these fields. Our goal here is then using the NP formalism \cite{Newman:1961qr} and Teukolsky formalism \cite{Teukolsky:1973ha} to solve Eq.~\eqref{eq:EE} perturbatively.

Following \cite{Li:2022pcy}, we introduce the dimensionless expansion parameter $\zeta$ to characterize the amplitude of geometrical deviations from BHs in vacuum GR (Schwarzschild or Kerr), which can be defined in terms of coupling constants in bGR theories (e.g., $\zeta\propto\alpha_{\dCS}^2$ in dCS gravity \cite{Alexander:2009tp, Wagle:2023fwl, Li:2025fci}) or characteristic density of surrounding matter in non-vacuum environments (e.g., $\zeta$ is quadratic in the characteristic density of the scalar cloud in \cite{Brito:2023pyl, Dyson:2025dlj, Li:2025ffh}\footnote{Reference \cite{Li:2025ffh} actually uses $\zeta$ for the amplitude of ultralight scalar field, so $\zeta$ here is quadratic in the one in \cite{Li:2025ffh}.}). In addition to $\zeta$, since EMRI is composed of a supermassive BH of mass $M$ and a stellar-mass companion of mass $m_p$, we also expand all the equations of motion in the mass ratio $\epsilon=m_p/M$. 

Using these two parameters, we then expand the metric $g_{\mu\nu}$ as follows:
\begin{equation} \label{eq:expansion_metric}
    g_{\mu\nu}=
    g^{(0,0)}_{\mu\nu}+\epsilon q^{(0,1)}_{\mu\nu}+\zeta h^{(1,0)}_{\mu\nu}
    +\zeta\epsilon Q^{(1,1)}_{\mu\nu}+\cdots,
\end{equation}
where $g_{\mu\nu}^{(0,0)}$ is the Schwarzschild or Kerr background metric, and $h_{\mu\nu}^{(1,0)}$ is the deviation of the background spacetime. The GW driven by the secondary in vacuum GR is encoded in $q_{\mu\nu}^{(0,1)}$, and the additional GW sourced by the background spacetime deviation is represented by $Q_{\mu\nu}^{(1,1)}$. In this work, we have used different symbols for metric perturbations of different orders, so we will drop their superscripts $(n,m)$ denoting the terms at $\mathcal{O}(\zeta^n,\epsilon^m)$ if possible for simplicity. In \cite{Li:2022pcy}, the authors further show that one can always construct a background tetrad such that all the NP quantities follow the same expansion in Eq.~\eqref{eq:expansion_metric}, e.g.,
\begin{align} \label{eq:expansion_Weyl}
    \Psi_0
    &=\Psi_0^{(0)}+\epsilon\Psi_0^{(1)} \nonumber\\
    &=\Psi_0^{(0,0)}+\zeta\Psi_{0}^{(1,0)}+\epsilon\Psi_{0}^{(0,1)}
    +\zeta\epsilon\Psi_{0}^{(1,1)}\,,
\end{align}
where we take the Weyl scalar $\Psi_0$ as an example. In the cases where we can hide the expansion in $\zeta$, we only use the superscript for an expansion in $\epsilon$, as shown in the first line of Eq.~\eqref{eq:expansion_Weyl}. 

\subsection{Derivation of the modified Teukolsky equation}
To derive the modified Teukolsky equation for an EMRI system, let us first define several operators:
\begin{subequations} \label{eq:NP_operators}
\begin{align}
    D_{[a,b,c,d]}
    &=D+a\varepsilon+b\bar{\varepsilon}+c\rho+d\bar{\rho}\,, \\
    \boldsymbol{\Delta}_{[a,b,c,d]}
    &=\boldsymbol{\Delta}+a\mu+b\bar{\mu}+c\gamma+d\bar{\gamma}\,, \\
    \delta_{[a,b,c,d]}
    &=\delta+a\bar{\alpha}+b\beta+c\bar{\pi}+d\tau\,, \\
    \bar{\delta}_{[a,b,c,d]}
    &=\bar{\delta}+a\alpha+b\bar{\beta}+c\pi+d\bar{\tau}\,,
\end{align}
\end{subequations}
where $\{D,\boldsymbol{\Delta},\delta,\bar{\delta}\}$ are the directional derivatives associated with the NP tetrad basis vectors $e_{a}^{\mu}$, and the remaining quantities are spin coefficients in the NP formalism. The definitions of all NP quantities are provided in \cite{Li:2022pcy}, or refer to \cite{Newman:1961qr, Chandrasekhar_1983, Pound:2021qin} for more comprehensive reviews of the NP formalism. As shown in the following, most parts of NP equations can be conveniently expressed in terms of the operators in Eq.~\eqref{eq:NP_operators}. Moreover, these operators can be easily translated into the GHP operators \cite{Geroch:1973am, Pound:2021qin} or the Chandrasekhar operators \cite{Chandrasekhar_1983, Dolan:2023enf, Ma:2024qcv}, which can help us simplify the source terms.

Similar to the original derivation of the Teukolsky equation in \cite{Teukolsky:1973ha}, we will work with the following two Ricci identities and one Bianchi identity,
\begin{subequations} \label{eq:Bianchi_simplified}
\begin{align} 
    & F_1\Psi_0-J_1\Psi_1-3\kappa\Psi_2=S_1\,, \label{eq:BianchiId_Psi0_1_simplify} \\
    & F_2\Psi_0-J_2\Psi_1-3\sigma\Psi_2=S_2\,, \label{eq:BianchiId_Psi0_2_simplify} \\
    & E_2\sigma-E_1\kappa-\Psi_0=0 \label{eq:RicciId_Psi0_simplify}\,,
\end{align}
\end{subequations}
where $\kappa$ and $\sigma$ are spin coefficients. The operators $F_{1,2}$, $J_{1,2}$, and $E_{1,2}$ are defined as 
\begin{align} \label{eq:auxiliary_operators}
    & F_1\equiv\bar{\delta}_{[-4,0,1,0]}\,,\quad
    && F_2\equiv\boldsymbol{\Delta}_{[1,0,-4,0]}\,, \nonumber\\
    & J_1\equiv D_{[-2,0,-4,0]}\,,\quad
    && J_2\equiv\delta_{[0,-2,0,-4]}\,, \nonumber\\
    & E_1\equiv\delta_{[-1,-3,1,-1]}\,,\quad
    && E_2\equiv D_{[-3,1,-1,-1]}\,.
\end{align}
The source terms $S_{1,2}$ are defined as
\begin{subequations} \label{eq:source_bianchi}
\begin{align}
    \label{eq:source_bianchi_1}
    \begin{split} 
        S_1\equiv& \;\delta_{[-2,-2,1,0]}\Phi_{00}
        -D_{[-2,0,0,-2]}\Phi_{01} \\
        & \;+2\sigma\Phi_{10}-2\kappa\Phi_{11}-\bar{\kappa}\Phi_{02}\,,
    \end{split} \\
    \label{eq:source_bianchi_2}
    \begin{split} 
        S_2\equiv& \;\delta_{[0,-2,2,0]}\Phi_{01}
        -D_{[-2,2,0,-1]}\Phi_{02} \\
        & \;-\bar{\lambda}\Phi_{00}+2\sigma\Phi_{11}-2\kappa\Phi_{12}\,,
    \end{split}
\end{align}
\end{subequations}
where $\Phi_{ij}$ are projections of the Ricci tensor $R_{\mu\nu}$ onto the NP tetrad basis $e^{\mu}_{a}$, that is,
\begin{equation}\label{eq:RicciScalars}
    \Phi_{ij}\sim R_{\mu\nu}e^{\mu}_{a}e^{\mu}_{b}\,,
\end{equation}
with complete expressions provided in \cite{Newman:1961qr, Chandrasekhar_1983, Pound:2021qin, Li:2022pcy}. In this work, we assume a vacuum BH spacetime in GR, so $\Phi_{ij}^{(0,0)}=0$.

For terms of order larger than $\mathcal{O}(\zeta^0,\epsilon^0)$ in $\Phi_{ij}$, there are two origins.
First, the effective stress-energy tensor $T_{\mu\nu}^{\eff}$ in Eq.~\eqref{eq:EE} due to bGR corrections or environmental effects will result in nonzero $\Phi_{ij}^{(1,0)}$ and $\Phi_{ij}^{(1,1)}$. For corrections only involving metric fields, such as higher-derivative gravity \cite{Cano:2019ore, Cano:2022wwo, Cano:2023tmv, Cano:2023jbk, Cano:2024ezp, Maenaut:2024oci}, $T_{\mu\nu}^{\eff}$ is always proportional to $\alpha_{\bGR}$ and $\zeta\propto\alpha_{\bGR}$, so after the linearization in Eq.~\eqref{eq:expansion_metric}, 
\begin{subequations} \label{eq:stress_hd_expand}
\begin{align}
    & T_{\mu\nu}^{\eff(1,0)}\sim\zeta[g_{\mu\nu}^{(0,0)}]\,, \\
    & T_{\mu\nu}^{\eff(1,1)}\sim\zeta[q_{\mu\nu}^{(0,1)}]\,,
\end{align}
\end{subequations}
where $[f]$ represents combinations of the field $f$ and its derivatives. In the presence of nonmetric fields, these extra fields usually enter at fractional powers of $\zeta$ due to hierarchical equations of motion: GWs in vacuum GR first drive these extra fields, which then source additional GWs \cite{Yunes:2009hc}. For example, in dCS gravity \cite{Yunes:2009hc, Cardoso:2009pk, Molina:2010fb, Pani:2011xj, Yagi:2012ya, Owen:2021eez, Wagle:2021tam, Srivastava:2021imr, Wagle:2023fwl, Chung:2025gyg, Li:2025fci}, the pseudoscalar $\vartheta$ satisfies
\begin{equation} \label{eq:scalar_dCS}
    \square\vartheta
    =-\frac{\alpha_{\dCS}}{4}R_{\nu\mu\rho\sigma}
    {}^{*} R^{\mu\nu\rho\sigma}\,,
\end{equation}
which, with Eq.~\eqref{eq:stress_dCS}, implies that $\vartheta$ enters at $\mathcal{O}(\alpha_{\dCS})$, and the background metric deformation enters at $\alpha_{\dCS}^2$, so $\zeta\propto\alpha_{\dCS}^{2}$. The stress-energy tensor $T_{\mu\nu}^{\eff}$ in Eq.~\eqref{eq:stress_dCS} after the expansion in Eq.~\eqref{eq:expansion_metric} then takes the form: 
\begin{subequations} \label{eq:stress_dcs_expand}
\begin{align}
    T_{\mu\nu}^{\eff(1,0)}
    \sim& \;\zeta^{\frac{1}{2}}[g_{\mu\nu}^{(0,0)}][\vartheta^{(\frac{1}{2},0)}]
    +[g_{\mu\nu}^{(0,0)}][\vartheta^{(\frac{1}{2},0)}]^2\,, \\
    T_{\mu\nu}^{\eff(1,1)}
    \sim& \;\zeta^{\frac{1}{2}}[q_{\mu\nu}^{(0,1)}][\vartheta^{(\frac{1}{2},0)}]
    +[q_{\mu\nu}^{(0,1)}][\vartheta^{(\frac{1}{2},0)}]^2 \nonumber\\
    & \;+[g_{\mu\nu}^{(0,0)}][\vartheta^{(\frac{1}{2},0)}]
    [\vartheta^{(\frac{1}{2},1)}]\,.   
\end{align}
\end{subequations}
We can observe that both the metric and nonmetric fields in Eqs.~\eqref{eq:stress_hd_expand} and \eqref{eq:stress_dcs_expand} enter at an order below $\mathcal{O}(\zeta^1,\epsilon^1)$. Similar arguments can also be made for such a perturbative treatment of environmental effects \cite{Brito:2023pyl, Rahman:2025mip, Dyson:2025dlj, Li:2025ffh, Datta:2025ruh, Polcar:2025yto}. The perturbed NP Ricci scalars are related to the perturbed stress-energy tensors as follows:
\begin{subequations}
\begin{align}
    & \Phi_{ij}^{\eff(1,0)}
    \sim T^{\eff(1,0)}_{\mu\nu}e_{a}^{\mu}e_{b}^{\nu}\,, \\
    & \Phi_{ij}^{\eff(1,1)}
    \sim T^{\eff(1,1)}_{\mu\nu}e_{a}^{\mu}e_{b}^{\nu}
    +T^{\eff(1,0)}_{\mu\nu}e_{a}^{\mu(0,1)}e_{b}^{\nu}\,,
\end{align}    
\end{subequations}
where we have dropped the superscript $(0,0)$ of all $\mathcal{O}(\zeta^0,\epsilon^0)$ quantities for simplicity.
Thus, $\Phi_{ij}^{\eff(1,1)}$ and $\Phi_{ij}^{\eff(1,0)}$ only depends on fields or order below $\mathcal{O}(\zeta^1,\epsilon^1)$, which is a crucial feature enabling the decoupling of $\Psi_{0,4}^{(1,1)}$ from other geometrical quantities, as shown later.

Second, the point-particle stress-energy tensor $T_{\mu\nu}^{p}$ associated with the secondary in Eq.~\eqref{eq:stress_particle} can contribute to $\Phi^{(0,1)}_{ij}$ and $\Phi^{(1,1)}_{ij}$. Since $T_{\mu\nu}^{p}$ is proportional to the particle's mass $m_p$, it always enter at $\mathcal{O}(\zeta^0,\epsilon^1)$. Since the particle's four-velocity can be modified on a deformed background spacetime, one generally gets the following after the expansion in Eq.~\eqref{eq:expansion_metric}:
\begin{equation} \label{eq:stress_particle_expand}
    T_{\mu\nu}^{p(1,1)}\sim T_{\mu\alpha}^{p(0,1)}h_{\nu}^{\alpha(1,0)}\,,
\end{equation}
where $T_{\mu\nu}^{p(0,1)}$ is Eq.~\eqref{eq:stress_particle} with $m_p$ factored out and evaluated on a Kerr background. Since the perturbed NP Ricci scalars driven by $T_{\mu\nu}^{p}$ take the form
\begin{subequations} \label{eq:NP_Ricci_particle_expand}
\begin{align}
    & \Phi_{ij}^{p(0,1)}
    \sim T^{p(0,1)}_{\mu\nu}e_{a}^{\mu}e_{b}^{\nu}\,, \\
    & \Phi_{ij}^{p(1,1)}
    \sim T^{p(1,1)}_{\mu\nu}e_{a}^{\mu}e_{b}^{\nu}
    +T^{\eff(0,1)}_{\mu\nu}e_{a}^{\mu(1,0)}e_{b}^{\nu}\,,
\end{align}
\end{subequations}
both $\Phi_{ij}^{p(0,1)}$ and $\Phi_{ij}^{p(1,1)}$ do not contain any fields at or above $\mathcal{O}(\zeta^1,\epsilon^1)$ given Eq.~\eqref{eq:stress_particle_expand}.

To derive the modified Teukolsky equation, \cite{Li:2022pcy} made some convenient gauge choices for both the background spacetime and dynamical perturbations following Chandrasekhar \cite{Chandrasekhar_1983}. Since we care about BH spacetimes perturbatively deviating from Petrov type D spacetimes in GR, we can set 
\begin{equation} \label{eq:tetrad_choice_00}
    \Psi_{0,1,3,4}^{(0,0)}=0\,.
\end{equation}
For dynamical perturbations, \cite{Li:2022pcy} showed that one can rotate the $\mathcal{O}(\zeta^0,\epsilon^1)$ and $\mathcal{O}(\zeta^1,\epsilon^1)$ parts of the tetrad such that
\begin{equation} \label{eq:tetrad_choice_01_11}
    \Psi_{1,3}^{(0,1)}=\Psi_{1,3}^{(1,1)}=0\,.
\end{equation}
Other than these gauge choices, we make no further assumptions about the Weyl scalars.

Under the gauge above, one can then easily decouple $\Psi_{0}^{(1,1)}$ from other NP quantities and derive a single equation for $\Psi_{0}^{(1,1)}$ (and similarly for $\Psi_{4}^{(1,1)}$) following \cite{Li:2022pcy}. The set of NP equations in Eq.~\eqref{eq:Bianchi_simplified} for EMRIs does not fundamentally differ from the one in \cite{Li:2022pcy} for ringdown, except for the additional contribution of the secondary's stress-energy tensor $T_{\mu\nu}^{p}$ to the NP Ricci scalars $\Phi_{ij}$. Therefore, similar procedures in \cite{Li:2022pcy} can be carried out here by first solving $\kappa$ and $\sigma$ in terms of $\Psi_{0,1}$ and $S_{1,2}$ from Eqs.~\eqref{eq:BianchiId_Psi0_1_simplify} and \eqref{eq:BianchiId_Psi0_2_simplify} and then plugging $\kappa$ and $\sigma$ back to Eq.~\eqref{eq:RicciId_Psi0_simplify}. Performing the two-parameter expansion in Eq.~\eqref{eq:expansion_Weyl} on the resulting equation and applying the gauge choices in Eqs.~\eqref{eq:tetrad_choice_00} and \eqref{eq:tetrad_choice_01_11}, we get \cite{Li:2022pcy, Li:2025ffh}
\begin{align} \label{eq:master_eqn_non_typeD_Psi0}
    H_{0}^{(0,0)}\Psi_0^{(1,1)}
    =\mathcal{S}_{\geo}^{(1,1)}+\mathcal{S}_{\eff}^{(1,1)}
    +\mathcal{S}_{p}^{(1,1)}\,,
\end{align}
where the operator $H_{0}$ is defined as 
\begin{equation} \label{eq:def_operators}
    \begin{aligned}
        & H_0=\mathcal{E}_2F_2-\mathcal{E}_1F_1-3\Psi_2\,, \\
        & \mathcal{E}_1=E_1-\Psi_2^{-1}\delta\Psi_2\,,\quad
        \mathcal{E}_2=E_2-\Psi_2^{-1}D\Psi_2\,,
    \end{aligned}
\end{equation}
and the operators $E_{1,2}$ and $F_{1,2}$ are defined in Eq.~\eqref{eq:auxiliary_operators}. At $\mathcal{O}(\zeta^0,\epsilon^0)$, $H_{0}^{(0,0)}$ reduces to the Teukolsky operator acting on the perturbed Weyl scalar $\Psi_0^{(1)}$ in GR \cite{Teukolsky:1973ha}. As we show below, the source terms in Eq.~\eqref{eq:master_eqn_non_typeD_Psi0} involve no geometrical quantities at or beyond $\mathcal{O}(\zeta^1,\epsilon^1)$, so Eq.~\eqref{eq:master_eqn_non_typeD_Psi0} is manifestly decoupled and serves as the modified Teukolsky equation to solve in this work.

The source terms on the right-hand side of Eq.~\eqref{eq:master_eqn_non_typeD_Psi0} are divided into three groups. The first source term $\mathcal{S}_{\geo}^{(1,1)}$ takes the form:
\begin{align} \label{eq:S_geo}
    & \mathcal{S}_{\geo}^{(1,1)}
    =-H_0^{(1,0)}\Psi_0^{(0,1)}-H_0^{(0,1)}\Psi_0^{(1,0)}
    +H_1^{(0,1)}\Psi_1^{(1,0)}\,, \nonumber\\
    & H_1=\mathcal{E}_2J_2-\mathcal{E}_1J_1\,.
\end{align}
This source term is regarded as purely ``geometrical'' since it only depends on the corrections to the central BH's geometry [i.e., the terms at $\mathcal{O}(\zeta^1,\epsilon^0)$] and GWs in vacuum GR [i.e., the terms at $\mathcal{O}(\zeta^0,\epsilon^1)$]. For the former contribution, we choose to keep $h_{\mu\nu}$ generic in this work, so we will express all the terms at $\mathcal{O}(\zeta^1,\epsilon^0)$ in terms of a generic $h_{\mu\nu}$ in the final equation. For the latter contribution, we choose to directly solve the RW and ZM equations associated with a secondary in circular equatorial orbits in Sec.~\ref{sec:perturbed_NP} and express $q_{\mu\nu}$ in terms of the RW and ZM functions.

The second source term $\mathcal{S}_{\eff}^{(1,1)}$ is directly driven by the effective stress-energy tensor $T_{\mu\nu}^{\eff}$ due to bGR corrections or external matter fields, where 
\begin{equation} \label{eq:S_bGR}
\begin{aligned}
    \mathcal{S}_{\eff}^{(1,1)}
    =& \;\mathcal{E}_2^{(0,0)}S_{2}^{\eff(1,1)}+
    \mathcal{E}_2^{(0,1)}S_{2}^{\eff(1,0)}
    -\mathcal{E}_1^{(0,0)}S_{1}^{\eff(1,1)} \\
    & \;-\mathcal{E}_1^{(0,1)}S_{1}^{\eff(1,0)}\,.
\end{aligned}
\end{equation}
The terms $S_{i}^{\eff}$ ($i=1,2$) are defined analogously to $S_{i}$ in Eq.~\eqref{eq:source_bianchi}, using the $\Phi_{ij}^{\eff}$ associated with $T_{\mu\nu}^{\eff}$. Since $T_{\mu\nu}^{\eff}$ begins at $\mathcal{O}(\zeta^1)$, $S_{i}^{\eff}$ likewise start at $\mathcal{O}(\zeta^1)$. As discussed above, $\Phi_{ij}^{\eff(1,1)}$ will only involve fields at orders below $\mathcal{O}(\zeta^1,\epsilon^1)$, so $S_{i}^{\eff(1,1)}$ do not depend on any geometrical quantities at $\mathcal{O}(\zeta^1,\epsilon^1)$. In this work, we do not consider $T_{\mu\nu}^{\eff}$ of any specific bGR theory or matter model, so we will be generic about the contribution from $T_{\mu\nu}^{\eff(1,1)}$. Nonetheless, $T_{\mu\nu}^{\eff(1,0)}$ can be calculated from the background metric corrections $h_{\mu\nu}$.

In comparison, the last source term $\mathcal{S}_{p}^{(1,1)}$ is directly driven by the stress-energy tensor $T_{\mu\nu}^{p(0,1)}$ associated with the secondary,
\begin{equation} \label{eq:S_p}
\begin{aligned}
    \mathcal{S}_{p}^{(1,1)}
    =& \;\mathcal{E}_2^{(0,0)}S_{2}^{p(1,1)}+
    \mathcal{E}_2^{(1,0)}S_{2}^{p(0,1)}
    -\mathcal{E}_1^{(0,0)}S_{1}^{p(1,1)}\\
    &-\mathcal{E}_1^{(1,0)}S_{1}^{p(0,1)}\,,
\end{aligned}
\end{equation}
where $S_{i}^{p}$ ($i=1,2$) are defined analogously to $S_{i}$ in Eq.~\eqref{eq:source_bianchi} with $\Phi_{ij}$ replaced by those associated with $T_{\mu\nu}^{p}$, respectively. Since $T_{\mu\nu}^{p}$ begins at $\mathcal{O}(\epsilon^1)$, $S_{i}^{p}$ start at $\mathcal{O}(\epsilon^1)$. Although $\mathcal{E}_{i}^{(1,0)}$ and $S_{i}^{p(1,1)}$ depend on corrections to the background BH spacetime [i.e., terms at $\mathcal{O}(\zeta^1,\epsilon^0)$], they do not directly depend on the effective stress-energy tensor $T_{\mu\nu}^{\eff}$ that generates these corrections. For the same reason as above, $S_{i}^{p(1,1)}$ do not depend on any geometrical quantities at or above $\mathcal{O}(\zeta^1,\epsilon^1)$.

The equation of $\Psi_4^{(1,1)}$ can be obtained from Eq.~\eqref{eq:master_eqn_non_typeD_Psi0} by the GHP transformation \cite{Geroch:1973am},
\begin{align} \label{eq:master_eqn_non_typeD_Psi4}
    & H_{4}^{(0,0)}\Psi_4^{(1,1)}
    =\mathcal{T}_{\geo}^{(1,1)}+\mathcal{T}_{\eff}^{(1,1)}
    +\mathcal{T}_{p}^{(1,1)}\,,
\end{align}
where $H_{4}^{(0,0)}$ is the Teukolsky operator in GR for $\Psi_4$. The source terms are defined as
\begin{subequations} \label{eq:source_Psi4}
\begin{align}
    \mathcal{T}_{\geo}^{(1,1)}
    =& \;-H_4^{(1,0)}\Psi_4^{(0,1)}-H_4^{(0,1)}\Psi_4^{(1,0)}
    +H_3^{(0,1)}\Psi_3^{(1,0)}\,, 
    \label{eq:T_geo} \\
    \mathcal{T}_{\eff}^{(1,1)}
    =& \;\mathcal{E}_4^{(0,0)}S_{4,T}^{\eff(1,1)}+
    \mathcal{E}_4^{(0,1)}S_{4}^{\eff(1,0)}
    -\mathcal{E}_3^{(0,0)}S_{3}^{\eff(1,1)} \nonumber\\
    &\;-\mathcal{E}_3^{(0,1)}S_{3}^{\eff(1,0)}\,, 
    \label{eq:T_bGR} \\
    \mathcal{T}_{p}^{(1,1)}
    =& \;\mathcal{E}_4^{(0,0)}S_{4}^{p(1,1)}+
    \mathcal{E}_4^{(1,0)}S_{4}^{p(0,1)}
    -\mathcal{E}_3^{(0,0)}S_{3}^{p(1,1)} \nonumber\\
    &\;-\mathcal{E}_3^{(1,0)}S_{3}^{p(0,1)}\,,
    \label{eq:T_p}
\end{align}
\end{subequations}
where $S_{3,T}$ and $S_{3,p}$ are defined in the same way as $S_3$ below, and similarly for $S_{4,T}$ and $S_{4,p}$, 
\begin{subequations} \label{eq:s3s4}
\begin{align}
    S_3\equiv& \;-\mathbf{\Delta}_{[0,2,2,0]}\Phi_{21}
    +\bar{\delta}_{[2,2,0,-1]}\Phi_{22} \nonumber\\
    & \;+2\nu\Phi_{11}+\bar{\nu}\Phi_{20}-2\lambda\Phi_{12}\,, \\
    S_4\equiv& \;-\mathbf{\Delta}_{[0,1,2,-2]}\Phi_{20}
    +\bar{\delta}_{[2,0,0,-2]}\Phi_{21}  \nonumber\\
    & \;+2\nu\Phi_{10}-2\lambda\Phi_{11}+\bar{\sigma}\Phi_{22}\,,
\end{align}
\end{subequations}
with $\lambda$ and $\nu$ being spin coefficients. The operators $H_{3,4}$ and $\mathcal{E}_{3,4}$ are defined as 
\begin{equation} \label{eq:H_in_Teuk_ORG}
\begin{aligned}
    & H_4=\mathcal{E}_4F_4-\mathcal{E}_3F_3-3\Psi_2\,,\quad
    H_3=\mathcal{E}_4J_4-\mathcal{E}_3J_3\,, \\
    & \mathcal{E}_3=E_3-\Psi_2^{-1}\bar{\delta}\Psi_2\,,\quad
    \mathcal{E}_4=E_4-\Psi_2^{-1}\mathbf{\Delta}\Psi_2\,,
\end{aligned}
\end{equation}
with
\begin{align}
    & F_3\equiv\delta_{[0,4,0,-1]}\,,\quad
    && F_4\equiv D_{[4,0,-1,0]}\,, \nonumber\\
    & J_3\equiv\mathbf{\Delta}_{[4,0,2,0]}\,,\quad
    && J_4\equiv\bar{\delta}_{[2,0,4,0]}\,, \nonumber\\
    & E_3\equiv\bar{\delta}_{[3,1,1,-1]}\,,\quad 
    && E_4\equiv\mathbf{\Delta}_{[1,1,3,-1]}\,.
\end{align}
In the rest of this paper, we will compute all the source terms appearing in the equations of $\Psi_0^{(1,1)}$ [i.e., Eq.~\eqref{eq:master_eqn_non_typeD_Psi0}] and $\Psi_4^{(1,1)}$ [i.e., Eq.~\eqref{eq:master_eqn_non_typeD_Psi4}], respectively.

\section{Background NP Quantities in Vacuum GR}
\label{sec:bacground_NP}

In this section, we briefly review the coordinate and tetrad choices for the background spacetime in vacuum GR [i.e., $\mathcal{O}(\zeta^0,\epsilon^0)$], while a more comprehensive review can be found in \cite{Pound:2021qin}. For simplicity, we drop the order-counting superscript $(0,0)$ of quantities at $\mathcal{O}(\zeta^0,\epsilon^0)$. We are interested in EMRI evolutions around deformed Schwarzschild and Kerr BHs, the latter of which has the line element in Boyer-Lindquist coordinates \cite{Boyer:1966qh} $(t,r,\theta,\phi)$ as
\begin{align}\label{eq:KerrMetric}
    ds^2=& \;-\left(1-\frac{2Mr}{\Sigma}\right)dt^2
    -\frac{4aMr\sin^2\theta}{\Sigma}dtd\phi
    +\frac{\Sigma}{\Delta}dr^2 \nonumber\\
    & \;+\Sigma d\theta^2+\left[\Delta
    +\frac{2Mr(r^2+a^2)}{\Sigma}\right]\sin^2\theta d\phi^2\,,
\end{align}
where $\Delta=r^2-2Mr+a^2$, $\Sigma=r^2+a^2\cos^2{\theta}$, $M$ is the BH mass, and $a$ is the BH spin. In the nonrotating limit, the Kerr metric becomes the Schwarzschild metric, i.e.,
\begin{equation}
    ds^2=-f(r) dt^2+\frac{dr^2}{f(r)}
    +r^2\left(d\theta^2+\sin^2\theta d\phi^2\right)\,,
\end{equation}
where $f(r)=1-2M/r$.

When studying Kerr BHs using the NP formalism, the two most common choices of tetrad are the Kinnersley tetrad \cite{Kinnersley:1969zza} and the Carter tetrad \cite{Carter:1987hk}. The original Teukolsky equation in \cite{Teukolsky:1973ha} was derived using the Kinnersley tetrad such that the Teukolsky equation naturally separates into second-order ordinary differential equations for the radial and angular parts of $\Psi_{0,4}$, respectively. However, certain symmetries of the Kerr spacetime are not preserved by the Kinnersley tetrad but are respected by the Carter tetrad \cite{Pound:2021qin}. For instance, the Carter tetrad $\{l^{\mu},n^{\mu},m^{\mu},\bar{m}^{\mu}\}$ transforms as $l^{\mu}\to-n^{\mu}$ and $m^{\mu}\to\bar{m}^{\mu}$ under the map $\{t,\phi\}\to\{-t,-\phi\}$. Although all results in this work could be obtained in the Kinnersley tetrad, we adopt the Carter tetrad for the remainder of this work, as its symmetry facilitates sanity checks of the source terms. In Boyer–Lindquist coordinates, the Carter tetrad is given by
\begin{subequations} \label{eq:Carter_tetrad}
\begin{align}
    & l^{\mu}
    =\frac{1}{\sqrt{2\Delta\Sigma}}\left[r^2+a^2,\Delta,0,a\right]\,, \\
    & n^{\mu}
    =\frac{1}{\sqrt{2\Delta\Sigma}}\left[r^2+a^2,-\Delta,0,a\right]\,, \\
    & m^{\mu}
    =\frac{1}{\sqrt{2\Sigma}}\left[ia\sin\theta,0,1,i\csc\theta\right]\,, \\ & \bar{m}^{\mu}
    =\frac{1}{\sqrt{2\Sigma}}\left[-ia\sin\theta,0,1,-i\csc\theta\right]\,,
\end{align}
\end{subequations}
and is related to the Kinnersley tetrad $\{l_{K}^{\mu},n_{K}^{\mu},m_{K}^{\mu},\bar{m}_{K}^{\mu}\}$ by
\begin{align}
    & l_{K}^{\mu}=\sqrt{\frac{2\Sigma}{\Delta}}l^{\mu}\,,\;\;\quad
    n_{K}^{\mu}=\sqrt{\frac{\Delta}{2\Sigma}}n^{\mu}\,, \nonumber\\
    & m_{K}^{\mu}=\frac{\sqrt{\Sigma}}{\Gamma}m^{\mu}\,,\quad
    \bar{m}_{K}^{\mu}=\frac{\sqrt{\Sigma}}{\bar{\Gamma}}\bar{m}^{\mu}\,,
\end{align}
where $\Gamma=r+ia\cos{\theta}$. In the Schwarzschild limit, Eq.~\eqref{eq:Carter_tetrad} reduces to
\begin{subequations} \label{eq:Carter_tetrad_Schw}
 \begin{align}
    & l^{\mu}=\frac{1}{\sqrt{2f(r)}}\left(1,f(r),0,0\right)\,, \\
    & n^{\mu}=\frac{1}{\sqrt{2f(r)}}\left(1,-f(r),0,0\right)\,, \\
    & m^{\mu}=\frac{1}{\sqrt{2}r}\left(0,0,1,i\csc{\theta}\right)\,, \\
    & \bar{m}^{\mu}=\frac{1}{\sqrt{2}r}\left(0,0,1,-i\csc{\theta}\right)\,.
\end{align}
\end{subequations}
From the tetrad in Eq.~\eqref{eq:Carter_tetrad}, the spin coefficients in Sec.~\ref{sec:MTF} can then be computed, where
\begin{align}\label{eq:SpinCoeiffsBackground}
    & \rho=\mu
    =-\frac{1}{\bar{\Gamma}}\sqrt{\frac{\Delta}{2\Sigma}}\,,\quad
    \tau=-\pi
    =-\frac{ia\sin\theta}{\bar{\Gamma}\sqrt{2\Sigma}}\,, \nonumber\\
    & \beta=-\alpha
    =-\frac{i}{\bar{\Gamma}}
    \frac{a+ir\cos\theta}{2\sin\theta\sqrt{2\Sigma}}\,, \nonumber\\
    & \varepsilon=\gamma
    =\frac{Mr-a^2-ia(r-M)\cos\theta}{2\bar{\Gamma}\sqrt{2\Sigma\Delta}}\,,
\end{align}
and the other spin coefficients vanish on the Kerr background. In the Schwarzschild limit, the above expressions significantly simplify so that the only nonzero spin coefficients are
\begin{align}
    & \rho=\mu
    =-\sqrt{\frac{r-2M}{2r^3}}\,,\quad
    \beta=-\alpha
    =\frac{\cot\theta}{2\sqrt{2}r}\,, \nonumber\\
    & \varepsilon=\gamma
    =\frac{M}{2\sqrt{2r^3(r-2M)}}\,.
\end{align}

One convenient formulation of the NP formalism we will use in this work is the GHP formalism \cite{Geroch:1973am}. Let us provide a concise review of the GHP formalism, while more details can be found in \cite{Geroch:1973am, Pound:2021qin}. In the GHP formalism, each quantity is assigned a type $\{p,q\}$, from which its spin-weight $s=(p-q)/2$, associated with rotations of the tetrad vectors $\{m^{\mu},\bar{m}^{\mu}\}$, and boost-weight $b=(p+q)/2$, associated with boosts of the tetrad vectors $\{l^{\mu},n^{\mu}\}$, are determined. By performing rotations and boosts, one can determine that the tetrad basis vectors have the following GHP type:
\begin{equation}
    l^{\mu}:\{1,1\}\,,\;n^{\mu}:\{-1,-1\}\,,\; 
    m^{\mu}:\{1,-1\}\,,\;\bar{m}^{\mu}:\{-1,1\}\,.
\end{equation}
Similarly, one can find the GHP type of the following spin coefficients to be
\begin{equation}
    \kappa:\{3,1\}\,,\;\sigma:\{3,-1\}\,,\;
    \rho:\{1,1\}\,,\;\tau:\{1,-1\}\,.
\end{equation}
The GHP type of other spin coefficients is given by the prime transformation, under which $\{l^{\mu},m^{\mu}\}\leftrightarrow\{n^{\mu},\bar{m}^{\mu}\}$, so the GHP type $\{p,q\}$ of an object is transformed to $\{-p,-q\}$. For example, the spin coefficient $\kappa=-m^{\mu}Dl_{\mu}$ becomes $-\nu=-\bar{m}^{\mu}\Delta n_{\mu}$ under the prime transformation. The other spin coefficients are related by the prime transformation as follows:
\begin{equation} \label{eq:spin_coeff_prime}
    \{\kappa',\sigma',\rho',\tau',\varepsilon',\beta'\}
    =-\{\nu,\lambda,\mu,\pi,\gamma,\alpha\}\,,
\end{equation}
which tells us the GHP type of $\{\nu,\lambda,\mu,\pi\}$. Other useful GHP transformations can be found in \cite{Geroch:1973am, Pound:2021qin}.

The remaining spin coefficients $\{\varepsilon,\gamma,\alpha,\beta\}$ do not possess definite GHP type \cite{Geroch:1973am, Pound:2021qin}, but they only appear as parts of the GHP operators below:
\begin{align} \label{eq:GHP_operators}
    & \text{\TH}=D-p\varepsilon-q\bar{\varepsilon}\,,\quad
    \text{\TH}'=\boldsymbol{\Delta}-p\gamma-q\bar{\gamma}\,, \nonumber\\
    & \cpartial=\delta-p\beta-q\bar{\alpha}\,,\quad
    \cpartial'=\bar{\delta}-p\alpha-q\bar{\beta}\,,
\end{align}
where $p$ and $q$ are determined by the GHP type $\{p,q\}$ of the object acted on by these operators. $\text{\TH}$ and $\text{\TH}'$ are boost-weight raising and lowering operators, as they shift the GHP type of the target by $\{1,1\}$ and $\{-1,-1\}$, respectively. $\cpartial$ and $\cpartial'$ are spin-weight raising and lowering operators, as they shift the GHP type of the target by $\{1,-1\}$ and $\{-1,1\}$, respectively. In the following sections, we will track the boost-weight $b$ and spin-weight $s$ of the quantities on which these operators act via subscripts, i.e., $\text{\TH}_b,\text{\TH}_b',\cpartial_s$, and $\cpartial_s'$. The operators defined in Eq.~\eqref{eq:NP_operators} can be expressed in terms of the GHP operators together with the remaining spin coefficient, and can also be translated into the Chandrasekhar operators of \cite{Chandrasekhar_1983}, as shown in \cite{Ma:2024qcv, Li:2025ffh}. Finally, the product of two objects with GHP type $\{p_1,q_1\}$ and $\{p_2,q_2\}$ produces an object of GHP type $\{p_1+p_2,q_1+q_2\}$, so the GHP type of remaining NP quantities can be deduced by counting the number of occurrences of each tetrad basis vector in their definition. Since only objects of the same GHP type can be added together, this language provides us with a useful self-consistency check on many equations.

\section{Perturbed NP Quantities at $\mathcal{O}(\zeta^1,\epsilon^0)$ and $\mathcal{O}(\zeta^0,\epsilon^1)$}
\label{sec:perturbed_NP}

To compute the relevant source terms up to $\mathcal{O}(\zeta^1,\epsilon^1)$ in Eqs.~\eqref{eq:master_eqn_non_typeD_Psi0} and \eqref{eq:master_eqn_non_typeD_Psi4}, we need to calculate all the spin coefficients, directional derivatives, and Weyl scalars at $\mathcal{O}(\zeta^1,\epsilon^0)$ and $\mathcal{O}(\zeta^0,\epsilon^1)$. In this section, we first calculate the metric perturbations $h_{\mu\nu}$ and $q_{\mu\nu}$, from which we then derive the corresponding NP quantities at $\mathcal{O}(\zeta^1,\epsilon^0)$ and $\mathcal{O}(\zeta^0,\epsilon^1)$.

\subsection{Perturbed metric}
\label{sec:perturbed_metric}

In this subsection, we prescribe how to calculate the metric perturbations at $\mathcal{O}(\zeta^1,\epsilon^0)$ and $\mathcal{O}(\zeta^0,\epsilon^1)$. Since the background metric deformation $h_{\mu\nu}$ at $\mathcal{O}(\zeta^1,\epsilon^0)$ depends on the specific bGR theory or matter model, we keep it generic in this work, assuming only that it is axisymmetric and stationary. Deformed BHs have been studied in the nonrotating and slow-rotation limits in various bGR theories \cite{Yunes:2009hc, Yunes:2011we, Yagi:2012ya, Yunes:2011we, Cano:2019ore} and astrophysical environments \cite{Herdeiro:2014goa, Brito:2023pyl, Cardoso:2022whc, Duque:2023seg, Destounis:2022obl, Gliorio:2025cbh, Mitra:2025tag, Babichev:2012sg, Kimura:2021dsa, Campos:2025zag}. For general rotating BHs, analytical \cite{Petrich:1988zz} and numerical solutions \cite{Kleihaus:2015aje} were only found in a few bGR theories or environmental models, though recent advances in spectral and pseudospectral methods allow solutions in a broader class of bGR theories and astrophysical environments \cite{Fernandes:2025osu, Lam:2025elw}. The procedures developed in this work can be applied once these background BH solutions are provided.

For the metric perturbation $q_{\mu\nu}$ at $\mathcal{O}(\zeta^0,\epsilon^1)$, one needs to solve the linearized Einstein equation
\begin{align}\label{eq:ppEqu}
    G_{\mu\nu}^{1}[q_{\alpha\beta}]
    =8\pi T_{\mu\nu}^{p}\,,
\end{align}
where $G_{\mu\nu}^{1}$ is the linearized Einstein operator, and $T_{\mu\nu}^{p}$ is the stress-energy tensor of a point particle. For Schwarzschild BHs, Regge and Wheeler \cite{Regge:1957td}, Zerilli \cite{Zerilli:1970se}, and Vishveshwara \cite{Vishveshwara:1970cc} developed the procedures to solve Eq.~\eqref{eq:ppEqu} in the RW gauge, where the even-parity metric perturbation $q_{\mu\nu}^{e}$ and odd-parity metric perturbation $q_{\mu\nu}^{o}$ decouple, i.e.,
\begin{equation}
    q_{\mu\nu}=q_{\mu\nu}^{o}+q_{\mu\nu}^{e}\,.
\end{equation}
After conducting a harmonic expansion of $q_{\mu\nu}^{e/o}$, i.e.,
\begin{equation}
    q^{e/o}_{\mu\nu}
    =\sum_{\ell m}\int e^{-i\omega t+im\phi}q^{e/o,\ell m}_{\mu\nu}(r,\theta)\,d\omega.
\end{equation}
and decomposing $q^{e/o,\ell m}_{\mu\nu}$ into scalar, vector, and tensor harmonics (see Appendix~\ref{sec:MetricReconstruction}), one can then find a master function to completely characterize the radial parts of each parity: the ZM function $Q^{e}_{\ell m}(r)$ and the RW function $Q^{o}_{\ell m}(r)$ for even and odd parity, respectively. The master functions $Q^{e/o}_{\ell m}(r)$ are related to the radial parts of $q^{e/o,\ell m}_{\mu\nu}$ by
\begin{align}
    Q^{o}_{\ell m}(r) 
    &=\frac{f(r)}{r}h^{\ell m}_{1}(r)\,, \\
    Q^{e}_{\ell m}(r) 
    &=\frac{r^2}{\lambda r+3M}K^{\ell m}(r)
    +\frac{rf(r)}{i\omega(\lambda r+3M)}H_{1}^{\ell m}(r)\,,
\end{align}
where $\lambda=(\ell+2)(\ell-1)/2$. The functions $h^{\ell m}_{1}(r)$, $K^{\ell m}(r)$, and $H^{\ell m}_{1}(r)$ are radial parts of $q^{e/o,\ell m}_{\mu\nu}$ in Eqs.~\eqref{eq:qEven} and \eqref{eq:qOdd}. For $\ell\geq 2$, it was found that $Q^{e/o}_{\ell m}(r)$ satisfies a one-dimensional Schrodinger-like equation \cite{Sago:2002fe}:
\begin{align} \label{eq:RWequs}
    {\left[\frac{d^2}{dr_*^2}+\omega^2-V^{e/o}_{\ell m}(r)\right]
    Q^{e/o}_{\ell m}(r)}=S^{e/o}_{\ell m}(r)\,,
\end{align}  
where $r_{*}$ is the tortoise coordinate $r_{*}=r+2M \log\left(r/2M-1\right)$, and the radial potentials $V^{e/o}_{\ell m}(r)$ take the form:
\begin{subequations}
\begin{align}
    V^{o}_{\ell m}(r)
    &=f(r)\left(\frac{\ell(\ell+1)}{r^2}-\frac{6 M}{r^3}\right)\,,\\
    V^{e}_{\ell m}(r)
    &=\frac{f(r)}{r^2\Lambda^2}\left[2\lambda^2
    \left(\lambda+1+\frac{3M}{r}\right)
    +\frac{18M^2}{r^2}\left(\lambda+\frac{M}{r}\right)\right]\,,
\end{align}    
\end{subequations}
where $\Lambda=\lambda+3M/r$. For a point particle in a circular equatorial orbit, the source terms $S^{e/o}_{\ell m}(r)$ associated with $T^p_{\mu\nu}$ are found to be \cite{Sago:2002fe}:
\begin{subequations} \label{eq:RW_Sources}
\begin{align}
    S^{o}_{\ell m}(r)
    &=\mathfrak{c}_1\delta(r-r_p)+\mathfrak{c}_2\frac{d}{dr}\delta(r-r_p)\,,\\
    S^{e}_{\ell m}(r)
    &=\mathfrak{c}_3\delta(r-r_p)+\mathfrak{c}_4\frac{d}{dr}\delta(r-r_p)\,,
\end{align} 
\end{subequations}
where $r_p$ is the radial position of the particle. The coefficients $\mathfrak{c}_i$, $1\leq i\leq4$, relying on $\{\ell,m,\omega\}$ are provided in Appendix~\ref{sec:MetricReconstruction} [i.e., Eq.~\eqref{eq:RW_SourcesCoeffs}].

For each $(\ell,m)$ mode, Eq.~\eqref{eq:RWequs} can be solved using a Green's function constructed from two independent homogeneous solutions with the following asymptotic behaviors:
\begin{subequations}
\begin{align}
    & Q_{\ell m}^{e/o,\mathrm{in}}(r_{*}\to-\infty)= e^{-i\omega r_{*}}\,,\\
    & Q_{\ell m}^{e/o,\mathrm{up}}(r_{*}\to\infty)= e^{i\omega r_{*}}\,.
\end{align}    
\end{subequations}
In this case, the inhomogeneous solutions to Eq.~\eqref{eq:RW_Sources} take the form 
\begin{equation}
    Q_{\ell m}^{e/o}(r)
    =c_{\ell m}^{e/o,\mathrm{in}}(r)Q_{\ell m}^{e/o,\mathrm{in}}(r)
    +c_{\ell m}^{e/o,\mathrm{up}}(r)Q_{\ell m}^{e/o,\mathrm{up}}(r)\,, 
\end{equation}
where  
\begin{align}
    c_{\ell m}^{e/o,\mathrm{in}}(r) 
    &= \frac{1}{W_{e/o}}\int_{r}^{\infty}\,dr'
    \frac{Q_{\ell m}^{e/o,\mathrm{up}}(r')S^{e/o}_{\ell m}(r')}{f(r')}\,,\\
    c_{\ell m}^{e/o,\mathrm{up}}(r) 
    &= \frac{1}{W_{e/o}}\int_{2M}^{r}\,dr'
    \frac{Q_{\ell m}^{e/o,\mathrm{in}}(r')S^{e/o}_{\ell m}(r')}{f(r')}\,,
\end{align}
and $W_{e/o}$ is the Wronskian, i.e.,
\begin{equation}
    W_{e/o}=
    Q_{\ell m}^{e/o,\mathrm{in}}(r)
    \partial_{r_*}Q_{\ell m}^{e/o,\mathrm{up}}(r)
    -Q_{\ell m}^{e/o,\mathrm{up}}(r)
    \partial_{r_*}Q_{\ell m}^{e/o,\mathrm{in}}(r)\,.
\end{equation}
The above approach was later formulated by Moncrief in a gauge-invariant way using a Hamiltonian description \cite{Moncrief:1974am}, from which Poisson and Martel found covariant master equations with covariant sources for each parity \cite{Martel:2005ir}.
For more generic orbits, one can use the method of extended homogeneous solutions to include eccentricity \cite{Hopper:2010uv, Barack:2008ms} and add higher multipole moments in the small mass-ratio expansion of the particle's stress-energy tensor \cite{Mathisson:1937zz} for a spinning secondary \cite{Mathews:2021rod}. 

Calculating metric perturbations of Kerr BHs is more challenging. Unlike the case for Schwarzschild spacetime, one cannot directly decouple the equations of different metric components and separate their radial and angular parts. The most widely used alternative is to solve the Teukolsky equation for the curvature perturbations characterized by the Weyl scalars $\Psi^{(1)}_0$ or $\Psi^{(1)}_4$ and reconstruct the corresponding metric perturbations. For vacuum perturbations, one main metric reconstruction method is the one developed by Chrzanowski, Cohen, and Kegeles (CCK) \cite{Chrzanowski:1975wv, Kegeles:1979an, Ori:2002uv}. Nonetheless, this approach is only valid in a radiation gauge \cite{Price:2006ke}, which restricts the structure of source terms that can appear for sourced perturbations. In particular, a radiation gauge is not generically allowed for the stress-energy tensor of a point particle \cite{Toomani:2021jlo}. Some progress has been made by matching the vacuum perturbations inside and outside the particle's orbit for circular equatorial orbits \cite{Keidl:2010pm, Shah:2010bi, Pound:2013faa, Merlin:2016boc} and precessing inclined orbits \cite{vandeMeent:2016pee, vanDeMeent:2017oet, Nasipak:2025tby}. However, the solution generated by this approach is usually singular along null rays extending from the particle's worldline to the horizon or infinity \cite{Green:2019nam}. Motivated to construct a more regular solution necessary for studying nonlinear contributions in self-force theory, several reconstruction techniques have been developed over the past few years, such as the Green, Hollands, and Zimmerman's extension of the CCK method to sourced perturbations via a corrector tensor \cite{Green:2019nam, Toomani:2021jlo, Bourg:2024vre}, the Lorenz gauge reconstruction method developed by \cite{Dolan:2021ijg, Dolan:2023enf} for point particles, and the extension of the latter to generic sources \cite{Wardell:2024yoi} based on the formalism by Aksteiner, Anderson, and Backd{\"a}hl \cite{Aksteiner:2016pjt, Nasipak:2025tby}. In our follow-up work, we will apply one of these techniques to study the EMRI evolution of a deformed Kerr BH.

\subsection{Perturbed NP quantities}

Given the perturbed metric $q_{\mu\nu}$ driven by the secondary calculated above and a generic background metric perturbation $h_{\mu\nu}$, we calculate the perturbed NP quantities at $\mathcal{O}(\zeta^1,\epsilon^0)$ and $\mathcal{O}(\zeta^0,\epsilon^1)$ in this subsection. The procedures to derive the spin coefficients, directional derivatives, and Weyl scalars at $\mathcal{O}(\zeta^1,\epsilon^0)$ and $\mathcal{O}(\zeta^0,\epsilon^1)$ are equivalent to a large extent, except for certain tetrad gauge chosen at $\mathcal{O}(\zeta^0,\epsilon^1)$, so we will use the quantities at $\mathcal{O}(\zeta^1,\epsilon^0)$ as a demonstration. All the results in this subsection are valid for rotating BHs in general. 

To compute the perturbed NP quantities, we follow the approach presented in \cite{Chandrasekhar_1983, Campanelli:1998jv, Wagle:2023fwl}. In the NP formalism, the null tetrad $e_{a}^{\mu}=\{l^{\mu},n^{\mu},m^{\mu},\bar{m}^{\mu}\}$ for a metric $g_{\mu\nu}$ satisfies
\begin{align} \label{eq:normalization}
    g_{\mu\nu}e^{\mu}_{a}e^{\nu}_{b}=\eta_{ab},
\end{align}
where $\eta_{ab}$ is the NP normalization tensor such that $l^{\mu}n_{\mu}=-1$ and $m^{\mu}\bar{m}_{\mu}=1$. Expanding Eq.~\eqref{eq:normalization} to $\mathcal{O}(\zeta^1,\epsilon^0)$, one gets \cite{Li:2022pcy}\footnote{We have added a minus sign to Eq.~(51) of \cite{Li:2022pcy} as the expansion of $g^{\mu\nu}$ in Eq.~(48) should be $g^{\mu\nu(0,0)}-\zeta h^{\mu\nu(1,0)}$ instead in the standard convention.}
\begin{equation} \label{eq:normalization_10}
    e_{a}^{\mu(1,0)}=A_{ab}^{(1,0)}e^{b\mu}\,,\;
    A_{(ab)}^{(1,0)}=\frac{1}{2}e_{a}^{\mu}e_{b}^{\nu}h_{\mu\nu}^{(1,0)}\,,
\end{equation}
where $A_{(ab)}=(A_{ab}+A_{ba})/2$ is the symmetric part of $A_{ab}$, and we have dropped the superscript of quantities at $\mathcal{O}(\zeta^0,\epsilon^0)$. Using tetrad rotations at $\mathcal{O}(\zeta^1,\epsilon^0)$, one can set the antisymmetric part $A_{[ab]}^{(1,0)}=0$, where $A_{[ab]}=(A_{ab}-A_{ba})/2$ \cite{Li:2022pcy}. In this case, we get 
\begin{equation} \label{eq:perturbed_tetrad}
\begin{aligned}
    l^{\mu(1,0)}
    &=-\frac{1}{2}\left(h_{ln}l^{\mu}+h_{ll}n^{\mu}
    -h_{l\bar{m}}m^{\mu}-h_{lm}\bar{m}^{\mu}\right)\,, \\
    n^{\mu(1,0)}
    &=-\frac{1}{2}\left(h_{nn}l^{\mu}+h_{ln}n^{\mu}
    -h_{n\bar{m}}m^{\mu}-h_{nm}\bar{m}^{\mu}\right)\,, \\
    m^{\mu(1,0)}
    &=-\frac{1}{2}\left(h_{nm}l^{\mu}+h_{lm}n^{\mu}
    -h_{m\bar{m}} m^{\mu}-h_{mm}\bar{m}^{\mu}\right)\,, \\
    \bar{m}^{\mu(1,0)}
    &=-\frac{1}{2}\left(h_{n\bar{m}}l^{\mu}+h_{l\bar{m}}n^{\mu}
    -h_{\bar{m}\bar{m}}m^{\mu}-h_{m\bar{m}}\bar{m}^{\mu}\right)\,.
\end{aligned}
\end{equation}
One advantage of the perturbed tetrad in Eq.~\eqref{eq:perturbed_tetrad} is that $\{l^{\mu},m^{\mu}\}\leftrightarrow\{n^{\mu},\bar{m}^{\mu}\}$ under the prime transformation up to $\mathcal{O}(\zeta^1,\epsilon^0)$. In this case, it provides a direct self-consistency check of the perturbed quantities which are prime transformations of each other [i.e., the spin coefficients in Eq.~\eqref{eq:spin_coeff_prime}]. For this reason, we also make the same choice for tetrad perturbations at $\mathcal{O}(\zeta^0,\epsilon^1)$ by replacing $h_{ab}$ with $q_{ab}$ in Eq.~\eqref{eq:spin_coeff_prime}.

To solve for the spin coefficients, one can linearize the commutation relations given by
\begin{align} \label{eq:NPScalarCommutation}
    \left[e_a^\mu,e_b^\mu\right]
    =\left(\gamma_{ba}^c-\gamma_{ab}^c\right)e_c^\mu
    \equiv C_{ab}{}^c e_c^\mu\,,
\end{align}
where $\gamma^{a}_{bc}$ are Ricci rotation coefficients and can be rewritten in terms of spin coefficients \cite{Chandrasekhar_1983}. Expanding Eq.~\eqref{eq:NPScalarCommutation} to $\mathcal{O}(\zeta^1,\epsilon^0)$, one gets \cite{Chandrasekhar_1983, Wagle:2023fwl}
\begin{equation} \label{eq:C_expand}
    \begin{aligned}
        C_{ab}{}^{c(1,0)}
        =& \;\partial_{a}A_{b}{}^{c(1,0)}
        -\partial_{b}A_{a}{}^{c(1,0)} \\
        &\;-\left(A_{a}{}^{d(1,0)}C_{bd}{}^{c}
        -A_{b}{}^{d(1,0)}C_{ad}{}^{c}
        +A_{d}{}^{c(1,0)}C_{ab}{}^{d}\right)\,,  
    \end{aligned}
\end{equation}
where the coefficients $A_{ab}^{(1,0)}$ are defined in Eq.~\eqref{eq:normalization_10} and can be read off from Eq.~\eqref{eq:perturbed_tetrad}. The relation between structure constants $C_{ab}{}^{c}$ and spin coefficients can be found in \cite{Chandrasekhar_1983} and Appendix B of \cite{Wagle:2023fwl}, from which one can calculate $C_{ab}{}^{c(0,0)}$ from the spin coefficients on the Kerr background and solve the spin coefficients at $\mathcal{O}(\zeta^1,\epsilon^0)$ in terms of $C_{ab}{}^{c(1,0)}$. Using Eqs.~\eqref{eq:perturbed_tetrad} and \eqref{eq:C_expand}, one can then express the spin coefficients at $\mathcal{O}(\zeta^1,\epsilon^0)$ in terms of $h_{ab}$, e.g.,
\begin{equation}
    \kappa^{(1,0)} 
    =-\frac{1}{2}(\text{\TH}_{1}-\rho )h_{lm}+\frac{1}{2}\cpartial_{0}h_{ll}\,,
\end{equation}
and the remaining spin coefficients are presented in Eq.~\eqref{eq:perturbedSpinCoeffs}. Here, we have used Eq.~\eqref{eq:GHP_operators} to replace directional derivatives and certain spin coefficients on the Kerr background with GHP operators. As a self-consistency check, one can see that the spin coefficients at $\mathcal{O}(\zeta^0,\epsilon^1)$ in Eq.~\eqref{eq:perturbedSpinCoeffs} satisfy the same relation in Eq.~\eqref{eq:spin_coeff_prime} under the prime transformation, as imposed by the tetrad chosen in Eq.~\eqref{eq:perturbed_tetrad}. 

The perturbed Weyl scalars $\Psi_{i}^{(1,0)}$ and NP Ricci scalars $\Phi_{ij}^{(1,0)}$ can be calculated from the perturbed spin coefficients found in the previous step by linearizing the Ricci identities, the latter of which are provided in \cite{Newman:1961qr, Chandrasekhar_1983} and Appendix A of \cite{Li:2022pcy}. As an example, one of the Ricci identities gives that
\begin{equation} \label{eq:psi0_Ricci}
    \Psi_0=D_{[-3,1,-1,-1]}\sigma-\delta_{[-1,-3,1,-1]}\kappa\,.
\end{equation}
Since $\sigma^{(0,0)}=\kappa^{(0,0)}=0$ for Kerr BHs, the linearization of Eq.~\eqref{eq:psi0_Ricci} gives
\begin{equation}
    \Psi_0^{(1,0)}=D_{[-3,1,-1,-1]}\sigma^{(1,0)}-\delta_{[-1,-3,1,-1]}\kappa^{(1,0)}\,.
\end{equation}
Finally, plugging Eqs.~\eqref{eq:kappa_10} and \eqref{eq:sigma_10} for $\kappa^{(1,0)}$ and $\sigma^{(1,0)}$, respectively, and replacing the directional derivatives at $\mathcal{O}(\zeta^0,\epsilon^0)$ by GHP operators according to Eq.~\eqref{eq:GHP_operators}, one gets
\begin{align}
    \Psi^{(1,0)}_0 
    =& \;-\frac{1}{2}\cpartial_{1}\cpartial_{0}h_{ll}
    +(\cpartial_{1}\text{\TH}_{1}-\rho\cpartial_{1})h_{lm} \nonumber\\
    & \;-\frac{1}{2}(\text{\TH}_{1}\text{\TH}_{0}-2\rho\text{\TH}_{0})h_{mm}\,.
\end{align}
Other Weyl scalars $\Psi_{i}^{(1,0)}$ found via this approach are presented in Eq.~\eqref{eq:perturbedWeylScalars} in terms of the background metric deformation $h_{ab}$. The NP Ricci scalars $\Phi_{ij}^{(1,0)}$ can also be calculated in the same way, with the results in terms of $h_{ab}$ listed in Eq.~\eqref{eq:Ricci10}. For the NP Ricci scalars $\Phi_{ij}^{(0,1)}$, one can directly project the corresponding Ricci tensor of $T_{\mu\nu}^{p(0,1)}$ in Eqs.~\eqref{eq:stress_particle} and \eqref{eq:stress_particle_circular} onto the NP basis. The results of $\Phi_{ij}^{(0,1)}$ are provided in Eq.~\eqref{eq:Ricci01}.

All the steps above to compute the spin coefficients and Weyl scalars at $\mathcal{O}(\zeta^1,\epsilon^0)$ are the same for the corresponding terms at $\mathcal{O}(\zeta^0,\epsilon^1)$, except for the additional tetrad choice in Eq.~\eqref{eq:tetrad_choice_01_11} setting $\Psi_{1,3}^{(0,1)}=0$. For this reason, we need to perform additional tetrad rotations at $\mathcal{O}(\zeta^0,\epsilon^1)$ \cite{Li:2022pcy} 
\begin{align} \label{eq:tetrad_rotations_10}
    & l^{\mu(0,1)}\rightarrow 
    l^{\mu(0,1)}+\bar{b}^{(0,1)}m^{\mu}+b^{(0,1)}\bar{m}^{\mu}\,, \nonumber\\
    & n^{\mu(0,1)}\rightarrow 
    n^{\mu(0,1)}+\bar{a}^{(0,1)}m^{\mu}+a^{(0,1)}\bar{m}^{\mu}\,, \nonumber\\
    & m^{\mu(0,1)}\rightarrow
    m^{\mu(0,1)}+a^{(0,1)}l^{\mu}+b^{(0,1)}n^{\mu}\,,
\end{align}		
where we have linearized and combined the type I and type II tetrad rotations in \cite{Chandrasekhar_1983}, given that the rotation parameters $a^{(1,0)}$ and $b^{(1,0)}$ are perturbative. Under the rotations in Eq.~\eqref{eq:tetrad_rotations_10}, the Weyl scalars $\Psi_i^{(0,1)}$ transform in the following way:
\begin{align}\label{eq:WeylScalarRotation}
    & \Psi^{(0,1)}_1\to\Psi^{(0,1)}_1+3b^{(0,1)}\Psi_2\,, \nonumber\\
    & \Psi^{(0,1)}_3\to\Psi^{(0,1)}_3+3\bar{a}^{(0,1)}\Psi_2\,,
\end{align}
and $\Psi_{0,2,4}^{(0,1)}$ remain unchanged. Thus, to remove $\Psi_{1,3}^{(0,1)}$, one can choose
\begin{align} \label{eq:TetradRotCoeffs}
    a^{(0,1)}=-\frac{\bar{\Psi}_3^{(0,1)}}{3\bar{\Psi}_2}\,,\quad
    b^{(0,1)}=-\frac{\Psi_1^{(0,1)}}{3\Psi_2}\,.
\end{align}
The transformation of the spin coefficients at $\mathcal{O}(\zeta^0,\epsilon^1)$ is given in Eq.~\eqref{eq:spin_coeffs_rotated}. Thus, to get the spin coefficients and Weyl scalars at $\mathcal{O}(\zeta^0,\epsilon^1)$, we first compute these quantities before any tetrad rotations by replacing every $h_{ab}$ in Eqs.~\eqref{eq:perturbedSpinCoeffs} and \eqref{eq:perturbedWeylScalars} with the corresponding $q_{ab}$, as we have chosen $e_{a}^{\mu(0,1)}$ to be in the same form as Eq.~\eqref{eq:perturbed_tetrad}. The resulting $\Psi_{1,3}^{(0,1)}$ give us the rotation parameters $a^{(0,1)}$ and $b^{(0,1)}$. We can then add the additional terms in Eq.~\eqref{eq:spin_coeffs_rotated} due to the tetrad rotations in Eqs.~\eqref{eq:tetrad_rotations_10} and \eqref{eq:TetradRotCoeffs} to the spin coefficients obtained in the previous step. The Weyl scalars $\Psi_{0,2,4}^{(0,1)}$ are still given by Eq.~\eqref{eq:perturbedWeylScalars} after replacing $h_{ab}$ with $q_{ab}$, while $\Psi_{1,3}^{(0,1)}$ are set to zero. The same gauge choice in Eq.~\eqref{eq:tetrad_choice_01_11} can be also made at $\mathcal{O}(\zeta^1,\epsilon^0)$ (i.e., set $\Psi_{1,3}^{(1,0)}=0$) by choosing the same $a^{(1,0)}$ and $b^{(1,0)}$ with $\Psi_{1,3}^{(0,1)}$ replaced by $\Psi_{1,3}^{(1,0)}$ in Eq.~\eqref{eq:TetradRotCoeffs}. In this work, we choose to follow the convention in \cite{Li:2022pcy}, so this gauge choice at $\mathcal{O}(\zeta^1,\epsilon^0)$ is not made.

\section{Constructing the source terms}
\label{sec:construct_source}

In this section, we compute the source terms in Eqs.~\eqref{eq:master_eqn_non_typeD_Psi0} and \eqref{eq:master_eqn_non_typeD_Psi4}. We will compute $\mathcal{S}_{\eff}^{(1,1)}$ in Eq.~\eqref{eq:S_bGR} as an example, while the other source terms can be derived similarly. In this work, we do not fix the form of $T^{\eff(1,1)}_{\mu\nu}$ in Eq.~\eqref{eq:EE} for all the source terms, as they have to be studied case by case. This also allows us to study how parametric deformations of GR BHs affect EMRIs \cite{Vigeland:2009pr, Vigeland:2011ji, McManus:2019ulj, Weller:2024qvo}. Our final results are organized by the dependence on directional derivatives at $\mathcal{O}(\zeta^0,\epsilon^0)$ and tabulated in Appendix~\ref{sec:SourceCoeffs}.

Notably, $\mathcal{S}_{\eff}^{(1,1)}$ has contributions from both $\Phi^{\eff(1,0)}_{ij}$ and $\Phi^{\eff(1,1)}_{ij}$, which we will label as type A and type B, respectively. For example, for $S_1^{\eff(1,1)}$ and $S_2^{\eff(1,1)}$ in Eq.~\eqref{eq:S_bGR}, we divide them into these two groups:
\begin{subequations}
\begin{align}
    S^{\eff(1,1)}_{1A}
    \equiv& \;\delta^{(0,1)}_{[-2,-2,1,0]}\Phi^{\eff(1,0)}_{00}
    -D^{(0,1)}_{[-2,0,0,-2]}\Phi^{\eff(1,0)}_{01} \nonumber\\
    & \;+2\sigma^{(0,1)}\Phi^{\eff(1,0)}_{10}
    -2\kappa^{(0,1)}\Phi^{\eff(1,0)}_{11} \nonumber\\
    & \;-\bar{\kappa}^{(0,1)}\Phi^{\eff(1,0)}_{02}\,, \\
    S^{\eff(1,1)}_{1B}
    \equiv& \;\delta_{[-2,-2,1,0]}\Phi^{\eff(1,1)}_{00}
    -D_{[-2,0,0,-2]}\Phi^{\eff(1,1)}_{01}\,, \\
    S^{\eff(1,1)}_{2A}
    \equiv& \;\delta^{(0,1)}_{[0,-2,2,0]}\Phi^{\eff(1,0)}_{01}
    -D^{(0,1)}_{[-2,2,0,-1]}\Phi^{\eff(1,0)}_{02} \nonumber\\
    & \;-\bar{\lambda}^{(0,1)}\Phi^{\eff(1,0)}_{00}
    +2\sigma^{(0,1)}\Phi^{\eff(1,0)}_{11} \nonumber\\
    & \;-2\kappa^{(0,1)}\Phi^{\eff(1,0)}_{12}\,,\\
    S^{\eff(1,1)}_{2B}
    \equiv& \;\delta_{[0,-2,2,0]}\Phi^{\eff(1,1)}_{01}
    -D_{[-2,2,0,-1]}\Phi^{\eff(1,1)}_{02}\,,
\end{align}    
\end{subequations}
where we have dropped the superscript $(0,0)$ for quantities at $\mathcal{O}(\zeta^0,\epsilon^0)$ and used that $\lambda=\sigma=\kappa=0$ at $\mathcal{O}(\zeta^0,\epsilon^0)$. We can similarly define the components $\mathcal{S}^{\eff(1,1)}_{A}$ and $\mathcal{S}^{\eff(1,1)}_{B}$, where 
\begin{subequations}
\begin{align}
    \mathcal{S}^{(1,1)}_\eff
    =& \;\mathcal{S}_{A}^{\eff(1,1)}+\mathcal{S}_{B}^{\eff(1,1)}\,, \\
    \mathcal{S}^{\eff(1,1)}_{A}
    =& \;\mathcal{E}_2^{(0,0)}S_{2A}^{\eff(1,1)}
    +\mathcal{E}_2^{(0,1)}S_{2}^{\eff(1,0)}
    -\mathcal{E}_1^{(0,0)}S_{1A}^{\eff(1,1)} \nonumber\\
    & \;-\mathcal{E}_1^{(0,1)}S_{1}^{\eff(1,0)}\,, \\
    \mathcal{S}^{\eff(1,1)}_{B}
    =& \;\mathcal{E}_2^{(0,0)}S_{2B}^{\eff(1,1)}
    -\mathcal{E}_1^{(0,0)}S_{1B}^{\eff(1,1)}\,.
\end{align}
\end{subequations}
To present the source terms explicitly, we expand the perturbed NP Ricci scalars in terms of the background metric deviation $h_{\mu\nu}$. Let us take $\Phi_{00}$ as an example, which is defined as
\begin{align}
    \Phi_{00}=\frac{1}{2}R_{\mu\nu}l^{\mu}l^{\nu}\,.
\end{align}
After perturbations, one then gets
\begin{subequations}
\begin{align}
    \Phi^{\eff(1,0)}_{00} 
    &=\frac{1}{2}R^{\eff(1,0)}_{\mu\nu}l^{\mu}l^{\nu}\,, \label{eq:pR10}\\
    \Phi^{\eff(1,1)}_{00} 
    &=R^{\eff(1,0)}_{\mu\nu}l^{\mu(0,1)}l^{\nu}
    +\frac{1}{2}R^{\eff(1,1)}_{\mu\nu}l^{\mu}l^{\nu}\,, \label{eq:pR11}
\end{align}    
\end{subequations}
where $R^{\eff}_{\mu\nu}$ refers to the Ricci tensor associated with $T_{\mu\nu}^{\eff}$ in Eq.~\eqref{eq:EE}. Using a tetrad at $\mathcal{O}(\zeta^0,\epsilon^1)$ similar to Eq.~\eqref{eq:perturbed_tetrad}, we can rewrite Eq.~\eqref{eq:pR11} as
\begin{align}
   \Phi^{\eff(1,1)}_{00}
   =& \;q_{ln}\Phi^{\eff(1,0)}_{00}+q_{l\bar{m}}\Phi^{\eff(1,0)}_{01}
   +q_{lm}\Phi^{\eff(1,0)}_{10}\nonumber\\
   & \;q_{ll}\left(\Phi^{\eff(1,0)}_{11}+3\Lambda^{\eff(1,0)}\right)
   +\frac{1}{2}R^{\eff(1,1)}_{\mu\nu}l^{\mu}l^{\nu}\,.
\end{align}
The results of other NP Ricci scalars at $\mathcal{O}(\zeta^1,\epsilon^1)$ are provided in Appendix~\ref{sec:RicciSourceTerms11}. This allows us to write the source term $S^{\eff(1,1)}_{1B}$ as an $\mathcal{O}(\zeta^0,\epsilon^1)$ differential operator acting on $\Phi^{\eff(1,0)}_{ij}$, i.e.,
\begin{align} \label{eq:STR11}
    S^{\eff(1,1)}_{1B}
    =\sum_{ij}{}_{1B}\mathcal{D}^{\eff(0,1)}_{ij}\Phi^{\eff(1,0)}_{ij}\,,
\end{align}  
where $i,j\in\{0,1,2\}$, and ${}_{1B}\mathcal{D}_{ij}^{\eff(0,1)}$ is a differential operator of first order. Similarly, one finds $\mathcal{S}^{\eff(1,1)}_{A}$ and $\mathcal{S}^{\eff(1,1)}_{B}$ to be in the form:
\begin{subequations}
\begin{align}
    \mathcal{S}^{\eff(1,1)}_{A} 
    &=\sum_{ij}{}_{A}\mathscr{D}^{\eff(0,1)}_{ij}\Phi^{(1,0)}_{ij},\\
    \mathcal{S}^{\eff(1,1)}_{B} 
    &=\sum_{ij}{}_{B}\mathscr{D}^{\eff(0,1)}_{ij}\Phi^{(1,0)}_{ij},
\end{align}
\end{subequations}
where ${}_{A}\mathscr{D}^{\eff(0,1)}_{ij}$ and ${}_{B}\mathscr{D}^{\eff(0,1)}_{ij}$ are differential operators of second order. Using Eq.~\eqref{eq:perturbed_tetrad} again, we can write directional derivatives at $\mathcal{O}(\zeta^0,\epsilon^1)$ in terms of the ones at $\mathcal{O}(\zeta^0,\epsilon^0)$, i.e.,
\begin{subequations} \label{eq:S_bGRExpansion}
\begin{align}
    \mathcal{S}^{\eff(1,1)}_{A} 
    &=-\sum_{ijkl}{}_{A}\mathcal{X}^{\eff(0,1)}_{ijkl}\mathcal{O}^{k}\mathcal{O}^{l}
    \Phi^{(1,0)}_{ij}\,, \label{eq:S_bGRAExpansion} \\
     \mathcal{S}^{\eff(1,1)}_{B} 
    &=-\sum_{ijkl}{}_{B}\mathcal{X}^{\eff(0,1)}_{ijkl}\mathcal{O}^{k}\mathcal{O}^{l}
    \Phi^{(1,0)}_{ij}\,, \label{eq:S_bGRBExpansion}
\end{align}    
\end{subequations}
where $\mathcal{O}^{k}$ is a directional derivative at $\mathcal{O}(\zeta^0,\epsilon^0)$ in the following set $\mathfrak{D} \equiv \{1,\text{\TH},\text{\TH '},\cpartial,\cpartial'\}$, with $k$ and $l$ being indices in $\mathfrak{D}$, while $i$ and $j$ are in the set $\{0,1,2\}$. The coefficients ${}_{A}\mathcal{X}^{\eff(0,1)}_{ijkl}$ and ${}_{B}\mathcal{X}^{\eff(0,1)}_{ijkl}$ are provided in Tables~\ref{tab:typeAS_bGR} and \ref{tab:typeBS_bGR} of Appendix~\ref{sec:PSI0SOURCES_NOND}, respectively.\footnote{We have added a minus sign to the definition of the coefficients ${}_{A}\mathcal{X}^{\eff(0,1)}_{ijkl}$ and ${}_{B}\mathcal{X}^{\eff(0,1)}_{ijkl}$ in Eq.~\eqref{eq:S_bGRExpansion} [and similarly in Eqs.~\eqref{eq:S_pGRAExpansion} and \eqref{eq:S_geoExpansion}] because they were originally computed in a perturbed tetrad with the opposite sign convention from Eq.~\eqref{eq:perturbed_tetrad}, following \cite{Li:2022pcy}.} We can similarly argue that the source terms $\mathcal{T}^{\eff(1,1)}_{A}$ and $\mathcal{T}^{\eff(1,1)}_{B}$ share the same expansion of Eq.~\eqref{eq:S_bGRExpansion}. To obtain the coefficients for $\mathcal{T}^{\eff(1,1)}_{A}$ and $\mathcal{T}^{\eff(1,1)}_{B}$, one can apply the prime transformation (discussed in Sec.~\ref{sec:bacground_NP}) to ${}_{A}\mathcal{X}^{\eff(0,1)}_{ijkl}$ and ${}_{B}\mathcal{X}^{\eff(0,1)}_{ijkl}$, the results of which are given in Tables~\ref{tab:typeAT_bGR} and \ref{tab:typeBT_bGR} of Appendix~\ref{sec:PSI4SOURCES_NOND}, respectively. 

By the same reasoning above, $\mathcal{S}^{(1,1)}_p$ and $\mathcal{T}^{(1,1)}_p$ have both type A and type B contributions driven by $R^{p(0,1)}_{\mu\nu}$ associated with $T^{p(0,1)}_{\mu\nu}$ in Eqs.~\eqref{eq:stress_particle} and \eqref{eq:stress_particle_circular}. These contributions share the same expansion in Eq.~\eqref{eq:S_bGRExpansion}, except the Ricci scalars and coefficients swap orders. For example, 
\begin{align} \label{eq:S_pGRAExpansion}
    \mathcal{S}^{p(1,1)}_{A} 
    &=-\sum_{ijkl}{}_{A}\mathcal{X}^{p(1,0)}_{ijkl}
    \mathcal{O}^{k}\mathcal{O}^{l}\Phi^{p(0,1)}_{ij}\,.
\end{align}
Since the NP Ricci scalars in Table~\ref{tab:typeAS_bGR} are generic, the coefficients ${}_{A}\mathcal{X}^{p(1,0)}_{ijkl}$ can be obtained from Table~\ref{tab:typeAS_bGR} by making a simultaneous transformation $q_{ab}\to h_{ab}$, $\Phi_{ij}^{\eff(1,0)}\to\Phi_{ij}^{p(0,1)}$ and setting the tetrad rotation parameters $a^{(0,1)}$ and $b^{(0,1)}$ to zero. The coefficients for $\mathcal{S}^{p(1,1)}_{B}$, $\mathcal{T}^{p(1,1)}_{A}$, and $\mathcal{T}^{p(1,1)}_{B}$ can be obtained similarly.

Lastly, we have the $\mathcal{O}(\zeta^1,\epsilon^1)$ geometrical source terms $\mathcal{S}_{\geo}$ and $\mathcal{T}_{\geo}$ given in Eqs.~\eqref{eq:S_geo} and \eqref{eq:T_geo}, respectively. Unlike previous source terms, the geometrical contribution is directly driven by the Weyl scalars at $\mathcal{O}(\zeta^1,\epsilon^0)$ and $\mathcal{O}(\zeta^0,\epsilon^1)$ instead of NP Ricci scalars. Adopting the same decomposition of Eq.~\eqref{eq:S_bGRExpansion}, we can expand $\mathcal{S}_{\geo}$ in terms of directional derivatives at $\mathcal{O}(\zeta^0,\epsilon^0)$, i.e.,
\begin{align}\label{eq:S_geoExpansion}
    \mathcal{S}^{(1,1)}_{\geo}
    =-\sum_{kl} 
    & \;\left[{}_{0}\mathcal{X}^{\geo(1,0)}_{kl}\mathcal{O}^{k}
    \mathcal{O}^{l}\Psi^{(0,1)}_0
    +{}_{0}\mathcal{X}^{\geo(0,1)}_{kl}\mathcal{O}^{k}
    \mathcal{O}^{l}\Psi^{(1,0)}_0 \right.\nonumber\\
    & \;\left.+{}_{1}\mathcal{X}^{\geo(0,1)}_{kl}\mathcal{O}^{k}
    \mathcal{O}^{l}\Psi^{(1,0)}_1\right]\,, 
\end{align}
where ${}_{0}\mathcal{X}^{\geo(1,0)}_{kl}$, ${}_{0}\mathcal{X}^{\geo(0,1)}_{kl}$, and ${}_{1}\mathcal{X}^{\geo(0,1)}_{kl}$ are the generating coefficients provided in Table~\ref{tab:S_geoTable}. To obtain the generating coefficients for $\mathcal{T}_{\geo}^{(1,1)}$, one can apply the prime transformation to Eq.~\eqref{eq:S_geoExpansion}, with the results presented in Table~\ref{tab:T_geoTable}. The perturbed Weyl scalars $\Psi_{0,1}^{(1,0)}$ are provided in Eq.~\eqref{eq:perturbedWeylScalars}, while $\Psi_{0}^{(0,1)}$ can be calculated by replacing every $h_{ab}$ in Eq.~\eqref{eq:perturbedWeylScalars} with $q_{ab}$, where $q_{ab}$ was solved in Sec.~\ref{sec:perturbed_metric}.

\section{Computing the Modified Flux}\label{sec:flux}

In this section, we derive the leading-order correction to the GW flux due to a generic background metric deformation in terms of perturbations of the Weyl scalars $\Psi_{0,4}$. We first review how the effective stress-energy tensor for linear gravitational radiation in GR at null infinity is related to the Weyl scalar $\Psi_4$, and then discuss how this relation can be extended to calculate the gravitational flux of an EMRI system in a deformed Kerr geometry. We briefly outline the limitations of this derivation and specify the subclass of theories in which it remains valid. In addition, we highlight the difficulties involved in computing the horizon flux in bGR theories or, more generally, beyond-Kerr geometries.

Within GR, the effective stress-energy tensor due to linear gravitational radiation was first computed by Issacson \cite{Isaacson:1968hbi, Isaacson:1968zza} 
\begin{align}\label{eq:IssacsonStress}
    T^{\text{I}}_{\mu\nu} = \frac{1}{32 \pi}\left\langle\left(\nabla_\mu q^{\alpha \beta}\right)\left(\nabla_\nu q_{\alpha \beta}\right)\right\rangle\,,
\end{align}
where the brackets $\langle \rangle$ denote a Brill-Hartle type averaging procedure \cite{Brill:1964zz}, and $q_{\alpha\beta}$ corresponds to GW in vacuum GR. This expression is only valid at future null infinity and in the Lorenz gauge. In GR, it has been shown that the transverse traceless gauge, equivalently the Lorenz gauge plus the trace-free condition, exists for any vacuum perturbation of a globally hyperbolic, vacuum solution \cite{Wald:1984rg}. This allows the GW energy flux at future null infinity in GR to be related to the perturbed Weyl scalar $\Psi^{(1)}_4$ \cite{Teukolsky:1973ha}
\begin{align} \label{eq:psi4flux}
    \frac{d^2E^{\infty}}{dt d\Omega}
    &=\lim_{r\rightarrow\infty}r^2 T^r{}_t \nonumber\\
    &=\lim_{r\rightarrow\infty}
    \frac{r^2\omega^2}{16\pi}\left[\left(A_{+}\right)^2
    +\left(A_{\times}\right)^2\right] \nonumber\\
    &=\lim_{r\rightarrow\infty}
    \frac{r^2}{4\pi\omega^2}
    \left|\Psi_4^{(1)}\right|^2\,,
\end{align}
where we have set $T_{\mu\nu}$ to be the Issacson stress-energy tensor and assume a decomposition of $q_{\mu\nu}$ in the form 
\begin{align}
    q_{\mu\nu}
    =\Re\left\{\left(A_{+}e^{+}_{\mu\nu}
    +A_{\times}e^{\times}_{\mu\nu}\right)
    e^{-i\omega(t-r)}\right\},
\end{align}
where $\Re$ takes the real component, while $e^{+}_{\mu \nu}$ and $e^{\times}_{\mu \nu}$ are the usual plus and cross polarization tensors, respectively. Using the tetrad in Eq.~\eqref{eq:Carter_tetrad} to compute the corresponding $\Psi_4^{(1)}$ of $q_{\mu\nu}$, we get the last line of Eq.~\eqref{eq:psi4flux}.

For a subset of bGR theories with dynamical scalar fields coupled to higher-order curvature terms, such as dCS and EdGB gravity, the effective stress–energy tensor of linear gravitational radiation has been shown to reduce to the Isaacson stress-energy tensor in Eq.~\eqref{eq:IssacsonStress} \cite{Stein:2010pn}. In particular, these additional higher-derivative terms do not modify the form of Eq.~\eqref{eq:IssacsonStress}. For this class of bGR theories, only the transverse, tensor degrees of freedom contribute to the infinity flux. Hence, using the expansion scheme in Eq.~\eqref{eq:expansion_metric}, we can compute the leading-order correction to the infinity energy flux by expanding Eq.~\eqref{eq:psi4flux} up to $\mathcal{O}(\zeta^1,\epsilon^1)$, i.e.,
\begin{align} \label{eq:ModFluxPsi4}
   \frac{d^2E^{\infty}}{dtd\Omega}
   =\lim_{r\rightarrow\infty}
   \frac{r^2}{4\pi\omega^2}\left(\Psi^{(0,1)}_4\bar{\Psi}^{(1,1)}_4 + \bar{\Psi}^{(0,1)}_4 \Psi^{(1,1)}_4\right)\,,
\end{align}
where $\Psi_4^{(0,1)}$ is the solution to the Teukolsky equation in vacuum GR driven by a point particle, while $\Psi_4^{(1,1)}$ is the solution to the modified Teukolsky equation we developed in this work and corresponds to additional GWs propagating to null infinity due to bGR or environmental effects.

However, in general, there are six propagating polarizations (two tensors, two scalars, and two vector types) in linear metric perturbations that can modify the stress-energy tensor in Eq.~\eqref{eq:IssacsonStress}. Some bGR theories can have additional propagating degrees of freedom, including de Rham–Gabadadze–Tolley (dRGT) massive gravity \cite{Hinterbichler:2011tt, Hassan:2011zd, Rham:2015mxa, Barack:2018yly}, Einstein-æther gravity \cite{Jacobson:2007veq}, pure metric $f(R,R_{\mu\nu}R^{\mu\nu})$ gravity \cite{Sotiriou:2006hs, Sotiriou:2008rp, Vilhena:2021bsx}, and other scalar-vector-tensor theories \cite{Bekenstein:2004ne, Bekenstein:2005nv}. For example, in dRGT massive gravity, the effective GW stress-energy tensor is given by \cite{Cardoso:2018zhm} 
\begin{align}
    T_{\mu\nu}=\frac{1}{32\pi}
    \left\langle q_{\alpha\beta,\mu}q_{,\nu}^{\alpha\beta}
    -q_{,\mu}q_{,\nu}\right\rangle\,,
\end{align}
where the trace $q = g^{\alpha \beta}q_{\alpha \beta}$ encodes a scalar breathing mode and contributes to the total energy flux. Nevertheless, a modified effective GW stress-energy tensor does not always change the form of Eq.~\eqref{eq:ModFluxPsi4}. Specifically, if modified equations of motion allow extra polarizations to decouple from the plus and cross polarizations, behaving as non-minimally coupled matter fields, Eq.~\eqref{eq:ModFluxPsi4} is still accurate up to order $\mathcal{O}(\zeta^1,\epsilon^1)$. In this case, the total energy flux at future null infinity should include not only gravitational radiation, but also contributions from dynamical matter fields (e.g., scalar or vector fields) \cite{Wald:1999wa, OuldElHadj:2023anu, Ashtekar:2024stm, Li:2025ffh}.

In addition, bGR theories or environmental effects could also modify the relation between the horizon flux and the Weyl scalar $\Psi_0$, which is more difficult to calculate and has not been systematically studied to the best of the authors' knowledge. Within GR, the energy flux to the horizon can be related to the infinitesimal change in the horizon’s surface area via the first law of BH thermodynamics. The surface area $\Sigma$ of a Kerr BH can be written in terms of its mass $M$ and angular momentum $J$,
\begin{align}\label{eq:KerrHorizonArea}
    \Sigma=4 \pi\left(r_{+}^2+a^2\right)
    =8 \pi\left[M^2+\sqrt{ \left(M^4-J^2\right)}\right]\,,
\end{align}
where $a = J/M$ and $r_{\pm} = M \pm \sqrt{M^2-a^2}$. We can use Eq.~\eqref{eq:KerrHorizonArea} to express the change in the horizon's area $d\Sigma$ in terms of the change in mass $dM$ and the change in angular momentum $dJ$:
\begin{align} \label{eq:differentialHorArea}
    \mathrm{d}\Sigma=\frac{8 \pi}{M \sqrt{ \left(M^2-a^2\right)}}\left(2 M^2 r_{+}\,dM-J\,dJ\right)\,.
\end{align}
Since a Kerr BH is stationary and axisymmetric, the perturbed quantities have a harmonic $t$- and $\phi$-dependence given by $e^{-i \omega t + i m \phi}$, where $m$ is an azimuthal quantum number and $\omega$ is a frequency. This permits a simple relationship between $dJ$ and $dM$:
\begin{align}\label{eq:JMRelation}
    dJ=\frac{m}{\omega}dM\,.
\end{align}
Then, inserting Eq.~\eqref{eq:JMRelation} into Eq.~\eqref{eq:differentialHorArea}, we obtain
\begin{align} \label{eq:area_law}
    \frac{d^2\Sigma}{dt d\Omega}
    =\frac{4\pi}{\varepsilon_0}\left(1-m\omega_{+}/\omega\right)
    \frac{d^2M}{dt d\Omega}\,,
\end{align}
where $\varepsilon_0=\sqrt{M^2-a^2}/(4Mr_{+})$, and $\omega_{+}=a/(2Mr_+)$ is the horizon frequency. On the other hand, the horizon area change $d\Sigma$ can be computed from the perturbed spin coefficient $\sigma^{(1)}$ in GR \cite{Hawking:1972hy, Teukolsky:1974yv, Chandrasekhar_1983}, i.e.,
\begin{align}\label{eq:HorizonFluxSigma}
    \frac{d^2\Sigma}{dtd\Omega}
    =\frac{2Mr_{+}}{\varepsilon_0}
    \left|\sigma^{\text{HH}(1)}(r_+)\right|^2\,,
\end{align}
where $\sigma^{\text{HH}(1)}$ is $\sigma^{(1)}$ in the Hartle–Hawking (HH) frame \cite{Hartle:1974gy}. Since $\sigma^{\text{HH}(1)}$ can be reconstructed from the perturbed Weyl scalar $\Psi^{\text{HH}(1)}_0$ on the horizon via \cite{Teukolsky:1974yv, Chandrasekhar_1983}
\begin{align}
    & \sigma^{\text{HH}(1)}(r_+)
    =-\frac{\Psi_0^{\text{HH}(1)}(r_+)}
    {i\left(\omega-m\omega_{+}\right)+2\varepsilon_0}\,,  
    \label{eq:simgaPsi0Relation} \\
    & \Psi_0^{\text{HH}(1)}(r_+)
    =\frac{\Delta^2\Psi_0^{(1)}(r_+)}{16M^2r_+^2}\,,
    \label{eq:psi0_HH}
\end{align}
where $\Psi_0^{\text{HH}(1)}$ is $\Psi_0^{(1)}$ in the HH frame, one obtains the horizon energy flux from Eqs.~\eqref{eq:area_law}, \eqref{eq:HorizonFluxSigma}, \eqref{eq:simgaPsi0Relation}, and \eqref{eq:psi0_HH} to be
\begin{align} \label{eq:horizon_flux}
    \frac{d^2E^{H}}{dt d\Omega}
    =& \;\lim_{r\rightarrow r_{+}}
    \bigg[\frac{1}{64\pi}\frac{\omega}{\omega-m\omega_{+}}
    \frac{1}{(\omega-m\omega_+)^2+4\varepsilon^2_0} \nonumber\\
    & \;\frac{\Delta^4}{(2Mr_+)^3}
    \left|\Psi^{(1)}_0\right|^2\bigg]\,.
\end{align}

In bGR theories or non-vacuum environments, several parts of this procedure to derive Eq.~\eqref{eq:horizon_flux} could fail and must be reestablished on a case-by-case basis. For instance, Eq.~\eqref{eq:HorizonFluxSigma} can be modified for non-Kerr BHs. Moreover, in Eq.~\eqref{eq:simgaPsi0Relation}, we have assumed the NP Ricci scalar $\Phi_{00}$ evaluated both on the background and to first-order vanishes. This may no longer be accurate up to $\mathcal{O}({\zeta^1,\epsilon^1})$ in bGR theories and beyond-Kerr geometries. For instance, certain classes of spinning ``bumpy'' BHs in \cite{Vigeland:2011ji} contain such source terms and must be taken into account when relating $\Psi^{(1,1)}_0$ and $\sigma^{(1,1)}$.

\section{Discussion}
\label{sec:discussion}

In this work, we have extended the MTF to compute the gravitational radiation from EMRI systems around nonrotating BHs with generic, axisymmetric, non-vacuum deformations. While previous work has focused on non-relativistic approximation schemes for matter environments \cite{Yunes:2011aa, Baumann:2018vus, Baumann:2019ztm, Baumann:2021fkf, Baumann:2022pkl, Takahashi:2023flk} and relativistic calculations for spherically symmetric metric deformations \cite{Brito:2023pyl, Duque:2023seg, Speeney:2024mas}, we have employed the Teukolsky formalism to establish essential foundations for modeling general, fully relativistic EMRI systems in non-vacuum environments and bGR theories. 

We have reviewed the MTF and utilized its two-parameter expansion scheme to separate contributions from the theory-agnostic background metric correction at $\mathcal{O}(\zeta^1,\epsilon^0)$ and the usual GW radiation produced by an EMRI system with a fixed trajectory in GR at $\mathcal{O}(\zeta^0,\epsilon^1)$. Given the background metric correction $h_{\mu\nu}$ and the GR GW $q_{\mu\nu}$ obtained from the RW and ZM equations, we constructed the corresponding tetrads $e^{\mu(1,0)}_a$ and $e^{\mu(0,1)}_a$. We then outline the procedure to compute the perturbed spin coefficients, directional derivatives, and Weyl scalars at  $\mathcal{O}(\zeta^1,\epsilon^0)$ and $\mathcal{O}(\zeta^0,\epsilon^1)$, following the prescription in \cite{Chandrasekhar_1983, Campanelli:1998jv, Wagle:2023fwl}. These quantities are required to compute the geometrical [Eqs.~\eqref{eq:S_geo} and \eqref{eq:T_geo}], particle stress-energy [Eqs.~\eqref{eq:S_p} and \eqref{eq:T_p}] and effective stress-energy [Eqs.~\eqref{eq:S_bGR} and \eqref{eq:T_bGR}] source terms. Although the form of the final source terms is complex, we have organized them using the operator notation of Sec.~\ref{sec:perturbed_NP} and tabulated them in Appendix~\ref{sec:SourceCoeffs}. Having obtained a solution to the master equation for the Weyl scalar $\Psi^{(1,1)}_4$ [i.e., Eq.~\eqref{eq:master_eqn_non_typeD_Psi4}], the modified flux at future null infinity can be computed using Eq.~\eqref{eq:ModFluxPsi4}. This assumes that the transverse tensor polarizations in GR decouple from any additional polarizations in a bGR theory or non-vacuum environments. To account for the additional propagating degrees of freedom, one may have to compute: i. the modification to the effective Issacson stress; ii. reconstruct (or partially reconstruct) the metric perturbation $Q_{\mu\nu}^{(1,1)}$. Obtaining $Q_{\mu\nu}^{(1,1)}$ would also aid the computation of the modified horizon flux. We leave a full analysis of these cases for future work.

One avenue for future research is to adopt a specific bGR theory or a prescribed matter configuration within GR, which would allow the computation of the Ricci tensor up to $\mathcal{O}(\zeta^1,\epsilon^1)$. At this order, the Ricci tensor contributes directly to the source term within the master equations of $\Psi^{(1,1)}_{0,4}$ and must be included for a complete calculation. In particular, this contribution captures both how GWs are evaluated with the modified gravitational equations of motion and how the $\mathcal{O}(\zeta^0,\epsilon^1)$ stress-energy tensor couples to the modified geometry at $\mathcal{O}(\zeta^1 ,\epsilon^0)$.
The latter is particularly important since geodesic motion for the Kerr metric is integrable. This feature is generically lost for BHs in bGR theories \cite{Deich:2022vna, Daskk} and other parametrized beyond-Kerr BHs \cite{Weller:2024qvo, Vigeland:2011ji, Glampedakis:2005cf, Cardoso:2014rha, Bambi:2015kza}. Even when modeling stationary matter environments around BHs within an EMRI system, it is crucial to account for the stress–energy tensor of the radiated matter distribution, since this contributes to $R^{(1,1)}_{\mu\nu}$ within our approximation scheme. Relevant examples where this contribution may become important include Dirac stars, where the stress–energy tensor of the environment is sourced from a self-interacting fermion field \cite{Herdeiro:2020jzx, Herdeiro:2019mbz}, BHs with bosonic hair \cite{Herdeiro:2014goa, Herdeiro:2015waa, Cunha:2015yba, Herdeiro:2015gia, Liebling:2012fv, Li:2025ffh, Dyson:2025dlj}, and BHs with Proca hair, in which a massive vector field forms through superradiance \cite{Herdeiro:2020jzx, Herdeiro:2016tmi, Santos:2020pmh, Santos:2020sut}. 

A key challenge for the MTF is to reconstruct the sourced metric perturbation and all NP quantities at $\mathcal{O}(\zeta^0,\epsilon^1)$ in GR. Currently, a diverse set of approaches is being developed to address this challenge \cite{Aksteiner:2016pjt, Green:2019nam, Loutrel:2020wbw, Toomani:2021jlo, Wardell:2024yoi, Osburn:2022bby, Dolan:2023enf, Nasipak:2025tby}. Building on this framework, future work can focus on studying more generic EMRI systems to include precession, inclination, and higher order self-force terms \cite{Barack:2009ux, Barack:2018yvs, Pound:2021qin, Nasipak:2025tby}. 
Even though we focused on the source terms in the MTF for a deformed Schwarzschild background in this work, the key steps we took allow for a perturbative deformation of the Kerr metric.

\section{Acknowledgments} 
We thank Chaoyi Yang for checking some results in Appendix~\ref{sec:NPQuantities}. C.~W. and Y.~C.’s research is supported by the Simons Foundation (Award No. 568762), the Brinson Foundation, and the National Science Foundation (via Grants No. PHY-2011961 and No. PHY-2011968). D.~L. acknowledges support from the Simons Foundation (via Award No. 896696), the Simons Foundation International (via Grant No. SFI-MPS-BH-00012593-01), the NSF (via Grants No. PHY-2207650 and PHY-2512423), and the National Aeronautics and Space Administration through award 80NSSC22K0806. H.~Y. is supported by the Natural Science Foundation of China (Grant 12573048). P.~B. acknowledges support from the Dutch Research Council (NWO) with file number OCENW.M.21.119. 

\label{sec:acknowledgements}
\appendix

\begin{widetext}
\section{Lists of perturbed NP quantities at $\mathcal{O}(\zeta^1,\epsilon^0)$ and $\mathcal{O}(\zeta^0,\epsilon^1)$}
\label{sec:NPQuantities}

In this appendix, we list the perturbed spin coefficients and Weyl scalars found by the procedures in Sec.~\ref{sec:perturbed_NP}. For the spin coefficients at $\mathcal{O}(\zeta^1,\epsilon^0)$, we get
\begin{subequations}\label{eq:perturbedSpinCoeffs}
\begin{align}
    \kappa^{(1,0)} 
    &=\frac{1}{2}(\text{\TH}_{1}-\rho )h_{lm}-\frac{1}{2}\cpartial_{0}h_{ll}\,,
    \label{eq:kappa_10} \\
    \nu^{(1,0)}
    &=-\frac{1}{2}(\text{\TH}'_{-1}+\mu)h_{n\bar{m}}+\frac{1}{2}\cpartial'_{0}h_{nn}\,, \\
    \sigma^{(1,0)} 
    &=-\frac{1}{2}\cpartial_{1}h_{lm}+\frac{1}{2}\text{\TH}_{0}h_{mm}\,, 
    \label{eq:sigma_10}\\
    \lambda^{(1,0)} 
    &=\frac{1}{2}\cpartial'_{-1}h_{n\bar{m}}-\frac{1}{2}\text{\TH}'_{0} h_{\bar{m}\bar{m}}\,, \\
    \rho^{(1,0)} 
    &=-\frac{1}{2}(\rho h_{ln}-\mu h_{ll})
    +\frac{1}{2}(\text{\TH}'_{0}h_{m\bar{m}}-\cpartial_{-1}h_{l\bar{m}})\,, \\
    \mu^{(1,0)} 
    &=-\frac{1}{2}(\mu h_{ln}-\rho h_{nn})
    -\frac{1}{2}(\text{\TH}'_{0} h_{m\bar{m}}-\cpartial'_{1}h_{nm})\,, \\
    \tau^{(1,0)} 
    &=\frac{1}{2}\mu h_{lm}
    +\frac{1}{2}(\text{\TH}_{-1}h_{nm}-\cpartial'_{0}h_{ln})\,, \\
    \pi^{(1,0)}
    &=\frac{1}{2}\rho h_{n\bar{m}}
    - \frac{1}{2}(\text{\TH}'_{1}h_{l\bar{m}}-\cpartial_{0}h_{ln})\,, \\
    \beta^{(1,0)} 
    &=-\frac{1}{4}\left[(\text{\TH}'_{1}+\mu)h_{lm}
    -(\text{\TH}'_{-1}+\rho)h_{nm}\right]
    -\frac{1}{4}(\cpartial_{0}h_{m\bar{m}}-\cpartial'_{2}h_{mm})
    -\frac{1}{2}(\varepsilon h_{nm}+\gamma h_{lm}
    -\beta h_{m\bar{m}}-\alpha h_{mm})\,, \\
    \alpha^{(1,0)} 
    &=-\frac{1}{4}\left[( \text{\TH}'_{1}+\mu)h_{l\bar{m}}
    -(\text{\TH}'_{-1}+\rho)h_{n\bar{m}}\right]
    +\frac{1}{4}(\cpartial'_{0}h_{m\bar{m}}-\cpartial_{-2}h_{\bar{m}\bar{m}})
    -\frac{1}{2}(\varepsilon h_{n\bar{m}}+\gamma h_{l\bar{m}}
    -\beta h_{m\bar{m}}-\alpha h_{\bar{m}\bar{m}})\,, \\
    \varepsilon^{(1,0)} 
    &=-\frac{1}{4}(\text{\TH}'_{2}h_{ll}-\text{\TH}_{0}h_{ln})
    -\frac{1}{4}(\cpartial_{-1}h_{l\bar{m}}-\cpartial'_{1}h_{lm})
    -\frac{1}{2}(\varepsilon h_{ln}+\gamma h_{ll}
    -\beta h_{l\bar{m}}-\alpha h_{lm})\,, \\
    \gamma^{(1,0)} 
    &=\frac{1}{4}(\text{\TH}'_{2}h_{ll}-\text{\TH}_{0}h_{ln})
    -\frac{1}{4}(\cpartial_{-1} h_{l\bar{m}}-\cpartial'_{1} h_{lm})
    -\frac{1}{2}(\varepsilon h_{nn}+\gamma h_{ln}
    -\beta h_{n\bar{m}}-\alpha h_{nm})\,.
\end{align}
\end{subequations}
For the Weyl scalars at $\mathcal{O}(\zeta^1,\epsilon^0)$, we obtain
\begin{subequations}\label{eq:perturbedWeylScalars}
\begin{align}
    \Psi^{(1,0)}_0 
    =& \;\frac{1}{2}\cpartial_{1}\cpartial_{0}h_{ll}
    -(\cpartial_{1}\text{\TH}_{1}-\rho\cpartial_{1})h_{lm}
    +\frac{1}{2}(\text{\TH}_{1}\text{\TH}_{0}-2\rho\text{\TH}_{0})h_{mm}\,, \\
    \Psi^{(1,0)}_1 
    =& \;\frac{1}{4}(\cpartial_{0}\text{\TH}'_{2}-2\mu\cpartial_{0})h_{ll}
    -\frac{1}{4}\cpartial_{0}\text{\TH}_{0}h_{ln}
    +\frac{1}{4}\cpartial'_{2}\text{\TH}_{0}h_{mm}
    -\frac{1}{4}\cpartial_{0}\text{\TH}_{0}h_{m\bar{m}}
    +\frac{1}{4}\cpartial_{0}\cpartial_{-1}h_{l\bar{m}}
    +\frac{1}{4}\text{\TH}_{0}\text{\TH}_{-1}h_{nm} \\
    & \;-\frac{1}{4}(\cpartial_{0}\cpartial'_{1}
    +\text{\TH}'_{2}\text{\TH}_{1}-\rho\text{\TH}'_{1}
    -\mu\text{\TH}_{1}+4\rho\mu-\Psi_2)h_{lm}\,, \nonumber\\
    \Psi^{(1,0)}_2 
    =& \;\frac{1}{12}(\text{\TH}'_{1}\text{\TH}'_{2}
    -2\mu\text{\TH}'_{2}+2\mu^2)h_{ll}
    -\frac{1}{6}(\cpartial_{-1}\cpartial'_{0}+\text{\TH}_{-1}\text{\TH}'_{0}
    +2\rho\mu+4\Psi_2)h_{ln} \nonumber\\
    & \;+\frac{1}{12}(\text{\TH}_{-1}\text{\TH}_{-2}
    +2\rho\text{\TH}'_{2}+2\rho^2)h_{nn}
    +\frac{1}{12}\cpartial'_{1}\cpartial'_{2}h_{mm}
    +\frac{1}{12}\cpartial_{-1}\cpartial_{-2}h_{\bar{m}\bar{m}}
    -\frac{1}{6}(\cpartial_{-1}\cpartial'_{0}+\text{\TH}_{-1}\text{\TH}'_{0}
    +2\rho\mu-2\Psi_2)h_{m\bar{m}} \nonumber\\
    & \;-\frac{1}{6}(\cpartial'_{1}\text{\TH}'_{1}-\mu\cpartial'_{1})h_{lm}
    -\frac{1}{6}(\cpartial_{-1}\text{\TH}_{-1}+\rho\cpartial_{-1})h_{n\bar{m}}
    +\frac{1}{3}(\cpartial_{-1}\text{\TH}'_{1}-\mu\cpartial_{-1})h_{l\bar{m}}
    +\frac{1}{3}(\cpartial'_{1}\text{\TH}_{-1}+\rho\cpartial'_{1})h_{nm}\,, \\ 
    \Psi^{(1,0)}_3
    =& \;-\frac{1}{4}(\cpartial'_{0}\text{\TH}_{-2}+2\rho\cpartial'_{0})h_{nn}
    \frac{1}{4}\cpartial'_{0}\text{\TH}'_{0}h_{ln}
    -\frac{1}{4}\cpartial_{-2}\text{\TH}'_{0}h_{\bar{m}\bar{m}}
    +\frac{1}{4}\cpartial'_{0}\text{\TH}'_{0}h_{m\bar{m}}
    -\frac{1}{4}\cpartial'_{0}\cpartial'_{1}h_{nm}
    -\frac{1}{4}\text{\TH}'_{0}\text{\TH}'_{1}h_{l\bar{m}} \nonumber\\
    & \;+\frac{1}{4}(\cpartial'_{0}\cpartial_{-1}
    +\text{\TH}_{-2}\text{\TH}'_{-1}+\mu\text{\TH}_{-1}
    +\rho\text{\TH}'_{-1}+4\rho\mu-\Psi_2)h_{n\bar{m}}\,, \\
    \Psi^{(1,0)}_4 
    =& \;-\frac{1}{2}\cpartial'_{-1}\cpartial'_{0}h_{nn}
    +(\cpartial'_{-1}\text{\TH}'_{-1}+\mu\cpartial'_{-1})h_{n\bar{m}}
    +\frac{1}{2}(\text{\TH}'_{-1}\text{\TH}'_{0}-2\mu\text{\TH}'_{0}) h_{\bar{m}\bar{m}}\,.
\end{align}
\end{subequations}
As discussed in Sec.~\ref{sec:perturbed_NP}, the perturbed spin coefficients and Weyl scalars at $\mathcal{O}(\zeta^0,\epsilon^1)$ can be obtained by simply replacing every $h_{ab}$ in Eqs.~\eqref{eq:perturbedSpinCoeffs} and \eqref{eq:perturbedWeylScalars} with the corresponding $q_{ab}$, with the additional tetrad rotations at $\mathcal{O}(\zeta^0,\epsilon^1)$ in Eqs.~\eqref{eq:tetrad_rotations_10} and \eqref{eq:TetradRotCoeffs} to set $\Psi_{1,3}^{(0,1)}=0$. Under these rotations, the spin coefficients at $\mathcal{O}(\zeta^0,\epsilon^1)$ transform in the following way \cite{Chandrasekhar_1983}:
\begin{align} \label{eq:spin_coeffs_rotated}
    & \kappa^{(0,1)}\to\kappa^{(0,1)}-(\text{\TH}_{1}-\rho)b^{(0,1)}\,,
    && \nu^{(0,1)}\to\nu^{(0,1)}+(\text{\TH}'_{-1}+\mu)\bar{a}^{(0,1)}\,, \nonumber\\
    & \sigma^{(0,1)}\to\sigma^{(0,1)}-\cpartial_{1}b^{(0,1)}\,,
    && \lambda^{(0,1)}\to\lambda^{(0,1)}
    +\cpartial'_{-1}\bar{a}^{(0,1)}\,, \nonumber\\
    & \rho^{(0,1)}\to\rho^{(0,1)}-\cpartial'_{1}b^{(0,1)}\,,
    && \mu^{(0,1)}\to\mu^{(0,1)}+\cpartial_{-1}\bar{a}^{(0,1)}\,, \nonumber\\
    & \tau^{(0,1)}\to\tau^{(0,1)}+\rho a^{(0,1)}
    -\text{\TH}'_{1}b^{(0,1)}\,,
    && \pi^{(0,1)}\to\pi^{(0,1)}+\mu\bar{b}^{(0,1)}
    +\text{\TH}_{-1}\bar{a}^{(0,1)}\,, \nonumber\\
    & \varepsilon^{(0,1)}\to\varepsilon^{(0,1)}
    +(b^{(0,1)}-\bar{b}^{(0,1)})\alpha\,,
    && \gamma^{(0,1)}\to\gamma^{(0,1)}
    +(a^{(0,1)}-\bar{a}^{(0,1)})\alpha\,, \nonumber\\
    & \alpha^{(0,1)}\to\alpha^{(0,1)}+\rho\bar{a}^{(0,1)}
    +\varepsilon\bar{a}^{(0,1)}+\gamma\bar{b}^{(0,1)}\,,
    && \beta^{(0,1)}\to\beta^{(0,1)}+\mu b^{(0,1)}
    +\varepsilon a^{(0,1)}+\gamma b^{(0,1)}\,.
\end{align}
After the tetrad rotations, we have $\Psi_{1,3}^{(0,1)}=0$, while the other Weyl scalars at $\mathcal{O}(\zeta^0,\epsilon^1)$ remain unchanged [i.e., Eq.~\eqref{eq:perturbedWeylScalars} with $h_{ab}$ replaced by $q_{ab}$]. The spin coefficients at $\mathcal{O}(\zeta^0,\epsilon^1)$ are then given by Eq.~\eqref{eq:perturbedSpinCoeffs} with $h_{ab}$ replaced by $q_{ab}$ plus the additional terms in Eq.~\eqref{eq:spin_coeffs_rotated}.

The NP Ricci scalars $\Phi^{\eff(1,0)}_{ij}$ can be found in the same way as how the perturbed Weyl scalars $\Psi^{(1,0)}_{i}$ are calculated from the Ricci identities, with the results given below:
\begin{subequations} \label{eq:Ricci10}
\begin{align}
    \Phi^{\eff(1,0)}_{00} 
    =& \;\frac{1}{2}(\cpartial'_{1}\cpartial_{0}
    +\rho\text{\TH}'_{2}+\mu\text{\TH}_{2}+2\Psi_2)h_{ll}
    -\frac{1}{2}(\cpartial_{-1}\text{\TH}_{1}-\rho\cpartial_{-1})h_{l\bar{m}}
    -\frac{1}{2}(\cpartial'_{1} \text{\TH}_{1}-\rho)\cpartial_{-1}h_{lm} \nonumber\\
    & \;+\frac{1}{2}(\text{\TH}_{1}\text{\TH}_{0}-2\rho\text{\TH}_{0})h_{m\bar{m}}
    -\rho\text{\TH}_{0}h_{ln}\,, \\
    \Phi^{\eff(1,0)}_{22} 
    =& \;\frac{1}{2}(\cpartial_{-1}\cpartial'_{0}
    -\mu\text{\TH}_{-2}-\rho\text{\TH}'_{-2}+2\Psi_2)h_{nn}
    -\frac{1}{2}(\cpartial'_{1}\text{\TH}'_{-1}+\mu\cpartial'_{1})h_{nm}
    -\frac{1}{2}(\cpartial_{-1}\text{\TH}'_{-1}+\mu\cpartial_{-1})h_{n\bar{m}} \nonumber\\
    & \;+\frac{1}{2}(\text{\TH}'_{-1}\text{\TH}'_{0}+2\mu\text{\TH}'_{0})h_{m\bar{m}}
    +\mu\text{\TH}'_{0}h_{ln}\,, \\
    \Phi^{\eff(1,0)}_{20} 
    =& \;\frac{1}{2}(\cpartial'_{-1}\text{\TH}'_{1}+\mu\cpartial'_{-1})h_{l\bar{m}}
    +\frac{1}{2}(\cpartial'_{-1}\text{\TH}_{-1}-\rho\cpartial'_{-1})h_{n\bar{m}}
    -\frac{1}{2}(\text{\TH}_{-1}\text{\TH}'_{0}
    -\rho\text{\TH}'_{0}+\mu\text{\TH}_{0})h_{\bar{m}\bar{m}}
    -\frac{1}{2}\cpartial'_{-1}\cpartial'_{0}h_{ln}\,, \\
    \Phi^{\eff(1,0)}_{02}
    =& \;\frac{1}{2}(\cpartial_{1}\text{\TH}_{-1}-\rho\cpartial_{1})h_{nm}
    +\frac{1}{2}(\cpartial_{1}\text{\TH}'_{1}+\mu\cpartial_{1})h_{lm}
    -\frac{1}{2}(\text{\TH}'_{1}\text{\TH}_{0}
    +\mu\text{\TH}_{0}-\rho\text{\TH}'_{0}) h_{mm}
    -\frac{1}{2}\cpartial_{1}\cpartial_{0}h_{ln}\,, \\
    \Phi^{\eff(1,0)}_{12} 
    =& \;\frac{1}{4}(\text{\TH}'_{0}\text{\TH}'_{1}
    +2\mu\text{\TH}'_{1}-2\mu^2)h_{lm}
    -\frac{1}{4}(-\cpartial_{0}\cpartial'_{1}+\text{\TH}'_{0}\text{\TH}_{-1}
    +\rho\text{\TH}'_{-1}+3\mu\text{\TH}_{1}-2\rho\mu+3\Psi_2)h_{nm} \nonumber\\
    & \;-\frac{1}{4}\cpartial_{0}\cpartial_{-1}h_{n\bar{m}}
    +\frac{1}{4}\cpartial_{0}\text{\TH}'_{0}h_{m\bar{m}}
    -\frac{1}{4}(\cpartial_{0}\text{\TH}'_{0}-2\mu\cpartial_{0})h_{ln}
    -\frac{1}{4}\cpartial'_{2}\text{\TH}'_{0}h_{mm}
    +\frac{1}{4}\cpartial_{0}\text{\TH}_{-2}h_{nn}\,, \\
    \Phi^{\eff(1,0)}_{10} 
    =& \;\frac{1}{4}(\text{\TH}_{0}\text{\TH}_{-1}
    -2\rho\text{\TH}_{-1}-2\rho^2)h_{n\bar{m}}
    -\frac{1}{4}(-\cpartial'_{0}\cpartial_{-1}+\text{\TH}_{0}\text{\TH}'_{1}
    -\mu\text{\TH}_{1}-3\rho\text{\TH}'_{-1}-2\rho\mu+3\Psi_2)h_{l\bar{m}} \nonumber\\
    & \;-\frac{1}{4}\cpartial'_{0}\cpartial'_{1}h_{lm}
    +\frac{1}{4}\cpartial'_{0}\text{\TH}_{0}h_{m\bar{m}}
    -\frac{1}{4}(\cpartial'_{0}\text{\TH}_{0}+2\rho\cpartial'_{0})h_{ln}
    -\frac{1}{4}\cpartial_{-2}\text{\TH}_{0}h_{\bar{m}\bar{m}}
    +\frac{1}{4}\cpartial'_{0}\text{\TH}'_{2}h_{ll}\,, \\
    \Phi^{\eff(1,0)}_{01} 
    =& \;\frac{1}{4}\cpartial_{0}\text{\TH}'_{2}h_{ll} 
    -\frac{1}{4}(\cpartial_{0}\text{\TH}_{0}+2\rho\cpartial_{0})h_{ln} 
    +\frac{1}{4}\cpartial_{2}\text{\TH}_{0}h_{m\bar{m}} 
    -\frac{1}{4}\cpartial'_{2}\text{\TH}_{0}h_{mm} 
    -\frac{1}{4}\cpartial_{0}\cpartial_{-1}h_{l\bar{m}} \nonumber\\
    & \;-\frac{1}{4}(-\cpartial_{0}\cpartial'_{1}
    +\text{\TH}_{0}\text{\TH}'_{1}-3\rho\text{\TH}'_{1}
    -\mu\text{\TH}_{1}-2\rho\mu + 3 \Psi_2)h_{lm} 
    +\frac{1}{4}(\text{\TH}_{0}\text{\TH}_{-1}
    -2\rho\text{\TH}_{-1}-2\rho^2)h_{n\bar{m}}\,, \\
    \Phi^{\eff(1,0)}_{21} 
    =& \;\frac{1}{4}\cpartial'_{0}\text{\TH}_{-2}h_{nn}
    -\frac{1}{4}(\cpartial'_{0}\text{\TH}'_{0}-2\mu\cpartial'_{0})h_{ln}
    +\frac{1}{4}\cpartial'_{2}\text{\TH}'_{0}h_{m\bar{m}}
    -\frac{1}{4}\cpartial_{-2}\text{\TH}'_{0}h_{\bar{m}\bar{m}}
    -\frac{1}{4}\cpartial'_{0}\cpartial'_{1}h_{nm} \nonumber\\
    & \;-\frac{1}{4}(-\cpartial'_{0}\cpartial_{-1}
    +\text{\TH}'_{0}\text{\TH}_{-1}+3\mu\text{\TH}_{-1}
    +\mu\text{\TH}'_{-1}-2\rho\mu+3\Psi_2)h_{n\bar{m}}
    +\frac{1}{4}(\text{\TH}'_{0}\text{\TH}'_{1}
    +2\mu\text{\TH}'_{1}-2\mu^2)h_{lm}\,, \\
    \Phi^{\eff(1,0)}_{11} 
    =& \;\frac{1}{8}(\text{\TH}'_{1}\text{\TH}'_{2}-2\mu^2)h_{ll}
    -\frac{1}{4}(\text{\TH}_{-1}\text{\TH}'_{0}-2\rho mu+2\Psi_2)h_{ln}
    +\frac{1}{8}(\text{\TH}_{-1}\text{\TH}_{-2}-2\rho^2)h_{nn} \nonumber\\
    & \;-\frac{1}{8}\cpartial'_{1}\cpartial'_{2}h_{mm}
    -\frac{1}{4}(-\cpartial_{-1}\cpartial'_{0}-\rho\text{\TH}'_{0}
    +\mu\text{\TH}_{0}-2\rho\mu+2\Psi_2)h_{m\bar{m}}
    -\frac{1}{8} \cpartial_{-1} \cpartial_{-2} h_{\bar{m}\bar{m}} \nonumber\\
    & \;+\frac{1}{4}\mu(\cpartial'_{1}h_{lm}+\cpartial_{-1}h_{l\bar{m}})
    -\frac{1}{4}\rho(\cpartial'_{1}h_{nm}+\cpartial_{-1} h_{n\bar{m}})\,, \\
    \Lambda^{\eff(1,0)} 
    =& \;-\frac{1}{24}(\text{\TH}'_{1}\text{\TH}'_{2}
    +4\mu\text{\TH}'_{2}+2\mu^2)h_{ll}
    -\frac{1}{12}(2 \cpartial_{-1}\cpartial'_{0}
    -\text{\TH}_{-1}\text{\TH}'_{0}-2\rho\mu+2\Psi_2)h_{ln}
    -\frac{1}{24}(\text{\TH}_{-1}\text{\TH}_{-2}
    -4\rho\text{\TH}'_{2}+2\rho^2)h_{nn} \nonumber\\
    & \;-\frac{1}{24}\cpartial'_{1}\cpartial'_{2}h_{mm}
    -\frac{1}{12}(-\cpartial_{-1}\cpartial'_{0}
    +2\text{\TH}_{-1}\text{\TH}'_{0}-3\rho\text{\TH}'_{0}
    +3\mu\text{\TH}_{0}-2\rho\mu+2\Psi_2)h_{m\bar{m}}
    -\frac{1}{24}\cpartial_{-1}\cpartial_{-2}h_{\bar{m}\bar{m}} \nonumber\\
    & \;+\frac{1}{12}(\cpartial'_{1}\text{\TH}'_{1}+2\mu\cpartial'_{1})h_{lm}
    +\frac{1}{12}(\cpartial_{-1}\text{\TH}'_{1}+2\mu\cpartial_{-1})h_{l\bar{m}}
    +\frac{1}{12}(\cpartial'_{1}\text{\TH}_{-1}-2\rho\cpartial'_{1})h_{nm}
    +\frac{1}{12}(\cpartial_{-1}\text{\TH}_{-1}-2\rho\cpartial_{-1})h_{n\bar{m}}\,.
\end{align}
\end{subequations}
On the other hand, the NP Ricci scalars at $\mathcal{O}(\zeta^0,\epsilon^1)$ can be directly found by projecting Eq.~\eqref{eq:stress_particle_circular} onto the NP tetrad such that
\begin{align} \label{eq:Ricci01}
    & \Phi^{p(0,1)}_{00}=\Phi^{p(0,1)}_{22}
    =\frac{2f(r)}{r^2}\pi m_p u_{(\delta)}^t\,,\quad
    \Phi^{p(0,1)}_{02}=\Phi^{p(0,1)}_{20}
    =-2\omega_z^2\pi m_p u_{(\delta)}^t\,, \nonumber\\
    & \Phi^{p(0,1)}_{21}=-\Phi^{p(0,1)}_{01}=\Phi^{p(0,1)}_{10}=-\Phi^{p(0,1)}_{12}
    =\frac{2i\sqrt{f(r)}\omega_z }{r}\pi m_p u_{(\delta)}^t\,, \nonumber\\
    & \Phi^{p(0,1)}_{11}=\left(\frac{f(r)}{r^2}
    +\omega_z^2\right)\pi m_pu^t_{(\delta)}\,,\quad
    \Lambda^{p(0,1)}
    =\frac{1}{3}\left(\frac{f(r)}{r^2}-\omega_z^2\right)\pi m_p u^t_{(\delta)}\,,
\end{align}
where we recall that $f(r)=1-2M/r$, $\omega_z=\sqrt{M/r_p^3}$, $r_p$ is the particle's radial position, $m_p$ is the particle's mass, and we define $u_{(\delta)}^t = u^t \delta^{(3)}(x^{i}_p - x^{i})=(1-3M/r_p)^{-1/2}\delta^{(3)}(x^{i}_p - x^{i})$, where $x^{i}_p = (r_p, \pi/2,\omega_z t)$ is the three-vector of the particle's position.

\section{NP Ricci scalars at $\mathcal{O}(\zeta^1,\epsilon^1)$}
\label{sec:RicciSourceTerms11}
In this appendix, we present the NP Ricci scalars $\Phi^{(1,1)}_{ij}$ with $i,j\in\{0,1,2\}$ in terms of metric perturbations and NP Ricci scalars at $\mathcal{O}(\zeta^1,\epsilon^0)$ (i.e., $h_{\mu\nu}$ and $\Phi^{\eff(1,0)}_{ij}$) and $\mathcal{O}(\zeta^0,\epsilon^1)$ (i.e., $q_{\mu\nu}$ and $\Phi^{p(0,1)}_{ij}$), respectively. We have divided up $\Phi_{ij}^{(1,1)}$ into two pieces based on whether they are driven by $\Phi^{\eff(1,0)}_{ij}$ or $\Phi^{p(0,1)}_{ij}$, i.e.,
\begin{equation}
    \Phi_{ij}^{(1,1)}=\Phi^{\eff(1,1)}_{ij}+\Phi^{p(1,1)}_{ij}\,,\quad
    i,j\in\{0,1,2\}\,,
\end{equation}
where
\begin{subequations}\label{eq:S006-SS001-EQ001}
\begin{align}
    \Phi^{\eff(1,1)}_{00}
    =& \;-q_{ln}\Phi^{\eff(1,0)}_{00}
    -q_{ll}\left(3\Lambda^{\eff(1,0)}+\Phi^{\eff(1,0)}_{11}\right)
    +\left(q_{lm}+2b\right)\Phi^{\eff(1,0)}_{10} \nonumber\\
    & \;+\left(q_{l\bar{m}}+2\bar{b}\right)\Phi^{\eff(1,0)}_{01}
    +\frac{1}{2}l^{\mu}l^{\nu}R^{\eff(1,1)}_{\mu\nu}\,, \\
    \Phi^{\eff(1,1)}_{22} 
    =& \;-q_{ln}\Phi^{\eff(1,0)}_{22}
    -q_{nn}\left(3\Lambda^{\eff(1,0)}+\Phi^{\eff(1,0)}_{11}\right)
    +\left(q_{n\bar{m}}+2\bar{a}\right)\Phi^{\eff(1,0)}_{12} \nonumber\\
    & \;+\left(q_{nm}+2a\right)\Phi^{\eff(1,0)}_{21}
    +\frac{1}{2}n^{\mu}n^{\nu}R^{\eff(1,1)}_{\mu\nu}\,, \\
    \Phi^{\eff(1,1)}_{02} 
    =& \;q_{m\bar{m}}\Phi^{\eff(1,0)}_{02}
    -q_{mm}\left(3\Lambda^{\eff(1,0)}-\Phi^{\eff(1,0)}_{11}\right)
    +\left(2a-q_{nm}\right)\Phi^{\eff(1,0)}_{01} \nonumber\\
    & \;+\left(2b-q_{lm}\right)\Phi^{\eff(1,0)}_{12}
    +\frac{1}{2}m^{\mu}m^{\nu}R^{\eff(1,1)}_{\mu\nu}\,, \\
    \Phi^{\eff(1,1)}_{20}
    =& \;q_{m\bar{m}}\Phi^{\eff(1,0)}_{20}
    -q_{\bar{m}\bar{m}}\left(3\Lambda^{\eff(1,0)}-\Phi^{\eff(1,0)}_{11}\right)
    +\left(2\bar{b}-q_{l\bar{m}}\right)\Phi^{\eff(1,0)}_{21} \nonumber\\
    & \;+\left(2\bar{a}-q_{n\bar{m}}\right)\Phi^{\eff(1,0)}_{10}
    +\frac{1}{2}\bar{m}^{\mu}\bar{m}^{\nu}R^{\eff(1,1)}_{\mu\nu}\,, \\
    \Phi^{\eff(1,1)}_{01} 
    =& \;\frac{1}{2}q_{mm}\Phi^{\eff(1,0)}_{10}
    -\frac{1}{2}\left(q_{nm}-2a\right)\Phi^{\eff(1,0)}_{00}
    +\frac{1}{2}\left(q_{m\bar{m}}-q_{ln}\right)\Phi^{\eff(1,0)}_{01}
    -\frac{1}{2}q_{ll}\Phi^{\eff(1,0)}_{12} \nonumber\\
    & \;+\frac{1}{2}(q_{l\bar{m}}+2\bar{b})\Phi^{\eff(1,0)}_{02} 
    -3q_{lm}\Lambda^{\eff(1,0)}-b\Phi^{\eff(1,0)}_{11}
    +\frac{1}{2}l^{\mu}m^{\nu}R^{\eff(1,1)}_{\mu\nu}\,, \\
    \Phi^{\eff(1,1)}_{21} 
    =& \;\frac{1}{2}q_{\bar{m}\bar{m}}\Phi^{\eff(1,0)}_{12}
    -\frac{1}{2}\left(q_{l\bar{m}}-2\bar{b}\right)\Phi^{\eff(1,0)}_{22}
    +\frac{1}{2}\left(q_{m\bar{m}}-q_{ln}\right)\Phi^{\eff(1,0)}_{21}
    -\frac{1}{2}q_{nn}\Phi^{\eff(1,0)}_{10} \nonumber\\
    & \;+\frac{1}{2}(q_{nm}+2a)\Phi^{\eff(1,0)}_{20}
    -3q_{n\bar{m}}\Lambda^{\eff(1,0)}-\bar{a}\Phi^{\eff(1,0)}_{11}
    +\frac{1}{2}n^{\mu}\bar{m}^{\nu}R^{\eff(1,1)}_{\mu\nu}\,.
\end{align}
\end{subequations}
and
\begin{subequations}
\begin{align}
    \Phi^{p(1,1)}_{00} 
    =& \;-h_{ln} \Phi^{p(0,1)}_{00}
    -h_{ll}\left(3\Lambda^{p(0,1)}+\Phi^{p(0,1)}_{11}\right)
    +h_{lm}\Phi^{p(0,1)}_{10}+h_{l\bar{m}}\Phi^{p(0,1)}_{01}\,, \\
    \Phi^{p(1,1)}_{22} 
    =& \;-h_{ln}\Phi^{p(0,1)}_{22}
    -h_{nn}\left(3\Lambda^{p(0,1)}+\Phi^{p(0,1)}_{11}\right)
    +h_{n\bar{m}}\Phi^{p(0,1)}_{12}+h_{nm}\Phi^{p(0,1)}_{21}\,, \\
    \Phi^{p(1,1)}_{02} 
    =& \;h_{m\bar{m}}\Phi^{p(0,1)}_{02}
    -h_{mm}\left(3\Lambda^{p(0,1)}-\Phi^{p(0,1)}_{11}\right)
    -h_{nm}\Phi^{p(0,1)}_{01}-h_{lm}\Phi^{p(0,1)}_{12}\,, \\
    \Phi^{p(1,1)}_{20} 
    =& \;h_{m\bar{m}}\Phi^{p(0,1)}_{20}
    -h_{\bar{m}\bar{m}}\left(3\Lambda^{p(0,1)}-\Phi^{p(0,1)}_{11}\right)
    -h_{l\bar{m}}\Phi^{p(0,1)}_{21}-h_{n\bar{m}}\Phi^{p(0,1)}_{10}\,, \\
    \Phi^{p(1,1)}_{01} 
    =& \;\frac{1}{2}h_{mm}\Phi^{p(0,1)}_{10}
    -\frac{1}{2}h_{nm}\Phi^{p(0,1)}_{00}
    +\frac{1}{2}\left(h_{m\bar{m}}-h_{ln}\right)\Phi^{p(0,1)}_{01}
    -\frac{1}{2}h_{ll}\Phi^{p(0,1)}_{12}
    +\frac{1}{2}h_{l\bar{m}}\Phi^{p(0,1)}_{02}
    -3h_{lm}\Lambda^{p(0,1)}\,, \\
    \Phi^{p(1,1)}_{21}
    =& \;\frac{1}{2}h_{\bar{m}\bar{m}}\Phi^{p(0,1)}_{12}
    -\frac{1}{2}h_{l\bar{m}}\Phi^{p(0,1)}_{22}
    +\frac{1}{2}\left(h_{m\bar{m}}-h_{ln}\right)\Phi^{p(0,1)}_{21}
    -\frac{1}{2}h_{nn}\Phi^{p(0,1)}_{10}
    +\frac{1}{2}h_{nm}\Phi^{p(0,1)}_{20}
    -3h_{n\bar{m}}\Lambda^{p(0,1)}\,.
\end{align}
\end{subequations}

\section{Calculation of $q_{\mu\nu}$ in GR}
\label{sec:MetricReconstruction}
In this appendix, we detail the procedure to reconstruct the metric perturbation $q_{\mu \nu}$ using the ZM and RW formalism. We begin by outlining our convention for decomposing the metric perturbation, master functions, and source terms in terms of spin-weighted spherical harmonics. Then we provide explicit formulas for the metric components in Eq.~\eqref{eq:PerturbedMetricComponets}. 

Due to the spherical symmetry of Schwarzschild BHs, we can write a generic metric perturbation using two classes of functions: the axial (odd type) and polar (even type) tensor spherical harmonics. We assume an ansatz for the metric perturbation $\mathbf{q}=q_{\mu\nu}e^{\mu}\otimes e^{\nu}$, where $\{e^{\mu}\}$ is the dual Boyer-Linquist vector basis, and $\mathbf{q}= \mathbf{q}^{e}+\mathbf{q}^{o}$ with
\begin{align}\label{eq:qOdd}
    \mathbf{q}^{o}
    =\sum_{\ell m}\frac{\sqrt{2\ell(\ell+1)}}{r}
    \left[ih_1^{\ell m}(t,r)\mathbf{c}_{\ell m}
    -h_0^{\ell m}(t,r)\mathbf{c}^{(0)}_{\ell m}\right]\,,
\end{align}
\begin{align}\label{eq:qEven}
    \mathbf{q}^{e}
    =\sum_{\ell m}f(r)H_0^{\ell m}(t,r)\mathbf{a}_{\ell m}^{(0)}
    -i\sqrt{2}H_1^{\ell m}(t,r)\mathbf{a}_{\ell m}^{(1)}
    +\frac{H_2^{\ell m}(t,r)}{f(r)}\mathbf{a}_{\ell m}
    +\sqrt{2}K^{\ell m}(t,r)\mathbf{g}_{\ell m}.
\end{align}
Similarly, we can decompose the stress-energy tensor of the test particle:
\begin{align}\label{eq:StressExansion}
    \mathbf{T}^p
    =\sum_{\ell m}&\bigg\{
    \mathcal{A}_{\ell m}^{(0)}\mathbf{a}_{\ell m}^{(0)}
    +\mathcal{A}_{\ell m}^{(1)}\mathbf{a}_{\ell m}^{(1)}
    +\mathcal{A}_{\ell m}\mathbf{a}_{\ell m} 
    +\mathcal{B}_{\ell m}^{(0)}\mathbf{b}_{\ell m}^{(0)}
    +\mathcal{B}_{\ell m}\mathbf{b}_{\ell m} \nonumber\\
    &+\mathcal{Q}_{\ell m}^{(0)}\mathbf{c}_{\ell m}^{(0)}
    +\mathcal{Q}_{\ell m}\mathbf{c}_{\ell m}
    +\mathcal{D}_{\ell m}\mathbf{d}_{\ell m}
    +\mathcal{G}_{\ell m}\mathbf{g}_{\ell m}
    +\mathcal{F}_{\ell m}\mathbf{f}_{\ell m}\bigg\}\,.
\end{align}
The explicit equations for the odd-type spin-weighted spherical harmonics $(\mathbf{c}_{\ell m},\mathbf{c}^{(0)}_{\ell m},\mathbf{d}_{\ell m})$ and even-type spin-weighted spherical harmonics $\left(\mathbf{a}_{\ell m}^{(0)}, \mathbf{a}_{\ell m}^{(1)}, \mathbf{a}_{\ell m}, \mathbf{b}_{\ell m}^{(0)}, \mathbf{b}_{\ell m}, \mathbf{g}_{\ell m}, \mathbf{f}_{\ell m}\right)$ are  given in Ref.~\cite{Sago:2002fe}. The expansion coefficients in Eq.~\eqref{eq:StressExansion} can be computed using the orthonormality properties of spin-weighted spherical harmonics, using the following inner product \cite{Cardoso:2018zhm},
\begin{align}
     (A, B)=\iint\gamma^{\mu\rho}\gamma^{\nu\sigma}
     \bar{A}_{\mu\nu}B_{\rho\sigma}\,d\Omega\,,
    \end{align}
where both $A$ and $B$ are rank-two tensors, $\gamma^{\mu\nu}$ is the Minkowski metric, $d\Omega$ is the differential solid angle on the two-sphere, and the bar denotes complex conjugation.

Using the Carter tetrad in Eq.~\eqref{eq:Carter_tetrad_Schw}, all nonvanishing NP projections $q^{e,\ell m}_{ab}$ are given by
\begin{align}\label{eq:S003-SS003-EQ002}
    q^{e,\ell m}_{ll} 
    =\frac{r}{r-2M}(H^{\ell m}_0+2H^{\ell m}_1+H^{\ell m}_2)\,,\quad
    q^{e,\ell m}_{ln} 
    =\frac{1}{2}(H^{\ell m}_0+H^{\ell m}_2)\,,\quad 
    q^{e,\ell m}_{nn} 
    =\frac{r-2M}{4r}(H^{\ell m}_0-2H^{\ell m}_1+H^{\ell m}_2)\,.
\end{align}
Similarly, the nonvanishing projections $q^{o,\ell m}_{ab}$ are given by
\begin{align}\label{eq:S003-SS003-EQ004}
   q^{o,\ell m}_{lm}
   =\frac{i}{\sqrt{2}r}\left(\frac{h^{\ell m}_0}{f(r)}
   +h^{\ell m}_1\right)\,,\quad
   q^{o,\ell m}_{nm}
   =-\frac{i}{2\sqrt{2}r}\left(h^{\ell m}_0-f(r)h^{\ell m}_1\right)\,,
\end{align}
where the other components can be obtained with complex conjugation. 

Using Eq.~\eqref{eq:StressExansion}, we can also decompose the source terms $S_{e/o}$ that appear in the RW and ZM equations, which are derived from the stress-energy tensor of a point particle [see Eq.~\eqref{eq:stress_particle}]. Using the tensor harmonic conventions in Ref.~\cite{Sago:2002fe}, it is useful to first define the following angular harmonics
\begin{subequations}
\begin{align}
    & X_{\ell m\omega}
    =2\frac{\partial}{\partial\phi}
    \left(\frac{\partial}{\partial\theta}
    -\cot\theta\right)Y_{\ell m\omega}\,, \\
    & W_{\ell m\omega}
    =\left(\frac{\partial^2}{\partial\theta^2}
    -\cot\theta\frac{\partial}{\partial\theta}
    -\frac{1}{\sin^2\theta}\frac{\partial^2}{\partial\phi^2}\right) Y_{\ell m \omega}\,,
\end{align}
\end{subequations}
where $Y_{\ell m\omega}$ are the ordinary scalar spherical harmonics in the Fourier domain evaluated at the worldline of the particle. For instance, 
\begin{align}
    Y_{\ell m \omega}
    =\int\,dt\,e^{i\omega t}Y_{\ell m}[\theta(t),\phi (t)]\,,
\end{align}
where $(\theta(t),\phi(t))$ parametrizes the angular coordinates of the particle's worldline, and $t$ is the coordinate time. The source term $S^{\ell m \omega}_{\mathrm{o}}$ can also be expressed in the Fourier domain \cite{Sago:2002fe}
\begin{align}
    S^{\ell m \omega}_o
    =-\frac{8\pi i(r-2M)}{g(\ell)}
    \frac{d}{dr}\left[\left(1-\frac{2M}{r}\right)
    D_{\ell m\omega}\right]\,,
\end{align}
where 
\begin{align}
    D_{\ell m \omega}
    =\frac{imu^{t}\delta(r-R(t))}{g(\ell)}
    \left(\frac{\omega^2}{2}\overline{X}_{\ell m\omega}\right)\,,
\end{align}
and $g(\ell)=[\frac{1}{2} l(\ell+1)(\ell-1)(\ell+2)]^{1/2}$. We have also assumed that $R(t)$ parametrizes the radial coordinate of the particle's worldline, where $t$ is still the coordinate time. The source term $S^{\ell m}_{\mathrm{e}}$ is given by
\begin{align}
    S^{\ell m \omega}_{e}
    =& \;\frac{r-2M}{ir}\frac{d}{dr}
    \left[\frac{(r-2M)^2}{r(\lambda r+3M)}
    \left(\widetilde{C}_{2\ell m\omega}
    -\frac{i16\pi r^3}{g(l)(r-2 M)}F_{\ell m}\right)\right]\nonumber\\
    & \;+i\frac{(r-2M)^2}{r(\lambda r+3M)^2} 
    \left[\frac{\lambda(\lambda+1)r^2+3\lambda Mr+6M^2}{r^2} \widetilde{C}_{2\ell m\omega}
    -i\frac{\lambda r^2-3\lambda Mr-3M^2}{g(\ell)(r-2M)}
    \left(16\pi rF_{\ell m\omega}\right)\right]\,,
\end{align}
where
\begin{align}
    F_{\ell m\omega}
    =-\frac{m u^{t}\delta(r-R(t))}{g(\ell)}
    \left(\frac{\omega}{2}\overline{W}_{\ell m\omega}\right)\,,\quad
    \widetilde{C}_{2\ell m\omega}
    =-\frac{8\pi r^2}{i\omega}
    \frac{\left[\frac{1}{2}\ell(\ell+1)\right]^{-1/2}}{r-2M}B_{\ell m \omega}^{(0)}
    +\frac{1}{g(\ell)}\frac{16\pi ir^3}{r-2M}F_{\ell m\omega}\,,
\end{align}
and
\begin{align}
    B_{\ell m\omega}^{(0)}
    =\left[\frac{1}{2}\ell(\ell+1)\right]^{-1/2}
    \frac{i mu^{t}}{\omega}\left(1-\frac{2M}{r}\right)
    r^{-1}\delta(r-R(t))\overline{Y}_{\ell m\omega}\,.
\end{align}
The coefficients of the delta distributions in Eq.~\eqref{eq:RW_Sources} can then be computed. We find them to be:
\begin{subequations} \label{eq:RW_SourcesCoeffs}
\begin{align}
    \mathfrak{c}_1 
    &=-\frac{4\pi mu^t{\omega}^2f}{g(\ell)^2}
    \overline{X}_{\ell m\omega}\,,\\
    \mathfrak{c}_2 
    &=\frac{4\pi mu^t{\omega}^2rf^2}{g(\ell)^2}
    \overline{X}_{\ell m\omega}\,,\\
    \mathfrak{c}_3 
    &=-\frac{8\pi imu^t}{r(\lambda r+3M)^2}
    (\lambda(\lambda+1)r^2+5\lambda Mr+12M^2)\overline{Y}_{\ell m\omega}
    -\frac{8\pi mu^trf\omega^2}{g(\ell)^2}\overline{W}_{\ell m\omega}\,,\\
    \mathfrak{c}_4 
    &=-\frac{8\pi imu^t}{r^2(\lambda r+3M)}
    (8M^3-12M^2+6M^2r+6Mr^2-r^3)\overline{Y}_{\ell m\omega}\,.
\end{align}   
\end{subequations}
After solving for $Q^{e/o}_{lm\omega}$ in Eq.~\eqref{eq:RWequs}, the metric components can be directly computed using the following: 
\begin{subequations} \label{eq:PerturbedMetricComponets}
\begin{align}
    K_{\ell m\omega}
    &= \frac{\lambda(\lambda+1)r^2+3\lambda Mr+6M^2}
    {r^2(\lambda r+3M)}Q_{\ell m\omega}^{e}
    +\frac{r-2M}{r}\frac{dQ_{\ell m\omega}^{e}}{dr}
    -\frac{r(r-2M)}{\lambda r+3M}
    \frac{16\pi r}{g(\ell)}F_{\ell m\omega}
    +\frac{i(r-2M)^2}{r(\lambda r+3M)}\widetilde{C}_{2\ell m\omega}\,, \\
    H_{1\ell m\omega}
    &=-i\omega\frac{\lambda r^2-3\lambda Mr-3M^2}
    {(r-2M)(\lambda r+3M)}Q_{\ell m\omega}^{e}
    -i\omega r\frac{dQ_{\ell m\omega}^{e}}{dr}
    +\frac{i\omega r^3}{\lambda r+3M}
    \frac{16\pi r}{g(\ell)}F_{\ell m\omega}
    +\frac{\omega r(r-2M)}{\lambda r+3M}
    \widetilde{C}_{2\ell m\omega}\,, \\
    H_{0\ell m\omega}
    &=\frac{\lambda r(r-2M)-\omega^2r^4+M(r-3M)}
    {(r-2M)(\lambda r+3M)}K_{\ell m\omega}
    +\frac{M(\lambda+1)-\omega^2r^3}
    {ir(\lambda r+3M)}H_{1\ell m\omega}\,, \\
    H_{2 \ell m\omega}
    &=H_{0\ell m\omega}-\frac{16\pi r^2}{g(\ell)}F_{\ell m\omega}\,, \\
    h_{1\ell m\omega}
    &=\frac{r^2}{r-2M}Q_{\ell m\omega}^{o}\,, \\
    h_{0\ell m\omega}
    &=\frac{i}{\omega}\frac{d}{dr^*}
    \left(rQ_{\ell m\omega}^{o}\right)
    -\frac{8\pi r(r-2M)}{\omega g(\ell)}D_{\ell m\omega}\,.
\end{align}
\end{subequations}

\newpage
\section{Source Coefficients}
\label{sec:SourceCoeffs}

\subsection{$\Psi_0$ Multiplicative Type Source Coefficients}\label{sec:PSI0SOURCES_NOND}
 
\captionof{table}{Table of type A generating coefficients for the source term $\mathcal{S}_{\eff}^{(1,1)}$ in Eq.~\eqref{eq:S_bGR}}\label{tab:typeAS_bGR}
\begin{adjustbox}{angle=90}
	\begin{tabular}{|| c ||  p{2cm} | p{2cm} | p{2cm} | p{2cm} | p{2cm} | p{2cm} | p{2cm} || } 
		 \hline
		 Operator & $\Phi^{(1,0)}_{00}$ & $\Phi^{(1,0)}_{01}$  & $\Phi^{(1,0)}_{02}$ & $\Phi^{(1,0)}_{12}$ & $\Phi^{(1,0)}_{10}$ & $\Phi^{(1,0)}_{11}$ & $\Lambda^{(1,0)}$ \\[0.5ex]
		 \hline\hline
		 $\cpartial\cpartial$ & $q_{ln}$ & $- q_{l\bar{m}} - 2\bar{b}$ &  $-q_{lm}-2b$ & $0$ & $0$ & $\frac{1}{2} q_{ll}$ & $3q_{ll}$ \\ [1ex] 
		 \hline
		 $\cpartial\text{\TH}$ & $-q_{nm}+2a$ & $-q_{ln}+q_{m\bar{m}}$ & $q_{l\bar{m}} + 2\bar{b}$ & $-q_{ll}$ & $q_{mm}$ & $2b$ & $-6q_{lm}$\\
		 \hline
		 $\cpartial\text{\TH}'$ & $0$ & $0$ & $0$ & $0$ & $0$ & $0$ & $0$ \\
		 \hline
		 $\cpartial$ & $ ( 2\text{\TH} - 6\rho ) a + (- \text{\TH} + 3\rho ) q_{nm} + 2 \cpartial q_{ln}$ & $ ( -\text{\TH} + 3\rho ) q_{ln} + ( \text{\TH} - 3 \rho ) q_{m\bar{m}} - 4 \cpartial \bar{b} - 2 \cpartial q_{l\bar{m}}$ & $ ( \text{\TH} - 3\rho ) q_{l\bar{m}} + ( 2\text{\TH} - 6\rho ) \bar{b} $ & $ ( - \cpartial + 3\rho )q_{ll} $ & $ -2\cpartial q_{lm} - 4 \cpartial b + ( \text{\TH} - 3\rho ) q_{mm} $ & $ \cpartial q_{ll} + ( 2 \text{\TH} - 6\rho ) b $ & $6\cpartial q_{ll} + ( - 6 \text{\TH} + 18 \rho )q_{lm} $ \\
		 \hline
		 $\text{\TH}\text{\TH}$ & $0$ & $q_{nm}-2a$ & $-q_{m\bar{m}}$ & $q_{lm}-2b$ & $0$ & $-\frac{1}{2} q_{mm}$ & $3q_{mm}$ \\
		 \hline
		 $\text{\TH}$ & $2\cpartial a - \cpartial q_{nm}$ & $- \cpartial q_{ln} + \cpartial q_{m\bar{m}} + ( 2 \text{\TH} - 6\rho ) q_{nm} + ( - 4 \text{\TH} + 12 \rho ) a$ & $ ( - 2\text{\TH} + 6\rho ) q_{m\bar{m}} + 2\cpartial \bar{b} + \cpartial q_{l\bar{m}} $ & $ - \cpartial q_{ll} + ( 2 \cpartial - 6\rho ) q_{lm} + ( - 4\cpartial + 12\rho ) b$ & $ \cpartial q_{mm} $ & $ 2 \cpartial b + ( - \text{\TH} + 3\rho ) q_{mm} $ & $ - 6 \cpartial q_{lm} + ( 6 \text{\TH} - 18\rho ) q_{mm} $  \\
		 \hline
		 $\text{\TH}\text{\TH}'$ & $0$ & $0$ & $0$ & $0$ & $0$ & $0$ & $0$ \\
		 \hline
		 $\cpartial\cpartial'$ & $0$ & $0$ & $0$ & $0$ & $0$ & $0$ & $0$ \\
		 \hline
		 $1$ & $ (- \cpartial \text{\TH} + 3\rho \cpartial ) q_{nm} + ( 2 \cpartial \text{\TH} - 6 \rho \cpartial ) a + \cpartial \cpartial q_{ln}$ & $(\text{\TH}\text{\TH} - 6\rho \text{\TH} + 4 \rho^2 )q_{nm} + ( - 2 \text{\TH}\text{\TH} + 12 \rho \text{\TH} - 8 \rho^2 ) a  - \cpartial \cpartial q_{l\bar{m}} - 2 \cpartial \cpartial \bar{b} + ( - \cpartial \text{\TH} + 3 \rho \cpartial ) q_{ln} + ( \cpartial \text{\TH} - 3\rho \cpartial ) q_{m\bar{m}} $ & $ (- \text{\TH}\text{\TH} + 6\rho \text{\TH} - 4\rho^2 ) q_{m\bar{m}} + ( \cpartial \text{\TH} - 3\rho \cpartial ) q_{l\bar{m}} + ( 2\cpartial \text{\TH} - 6\rho \cpartial )\bar{b}$ & $ ( - \cpartial \cpartial + 3\rho \cpartial ) q_{ll} + ( \cpartial\cpartial - 6\rho \cpartial + 4\rho^2 ) q_{lm} + ( - 2 \cpartial\cpartial + 12 \rho \cpartial - 8 \rho^2 ) b$ & $ ( \cpartial \text{\TH} - 3\rho \cpartial ) q_{mm} - \cpartial \cpartial q_{lm} - 2 \cpartial \cpartial b $ & $ ( - \frac{1}{2} \text{\TH}\text{\TH} + 3\rho \text{\TH} - 2 \rho^2 ) q_{mm} + ( 2 \cpartial \text{\TH} - 6\rho \cpartial ) b + \frac{1}{2} \cpartial \cpartial q_{ll}$ & $ ( 3 \text{\TH}\text{\TH} - 18 \rho \text{\TH} + 12 \rho^2 ) q_{mm} + ( - 6\cpartial \text{\TH} + 18 \rho \cpartial ) q_{lm} + 3 \cpartial \cpartial q_{ll}$ \\
		 \hline
	\end{tabular}
\end{adjustbox}

\newpage
\captionof{table}{Table of type B generating coefficients for the source term $\mathcal{S}_{\eff}^{(1,1)}$ in Eq.~\eqref{eq:S_bGR}}\label{tab:typeBS_bGR}
\begin{adjustbox}{angle=90}
	\begin{tabular}{|| c || p{4cm} | p{4cm} | p{4cm} | p{2cm} | p{2cm} | p{2cm} || } 
		 \hline
		 Operator & $\Phi^{(1,0)}_{00}$ & $\Phi^{(1,0)}_{01}$ & $\Phi^{(1,0)}_{02}$ & $\Phi^{(1,0)}_{11}$ & $\Phi^{(1,0)}_{12}$ & $\Phi^{(1,0)}_{10}$ \\[0.5ex]
		 \hline\hline
		 $\cpartial\cpartial$ & $q_{m\bar{m}}$ & $q_{l\bar{m}}+2\bar{b}$ & $0$ & $0$ & $0$ & $0$ \\ [1ex] 
		 \hline
		 $\cpartial\text{\TH}$ & $-q_{nm}-2a$ & $q_{ln}-q_{m\bar{m}}$ & $q_{l\bar{m}} - 2 \bar{b} $ & $0$ & $0$ & $0$  \\
		 \hline
		 $\cpartial\text{\TH}'$ & $-q_{lm}-2b$ & $q_{ll}$ & $0$ & $0$ & $0$ & $0$  \\
		 \hline
		 $\cpartial$ & $ \frac{3}{2} \cpartial q_{ln} + \cpartial q_{m\bar{m}} - \frac{1}{2} \cpartial' q_{mm} + ( - \frac{5}{2} \text{\TH} + \frac{3}{2} \rho ) q_{nm} + ( \text{\TH}' - \mu ) q_{lm} + \frac{1}{\Psi_2} \cpartial \Psi^{(0,1)}_{2} + ( - 2 \text{\TH} - 6 \rho ) a + ( - \text{\TH}' + \mu ) b $ & $ - \frac{5}{2} \cpartial q_{l\bar{m}} + ( \text{\TH} - 3\rho ) q_{ln} + ( \frac{3}{2} \text{\TH} + 3 \rho ) q_{m\bar{m}} + ( - \text{\TH}' + 3\mu ) q_{ll} - \frac{1}{\Psi_2} ( \text{\TH} - 3\rho ) \Psi^{(0,1)}_{2} + \cpartial' b + 4 \cpartial \bar{b} $ & $ \frac{1}{2} \cpartial' q_{ll} - 2 \rho q_{l\bar{m}} + ( -2 \text{\TH} + 6 \rho  ) \bar{b} $ & $ ( - \text{\TH} + \rho ) q_{lm} + \cpartial q_{ll} + ( - 2 \text{\TH} + 2 \rho ) b $ & $0$ & $ - \cpartial q_{lm} + \text{\TH} q_{mm} + 2 \cpartial b $ \\
		 \hline
		 $\text{\TH}\text{\TH}$ & $0$ & $q_{nm}+2a$ & $q_{ln}$ & $0$ & $0$ & $0$  \\
		 \hline
		 $\text{\TH}$ & $ - \frac{1}{2} \text{\TH}' q_{mm} - 2 \cpartial a$ & $ \mu q_{lm} + ( \frac{5}{2} \text{\TH} - \frac{11}{2} \rho ) q_{nm} + \cpartial' q_{mm} - \frac{3}{2} \cpartial q_{ln} - \cpartial q_{m\bar{m}} - \frac{1}{\Psi_2} \cpartial \Psi^{(0,1)}_{2} + ( \text{\TH}' + \mu ) b + ( 4 \text{\TH} - 12 \rho ) a $ & $ ( \frac{1}{2} \text{\TH}' - 3 \mu ) q_{ll} + ( - \text{\TH} + 6 \rho ) q_{ln} - \cpartial' q_{lm} + \frac{5}{2} \cpartial q_{l\bar{m}} - \frac{3}{2} \text{\TH} q_{m\bar{m}} + \frac{1}{\Psi_2} ( \text{\TH} - 3\rho ) \Psi^{(0,1)}_{2} - \cpartial' b -  2 \cpartial \bar{b}$ & $ - \text{\TH} q_{mm} + \cpartial q_{lm} - 2 \cpartial b$ & $ ( \text{\TH} - \rho ) q_{lm} - \cpartial q_{ll} + ( 2 \text{\TH} - 2 \rho ) b $ &  \\
		 \hline
		 $\text{\TH}\text{\TH}'$ & $0$ & $q_{lm}+2b$ & $0$ & $0$ & $0$ & $0$  \\
		 \hline
		 $\cpartial\cpartial'$ & $0$ & $-q_{lm}+2b$ & $0$ & $0$ & $0$ & $0$ \\
		 \hline
		 $1$ & $ \frac{1}{2} ( - \text{\TH} \text{\TH}' + 5 \rho \text{\TH}' ) q_{mm} + \frac{1}{2} \cpartial \cpartial q_{ln} + \frac{1}{2} \cpartial \text{\TH}' q_{lm} + ( - \frac{1}{2} \cpartial \text{\TH} - \frac{3}{2} \rho \cpartial ) q_{nm} + \mu \cpartial b + ( - 2 \cpartial \text{\TH} + 6 \rho \cpartial ) a  $ & $ ( \frac{1}{2} \cpartial \text{\TH} + 3 \rho \cpartial ) q_{m\bar{m}} + ( - \frac{1}{2} \cpartial \text{\TH} + 5\rho \cpartial ) q_{ln} + ( - \frac{1}{2} \cpartial \text{\TH}' + \mu \cpartial ) q_{ll} + ( \frac{1}{2} \cpartial' \text{\TH} - 3\rho \cpartial'  ) q_{mm} - \frac{1}{2} \cpartial \cpartial q_{l\bar{m}} + ( - \frac{1}{2} \cpartial \cpartial' + \frac{1}{2} \text{\TH} \text{\TH}' - \frac{3}{2} \rho \text{\TH}' - \frac{1}{2} \mu \text{\TH} - \rho \mu + \frac{5}{2} \Psi_2 ) q_{lm} + ( \frac{1}{2} \text{\TH} \text{\TH} - 6 \rho \text{\TH} + 4 \rho^2 ) q_{nm} + \frac{2\rho}{\Psi_2} \cpartial \Psi^{(0,1)}_{2} + ( 2 \text{\TH} \text{\TH} - 12 \rho \text{\TH} + 8 \rho^2 ) a + ( 2 \cpartial \cpartial - 2 \rho \text{\TH}' + 2 \rho \mu + 2 \Psi_2 ) b $ & $ ( - \frac{1}{2} \text{\TH} \text{\TH} + \frac{7}{2} \rho \text{\TH} ) q_{m\bar{m}} + ( \frac{1}{2} \cpartial' \cpartial - \frac{1}{2} \rho \text{\TH}' - \frac{1}{2} \mu \text{\TH} + 4 \rho \mu - \Psi_2 ) q_{ll} + ( \mu \text{\TH} - 4 \rho mu ) q_{ln} + ( - \frac{1}{2} \cpartial' \text{\TH} + \frac{5}{2} \rho \cpartial' ) q_{lm} + ( \frac{1}{2} \cpartial \text{\TH} - 5 \rho \cpartial ) q_{l\bar{m}} - \frac{1}{\Psi_2} ( \rho \text{\TH} - 3 \rho^2 ) \Psi^{(0,1)}_{2}+ \rho \cpartial' b + ( - 2 \cpartial \text{\TH} + 6 \rho \cpartial ) \bar{b} $ & $ ( - \text{\TH} \text{\TH} + 5 \rho \text{\TH} ) q_{mm} + \cpartial \cpartial q_{ll} - 3 \rho \cpartial q_{lm} + ( - 4 \cpartial \text{\TH} + 10 \rho \cpartial ) b $ & $ ( \text{\TH} \text{\TH} - 6 \rho + 4 \rho^2 ) q_{lm} + ( - \cpartial + 4 \rho ) q_{ll} + ( 2 \text{\TH} \text{\TH} - 12 \rho + 8 \rho^2 ) b $ & $ - \cpartial \cpartial q_{lm} + \cpartial \text{\TH} q_{mm} + 2 \cpartial \cpartial b$ \\
		 \hline
		 $\cpartial'\cpartial$ & $0$ & $0$ & $0$ & $0$ & $0$ & $0$ \\
		 \hline
		 $\text{\TH}'\text{\TH}$ & $0$ & $0$ & $-q_{ll}$ & $0$ & $0$ & $0$ \\
		 \hline
		 $\text{\TH}'$ & $ - \frac{1}{2} \cpartial q_{lm} - \cpartial b $ & $ \frac{1}{2} \cpartial q_{ll} + ( \frac{1}{2} \text{\TH} - \frac{7}{2} \rho ) q_{lm} + ( \text{\TH} - 7\rho ) b $ & $ ( - \frac{1}{2} \text{\TH} + 3 \rho ) q_{ll} $ & $0$ & $0$ & $0$ \\
		 \hline
		 $\text{\TH}'\text{\TH}'$ & $0$ & $0$ & $0$ & $0$ & $0$ & $0$ \\
		 \hline
		 $\cpartial'$ & $\frac{1}{2} \cpartial q_{mm}$ & $ - \frac{1}{2} \cpartial q_{lm} + ( -\frac{1}{2} \text{\TH} + 3 rho ) q_{mm} + \cpartial b $ & $ ( \frac{1}{2} \text{\TH} - \frac{5}{2} \rho ) q_{lm} + ( - \text{\TH} + 5\mu ) b $ & $0$ & $0$ & $0$ \\
		 \hline
		 $\cpartial'\text{\TH}$ & $0$ & $-q_{mm}$ & $q_{lm}-2b$ & $0$ & $0$ & $0$ \\
		 \hline
		 $\cpartial'\text{\TH}'$ & $0$ & $0$ & $0$ & $0$ & $0$ & $0$ \\
		 \hline
		 $\cpartial'\cpartial'$ & $0$ & $0$ & $0$ & $0$ & $0$ & $0$ \\ [1ex] 
		 \hline
	\end{tabular}
\end{adjustbox}

\newpage
\captionof{table}{Table of generating coefficients for the source term $\mathcal{S}_{\geo}^{(1,1)}$ in Eq.~\eqref{eq:S_geo}}\label{tab:S_geoTable}
\begin{adjustbox}{angle=90}
        \begin{tabular}{|| c || p{6cm} | p{6cm} | p{6cm} || } 
		 \hline
		 Operator & $\Psi^{(0,1)}_0$ & $\Psi^{(1,0)}_0$ &  $\Psi^{(1,0)}_1$ \\[0.5ex]
		 \hline\hline
		 $\cpartial\cpartial$ & $-\frac{1}{2}h_{\bar{m}\bar{m}}$ & $-\frac{1}{2}q_{\bar{m}\bar{m}}$ & $0$ \\ [1ex] 
		 \hline
		 $\cpartial\text{\TH}$ & $h_{n\bar{m}}$ & $q_{n\bar{m}}$ & $0$ \\ 
		 \hline
		 $\cpartial\text{\TH}'$ & $h_{l\bar{m}}$ & $q_{l\bar{m}}$ & $0$ \\ 
		 \hline
		 $\cpartial$ & $- \frac{1}{2} \cpartial' h_{ln} + ( - \frac{1}{2} \text{\TH}' + \frac{3}{2} \mu ) h_{l\bar{m}} + ( \frac{3}{2} \text{\TH} - \frac{7}{2} \rho )h_{n\bar{m}} + \cpartial' h_{m\bar{m}} - \frac{3}{2} \cpartial h_{\bar{m}\bar{m}}$ & $ - \frac{1}{2} \cpartial' q_{ln} + ( - \frac{1}{2} \text{\TH}' + \frac{3}{2} \mu ) q_{l\bar{m}} + ( \frac{3}{2} \text{\TH} - \frac{7}{2} \rho )q_{n\bar{m}} + \cpartial' q_{m\bar{m}} - \frac{3}{2} \cpartial q_{\bar{m}\bar{m}}$ & $- \frac{3}{2}  \text{\TH} q_{m\bar{m}} + \frac{3}{2} \cpartial q_{l\bar{m}} + \frac{1}{\Psi_2} (  - \text{\TH} + 3 \rho ) \Psi^{(1,0)}_2 - 3 \cpartial' b$ \\ 
		 \hline
		 $\text{\TH}\text{\TH}$ & $-\frac{1}{2}h_{nn}$ & $-\frac{1}{2}q_{nn}$ & $0$ \\ 
		 \hline
		 $\text{\TH}$ & $ ( \text{\TH}' - \mu ) h_{ln} + ( - \frac{3}{2} \text{\TH}' + 3 \rho ) h_{nn} + \frac{3}{2} \cpartial h_{n\bar{m}} - \frac{1}{2} \cpartial' h_{nm} - \text{\TH}' h_{m\bar{m}}$ & $ ( \text{\TH} - \mu ) q_{ln} + ( - \frac{3}{2} \text{\TH} + 3 \rho ) q_{nn} + \frac{3}{2} \cpartial q_{n\bar{m}} - \frac{1}{2} \cpartial' q_{nm} - \text{\TH}' q_{m\bar{m}}$ & $ ( \frac{3}{2} \text{\TH} + \frac{3}{2} \rho ) q_{nm} - \frac{3}{2} \cpartial q_{ln} + \frac{1}{\Psi_2} \cpartial \Psi^{(1,0)}_2 + ( 3 \text{\TH}' - 3 \mu ) b$ \\ 
		 \hline
		 $\text{\TH}\text{\TH}'$ & $-\frac{1}{2}h_{ln}$ & $-\frac{1}{2}q_{ln}$ & $0$ \\ 
		 \hline
		 $\cpartial\cpartial'$ & $-\frac{1}{2}h_{m\bar{m}}$ & $-\frac{1}{2}q_{m\bar{m}}$ & $0$ \\ 
		 \hline
		 $1$ & $  ( \frac{1}{2} \mu \text{\TH}' - 2 \mu^2 ) h_{ll} +  ( - \frac{1}{2} \cpartial \cpartial' + \text{\TH} \text{\TH}' - 5\rho \text{\TH}' - \mu \text{\TH} + 4 \rho \mu + \Psi_2) h_{ln} + ( \text{\TH} \text{\TH} + \frac{11}{2} \rho \text{\TH} - 2 \rho^2 ) h_{nn} - \frac{1}{2} \mu \cpartial' h_{lm} + ( - \frac{1}{2} \cpartial \text{\TH}' + \frac{5}{2} \mu \cpartial ) h_{l\bar{m}} + ( - \frac{1}{2} \cpartial' \text{\TH} + 2 \rho \cpartial' ) h_{nm} + (2 \cpartial \text{\TH} - \frac{11}{2} \rho \cpartial )h_{n\bar{m}} - \cpartial \cpartial h_{\bar{m}\bar{m}} + ( \cpartial' \cpartial - \frac{1}{2} \text{\TH}' \text{\TH} + \frac{5}{2} \rho \text{\TH}' - \mu \text{\TH} ) h_{m\bar{m}} + \frac{1}{\Psi_2} ( \mu \text{\TH} - 3 \rho \mu + 3 \Psi_2 ) \Psi^{(1,0)}_2$ & $ ( \frac{1}{2} \mu \text{\TH}' - 2 \mu^2 ) q_{ll} + ( - \frac{1}{2} \cpartial \cpartial' + \text{\TH} \text{\TH}' - 5\rho \text{\TH}' - \mu \text{\TH} + 4 \rho \mu + \Psi_2) q_{ln} + ( \text{\TH} \text{\TH} + \frac{11}{2} \rho \text{\TH} - 2 \rho^2 ) q_{nn} - \frac{1}{2} \mu \cpartial' q_{lm} + ( - \frac{1}{2} \cpartial \text{\TH}' + \frac{5}{2} \mu \cpartial ) q_{l\bar{m}} + ( - \frac{1}{2} \cpartial' \text{\TH} + 2 \rho \cpartial' ) q_{nm} + (2 \cpartial \text{\TH} - \frac{11}{2} \rho \cpartial ) q_{n\bar{m}} - \cpartial \cpartial q_{\bar{m}\bar{m}} + ( \cpartial' \cpartial - \frac{1}{2} \text{\TH}' \text{\TH} + \frac{5}{2} \rho \text{\TH}' - \mu \text{\TH} ) q_{m\bar{m}} + \frac{1}{\Psi_2} ( \mu \text{\TH} - 3 \rho \mu + 3 \Psi_2 ) \Psi^{(0,1)}_2 - \mu \cpartial' b$ & $ ( \frac{1}{2} \cpartial \text{\TH}' - 2 \mu \cpartial ) q_{ll} + ( - \frac{5}{2} \cpartial \text{\TH} + 6 \rho \cpartial ) q_{ln} + \frac{1}{2} \cpartial' \text{\TH} q_{mm} - \frac{5}{2} \cpartial \text{\TH} q_{m\bar{m}} + ( - \frac{1}{2} \cpartial \cpartial' - \frac{1}{2} \text{\TH}' \text{\TH} + \frac{3}{2} \mu \text{\TH} + \frac{1}{2} \rho \text{\TH}' - 3 \rho \mu + \frac{1}{2} \Psi_2 ) q_{lm} + \frac{5}{2} \cpartial \cpartial q_{l\bar{m}} + ( \frac{5}{2} \text{\TH} \text{\TH}  - 6 \rho \text{\TH} - 6 \rho^2 ) q_{nm} - \frac{4\rho}{\Psi_2} \cpartial \Psi^{(1,0)}_2 + ( - 4 \cpartial' \cpartial + 4 \text{\TH} \text{\TH}' - 2 \mu \text{\TH} - 16 \rho \text{\TH}' + 18 \rho \mu - 18 \Psi_2 ) b$ \\ 
		 \hline
		 $\cpartial'\cpartial$ & $-\frac{1}{2}h_{m\bar{m}}$ & $-\frac{1}{2}q_{m\bar{m}}$ & $0$ \\ 
		 \hline
		 $\text{\TH}'\text{\TH}$ & $-\frac{1}{2}h_{ln}$ & $-\frac{1}{2}q_{ln}$ & $0$ \\ 
		 \hline
		 $\text{\TH}'$ & $ ( - \text{\TH} + 5\rho ) h_{ln} + ( \frac{1}{2} \text{\TH}' - 3 \mu ) h_{ll} - \frac{1}{2} \cpartial' h_{lm} + 2 \cpartial h_{l\bar{m}} - \text{\TH} h_{m\bar{m}} + \frac{1}{\Psi_2} ( \text{\TH} - 3 \rho ) \Psi^{(1,0)}_2$ & $ ( - \text{\TH} + 5\rho ) q_{ln} +( \frac{1}{2} \text{\TH}' - 3 \mu ) q_{ll} - \frac{1}{2} \cpartial' q_{lm} + 2 \cpartial q_{l\bar{m}} + \text{\TH} q_{m\bar{m}} + \frac{1}{\Psi_2} ( \text{\TH} - 3 \rho ) \Psi^{(0,1)}_2 - \cpartial' b$ & $( \frac{1}{2} \text{\TH} - \frac{1}{2} \rho ) q_{lm} - \frac{1}{2} \cpartial q_{ll} + ( \text{\TH} - \rho ) b$ \\ 
		 \hline
		 $\text{\TH}'\text{\TH}'$ & $-\frac{1}{2}h_{ll}$ & $-\frac{1}{2}q_{ll}$ & $0$ \\ 
		 \hline
		 $\cpartial'$ & $- \cpartial h_{ln} + ( 2 \text{\TH} - 3\rho ) h_{nm} + ( - \frac{1}{2} \text{\TH}' + \frac{3}{2} \mu ) h_{lm} - \cpartial h_{m\bar{m}} + \frac{1}{2} \cpartial' h_{mm} - \frac{1}{\Psi_2} \cpartial \Psi^{(1,0)}_2$ & $- \cpartial q_{ln} + ( 2 \text{\TH} - 3\rho ) q_{nm} + ( - \frac{1}{2} \text{\TH}' + \frac{3}{2} \mu ) q_{lm} - \cpartial q_{m\bar{m}} + \frac{1}{2} \cpartial' q_{mm} - \frac{1}{\Psi_2} \cpartial \Psi^{(0,1)}_2 + ( \text{\TH}' - \mu ) b$ & $- \frac{1}{2}  \text{\TH} q_{mm} + \frac{1}{2} \cpartial q_{lm} - \cpartial b$ \\ 
		 \hline
		 $\cpartial'\text{\TH}$ & $h_{nm}$ & $q_{nm}$ & $0$ \\ 
		 \hline
		 $\cpartial'\text{\TH}'$ & $h_{lm}$ & $q_{lm}$ & $0$ \\ 
		 \hline
		 $\cpartial'\cpartial'$ & $-\frac{1}{2}h_{mm}$ & $-\frac{1}{2}q_{mm}$ & $0$ \\ [1ex] 
		 \hline
        \end{tabular}
\end{adjustbox}   

\newpage
\subsection{$\Psi_4$ Multiplicative Type Source Coefficients}
\label{sec:PSI4SOURCES_NOND}

\captionof{table}{Table of type A generating coefficients for the source term $\mathcal{T}_{\eff}^{(1,1)}$ in Eq.~\eqref{eq:T_bGR}}\label{tab:typeAT_bGR}
\begin{adjustbox}{angle=90}
	\begin{tabular}{|| c ||  p{2cm} | p{2cm} | p{2cm} | p{2cm} | p{2cm} | p{2cm} | p{2cm} || } 
		 \hline
		 Operator & $\Phi^{(0,1)}_{22}$ & $\Phi^{(0,1)}_{21}$  & $\Phi^{(0,1)}_{20}$ & $\Phi^{(0,1)}_{10}$ & $\Phi^{(0,1)}_{12}$ & $\Phi^{(0,1)}_{11}$ & $\Lambda^{(0,1)}$ \\[0.5ex]
		 \hline\hline
		 $\cpartial'\cpartial'$ & $q_{ln}$ & $- q_{nm} - 2a$ &  $-q_{n\bar{m}}-2\bar{a}$ & $0$ & $0$ & $\frac{1}{2} q_{nn}$ & $3q_{nn}$ \\ [1ex] 
		 \hline
		 $\cpartial'\text{\TH}'$ & $-q_{l\bar{m}}+2\bar{b}$ & $-q_{ln}+q_{m\bar{m}}$ & $q_{nm} + 2a$ & $-q_{nn}$ & $q_{\bar{m}\bar{m}}$ & $2\bar{a}$ & $-6q_{n\bar{m}}$\\
		 \hline
		 $\cpartial'\text{\TH}$ & $0$ & $0$ & $0$ & $0$ & $0$ & $0$ & $0$ \\
		 \hline
		 $\cpartial'$ & $ ( 2\text{\TH}' + 6\mu ) \bar{b} + (- \text{\TH}' - 3\mu ) q_{l\bar{m}} + 2 \cpartial' q_{ln}$ & $ ( -\text{\TH}' - 3\mu ) q_{ln} + ( \text{\TH}' + 3 \mu ) q_{m\bar{m}} - 4 \cpartial' a - 2 \cpartial' q_{nm}$ & $ ( \text{\TH}' + 3\mu ) q_{nm} + ( 2\text{\TH}' + 6\mu ) a $ & $ ( - \cpartial' - 3\mu )q_{nn} $ & $ -2\cpartial' q_{n\bar{m}} - 4 \cpartial' \bar{a} + ( \text{\TH}' + 3\mu ) q_{\bar{m}\bar{m}} $ & $ \cpartial' q_{nn} + ( 2 \text{\TH}' + 6\mu ) \bar{a} $ & $6\cpartial' q_{nn} + ( - 6 \text{\TH}' - 18 \mu )q_{n\bar{m}} $ \\
		 \hline
		 $\text{\TH}'\text{\TH}'$ & $0$ & $q_{l\bar{m}}-2\bar{b}$ & $-q_{m\bar{m}}$ & $q_{n\bar{m}}-2\bar{a}$ & $0$ & $-\frac{1}{2} q_{\bar{m}\bar{m}}$ & $3q_{\bar{m}\bar{m}}$ \\
		 \hline
		 $\text{\TH}'$ & $2\cpartial' \bar{b} - \cpartial' q_{l\bar{m}}$ & $- \cpartial' q_{ln} + \cpartial' q_{m\bar{m}} + ( 2 \text{\TH}' + 6\mu ) q_{l\bar{m}} + ( - 4 \text{\TH}' - 12 \mu ) \bar{b} $ & $ ( - 2\text{\TH}' - 6\mu ) q_{m\bar{m}} + 2\cpartial' a + \cpartial' q_{nm} $ & $ - \cpartial' q_{nn} + ( 2 \cpartial' + 6\mu ) q_{n\bar{m}} + ( - 4\cpartial' - 12\mu ) \bar{a}$ & $ \cpartial' q_{\bar{m}\bar{m}} $ & $ 2 \cpartial' \bar{a} + ( - \text{\TH}' - 3\mu ) q_{\bar{m}\bar{m}} $ & $ - 6 \cpartial' q_{n\bar{m}} + ( 6 \text{\TH}' + 18\mu ) q_{\bar{m}\bar{m}} $  \\
		 \hline
		 $\text{\TH}\text{\TH}'$ & $0$ & $0$ & $0$ & $0$ & $0$ & $0$ & $0$ \\
		 \hline
		 $\cpartial\cpartial'$ & $0$ & $0$ & $0$ & $0$ & $0$ & $0$ & $0$ \\
		 \hline
		 $1$ & $ (- \cpartial' \text{\TH}' - 3\mu \cpartial' ) q_{l\bar{m}} + ( 2 \cpartial' \text{\TH}' + 6 \mu \cpartial' ) \bar{b} + \cpartial' \cpartial' q_{ln}$ & $(\text{\TH}'\text{\TH}' + 6\mu \text{\TH}' + 4 \mu^2 )q_{l\bar{m}} + ( - 2 \text{\TH}'\text{\TH}' - 12 \mu \text{\TH}' - 8 \mu^2 ) \bar{b}  - \cpartial' \cpartial' q_{nm} - 2 \cpartial' \cpartial' a + ( - \cpartial' \text{\TH}' - 3 \mu \cpartial' ) q_{ln} + ( \cpartial' \text{\TH}' + 3\mu \cpartial' ) q_{m\bar{m}} $ & $ (- \text{\TH}\text{\TH} + 6\rho \text{\TH} - 4\rho^2 ) q_{m\bar{m}} + ( \cpartial \text{\TH} - 3\rho \cpartial ) q_{l\bar{m}} + ( 2\cpartial \text{\TH} - 6\rho \cpartial )\bar{b}$ & $ ( - \cpartial \cpartial + 3\rho \cpartial ) q_{ll} + ( \cpartial\cpartial - 6\rho \cpartial + 4\rho^2 ) q_{lm} + ( - 2 \cpartial\cpartial + 12 \rho \cpartial - 8 \rho^2 ) b$ & $ ( \cpartial \text{\TH} - 3\rho \cpartial ) q_{mm} - \cpartial \cpartial q_{lm} - 2 \cpartial \cpartial b $ & $ ( - \frac{1}{2} \text{\TH}'\text{\TH}' + 3\rho' \text{\TH}' - 2 \rho^{\prime2} ) q_{\bar{m}\bar{m}} + ( 2 \cpartial' \text{\TH} - 6\rho' \cpartial ) b + \frac{1}{2} \cpartial' \cpartial' q_{nn}$ & $ ( 3 \text{\TH}'\text{\TH}' - 18 \rho' \text{\TH}' + 12 \rho^{\prime 2} ) q_{\bar{m}\bar{m}} + ( - 6\cpartial \text{\TH} + 18 \rho' \cpartial ) q_{lm} + 3 \cpartial' \cpartial' q_{nn}$ \\
		 \hline
	\end{tabular}
\end{adjustbox}

\newpage
\captionof{table}{Table of type B generating coefficients for the source term $\mathcal{T}_{\eff}^{(1,1)}$ in Eq.~\eqref{eq:T_bGR}}\label{tab:typeBT_bGR}
\begin{adjustbox}{angle=90}
	\begin{tabular}{|| c || p{4cm} | p{4cm} | p{4cm} | p{2cm} | p{2cm} | p{2cm} || } 
		 \hline
		 Operator & $\Phi^{(1,0)}_{22}$ & $\Phi^{(1,0)}_{21}$ & $\Phi^{(1,0)}_{20}$ & $\Phi^{(1,0)}_{11}$ & $\Phi^{(1,0)}_{10}$ & $\Phi^{(1,0)}_{12}$ \\[0.5ex]
		 \hline\hline
		 $\cpartial\cpartial$ & $0$ & $0$ & $0$ & $0$ & $0$ & $0$ \\ [1ex] 
		 \hline
		 $\cpartial\text{\TH}$ & $0$ & $0$ & $0$ & $0$ & $0$ & $0$ \\
		 \hline
		 $\cpartial\text{\TH}'$ & $0$ & $-q_{\bar{m}\bar{m}}$ & $q_{n\bar{m}}-2\bar{a}$ & $0$ & $0$ & $0$ \\
		 \hline
		 $\cpartial$ &  $\frac{1}{2} \cpartial' q_{\bar{m}\bar{m}}$ & $ - \frac{1}{2} \cpartial' q_{n\bar{m}} + ( -\frac{1}{2} \text{\TH}' - 3 \mu ) q_{\bar{m}\bar{m}} + \cpartial' \bar{a} $ & $ ( \frac{1}{2} \text{\TH}' + \frac{5}{2} \mu ) q_{n\bar{m}} + ( - \text{\TH}' - 5\rho ) \bar{a} $ & $0$ & $0$ & $0$\\
		 \hline
		 $\text{\TH}\text{\TH}$ & $0$ & $0$ & $0$ & $0$ & $0$ & $0$ \\
		 \hline
		 $\text{\TH}$ & $ - \frac{1}{2} \cpartial' q_{n\bar{m}} - \cpartial' \bar{a} $ & $ \frac{1}{2} \cpartial' q_{nn} + ( \frac{1}{2} \text{\TH}' + \frac{7}{2} \mu ) q_{n\bar{m}} + ( \text{\TH}' + 7\mu ) \bar{a} $ & $ ( - \frac{1}{2} \text{\TH}' - 3 \mu ) q_{nn} $ & $0$ & $0$ & $0$ \\
		 \hline
		 $\text{\TH}\text{\TH}'$ & $0$ & $0$ & $-q_{nn}$ & $0$ & $0$ & $0$\\
		 \hline
		 $\cpartial\cpartial'$ & $0$ & $0$ & $0$ & $0$ & $0$ & $0$ \\
		 \hline
		 $1$ & $ \frac{1}{2} ( - \text{\TH}' \text{\TH} - 5 \mu \text{\TH} ) q_{\bar{m}\bar{m}} + \frac{1}{2} \cpartial' \cpartial' q_{ln} + \frac{1}{2} \cpartial' \text{\TH} q_{n\bar{m}} + ( - \frac{1}{2} \cpartial' \text{\TH}' + \frac{3}{2} \mu \cpartial' ) q_{l\bar{m}} - \rho \cpartial' \bar{a} + ( - 2 \cpartial' \text{\TH}' - 6 \mu \cpartial' ) \bar{b}  $ & $ ( \frac{1}{2} \cpartial' \text{\TH}' - 3 \mu \cpartial' ) q_{m\bar{m}} + ( - \frac{1}{2} \cpartial' \text{\TH}' - 5\mu \cpartial' ) q_{ln} + ( - \frac{1}{2} \cpartial' \text{\TH} - \rho \cpartial' ) q_{nn} + ( \frac{1}{2} \cpartial \text{\TH}' + 3\mu \cpartial  ) q_{\bar{m}\bar{m}} - \frac{1}{2} \cpartial' \cpartial' q_{nm} + ( - \frac{1}{2} \cpartial' \cpartial + \frac{1}{2} \text{\TH}' \text{\TH} + \frac{3}{2} \mu \text{\TH} + \frac{1}{2} \rho \text{\TH}' - \rho \mu + \frac{5}{2} \Psi_2 ) q_{n\bar{m}} + ( \frac{1}{2} \text{\TH}' \text{\TH}' + 6 \mu \text{\TH}' + 4 \mu^2 ) q_{l\bar{m}} - \frac{2\mu}{\Psi_2} \cpartial' \Psi^{(0,1)}_{2} + ( 2 \text{\TH}' \text{\TH}' + 12 \mu \text{\TH}' + 8 \mu^2 ) \bar{b} + ( 2 \cpartial' \cpartial' + 2 \mu \text{\TH} + 2 \rho \mu + 2 \Psi_2 ) \bar{a} $ & $ ( - \frac{1}{2} \text{\TH}' \text{\TH}' - \frac{7}{2} \mu \text{\TH}' ) q_{m\bar{m}} + ( \frac{1}{2} \cpartial \cpartial' + \frac{1}{2} \mu \text{\TH} + \frac{1}{2} \rho \text{\TH}' + 4 \rho \mu - \Psi_2 ) q_{nn} + ( - \rho \text{\TH}' - 4 \rho \mu ) q_{ln} + ( - \frac{1}{2} \cpartial \text{\TH}' - \frac{5}{2} \mu \cpartial ) q_{n\bar{m}} + ( \frac{1}{2} \cpartial' \text{\TH}' + 5 \mu \cpartial' ) q_{nm} + \frac{1}{\Psi_2} ( \mu \text{\TH}' + 3 \mu^2 ) \Psi^{(0,1)}_{2} - \mu \cpartial \bar{a} + ( - 2 \cpartial' \text{\TH}' - 6 \mu \cpartial' ) a $ & $ ( - \text{\TH}' \text{\TH}' - 5 \mu \text{\TH}' ) q_{\bar{m}\bar{m}} + \cpartial' \cpartial' q_{nn} + 3 \mu \cpartial' q_{n\bar{m}} + ( - 4 \cpartial' \text{\TH}' - 10 \mu \cpartial' ) \bar{a} $ & $ ( \text{\TH}' \text{\TH}' + 6 \mu + 4 \mu^2 ) q_{n\bar{m}} + ( - \cpartial' - 4 \mu ) q_{nn} + ( 2 \text{\TH}' \text{\TH}' + 12 \mu + 8 \mu^2 ) \bar{a} $ & $ - \cpartial' \cpartial' q_{n\bar{m}} + \cpartial' \text{\TH}' q_{\bar{m}\bar{m}} + 2 \cpartial' \cpartial' \bar{a}$ \\
		 \hline
		 $\cpartial'\cpartial$ & $0$ & $-q_{n\bar{m}}+2\bar{a}$ & $0$ & $0$ & $0$ & $0$ \\
		 \hline
		 $\text{\TH}'\text{\TH}$ & $0$ & $q_{n\bar{m}}+2\bar{a}$ & $0$ & $0$ & $0$ & $0$ \\
		 \hline
		 $\text{\TH}'$ & $ - \frac{1}{2} \text{\TH} q_{\bar{m}\bar{m}} - 2 \cpartial' \bar{b}$ & $ - \rho q_{n\bar{m}} + ( \frac{5}{2} \text{\TH}' + \frac{11}{2} \mu ) q_{l\bar{m}} + \cpartial q_{\bar{m}\bar{m}} - \frac{3}{2} \cpartial' q_{ln} - \cpartial' q_{m\bar{m}} - \frac{1}{\Psi_2} \cpartial' \Psi^{(0,1)}_{2} + ( \text{\TH} - \rho ) \bar{a} + ( 4 \text{\TH}' + 12 \mu ) \bar{b} $ & $ ( \frac{1}{2} \text{\TH} + 3 \rho ) q_{nn} + ( - \text{\TH}' - 6 \mu ) q_{ln} - \cpartial q_{n\bar{m}} + \frac{5}{2} \cpartial' q_{nm} - \frac{3}{2} \text{\TH}' q_{m\bar{m}} + \frac{1}{\Psi_2} ( \text{\TH}' + 3\mu ) \Psi^{(0,1)}_{2} - \cpartial \bar{a} -  2 \cpartial' a$ & $ - \text{\TH}' q_{\bar{m}\bar{m}} + \cpartial' q_{n\bar{m}} - 2 \cpartial' \bar{a}$ & $ ( \text{\TH}' + \mu ) q_{n\bar{m}} - \cpartial' q_{nn} + ( 2 \text{\TH}' + 2 \mu ) \bar{a} $ & $0$ \\
		 \hline
		 $\text{\TH}'\text{\TH}'$ & $0$ & $q_{l\bar{m}}+2\bar{b}$ & $q_{ln}$ & $0$ & $0$ & $0$  \\
		 \hline
		 $\cpartial'$ & $ \frac{3}{2} \cpartial' q_{ln} + \cpartial' q_{m\bar{m}} - \frac{1}{2} \cpartial q_{\bar{m}\bar{m}} + ( - \frac{5}{2} \text{\TH}' - \frac{3}{2} \mu ) q_{l\bar{m}} + ( \text{\TH} + \rho ) q_{n\bar{m}} + \frac{1}{\Psi_2} \cpartial' \Psi^{(0,1)}_{2} + ( - 2 \text{\TH}' + 6 \mu ) \bar{b} + ( - \text{\TH} - \rho ) \bar{a} $ & $ - \frac{5}{2} \cpartial' q_{nm} + ( \text{\TH}' + 3\mu ) q_{ln} + ( \frac{3}{2} \text{\TH}' - 3 \mu ) q_{m\bar{m}} + ( - \text{\TH} - 3\rho ) q_{nn} - \frac{1}{\Psi_2} ( \text{\TH}' + 3\mu ) \Psi^{(0,1)}_{2} + \cpartial \bar{a} + 4 \cpartial' a $ & $ \frac{1}{2} \cpartial q_{nn} + 2 \mu q_{nm} + ( -2 \text{\TH}' - 6 \mu  ) a $ & $ ( - \text{\TH}' - \mu ) q_{n\bar{m}} + \cpartial' q_{nn} + ( - 2 \text{\TH}' - 2 \mu ) \bar{a} $ & $0$ & $ - \cpartial' q_{n\bar{m}} + \text{\TH}' q_{\bar{m}\bar{m}} + 2 \cpartial' \bar{a} $ \\
		 \hline
		 $\cpartial'\text{\TH}$ & $-q_{n\bar{m}}-2b$ & $q_{nn}$ & $0$ & $0$ & $0$ & $0$ \\
		 \hline
		 $\cpartial'\text{\TH}'$ & $-q_{l\bar{m}}-2\bar{b}$ & $q_{ln}-q_{m\bar{m}}$ & $q_{nm} - 2 a $ & $0$ & $0$ & $0$ \\
		 \hline
		 $\cpartial'\cpartial'$ & $q_{m\bar{m}}$ & $q_{nm}+2a$ & $0$ & $0$ & $0$ & $0$  \\ [1ex] 
		 \hline
	\end{tabular}
\end{adjustbox}

\newpage
\captionof{table}{Table of generating coefficients for the source term $\mathcal{T}_{\geo}^{(1,1)}$ in Eq.~\eqref{eq:T_geo}}\label{tab:T_geoTable}
\begin{adjustbox}{angle=90}
	\begin{tabular}{|| c || p{6cm} | p{6cm} | p{6cm} || } 
		 \hline
		 Operator & $\Psi^{(0,1)}_4$ & $\Psi^{(1,0)}_4$ &  $\Psi^{(1,0)}_3$ \\[0.5ex]
		 \hline\hline
		 $\cpartial\cpartial$ & $-\frac{1}{2}h_{\bar{m}\bar{m}}$ & $-\frac{1}{2}q_{\bar{m}\bar{m}}$ & $0$ \\ [1ex] 
		 \hline
		 $\cpartial\text{\TH}$ & $h_{n\bar{m}}$ & $q_{n\bar{m}}$ & $0$ \\ 
		 \hline
		 $\cpartial\text{\TH}'$ & $h_{l\bar{m}}$ & $q_{l\bar{m}}$ & $0$ \\ 
		 \hline
		 $\cpartial$ & $ - \cpartial' h_{ln} + ( 2 \text{\TH}' + 3\mu ) h_{l\bar{m}} + ( - \frac{1}{2} \text{\TH} - \frac{3}{2} \rho ) h_{n\bar{m}} - \cpartial' h_{m\bar{m}} + \frac{1}{2} \cpartial h_{\bar{m}\bar{m}} - \frac{1}{\Psi_2} \cpartial' \Psi^{(1,0)}_2$ & $ - \cpartial' q_{ln} + ( 2 \text{\TH}' + 3\mu ) q_{l\bar{m}} + ( - \frac{1}{2} \text{\TH} - \frac{3}{2} \rho ) q_{n\bar{m}} - \cpartial' q_{m\bar{m}} + \frac{1}{2} \cpartial q_{\bar{m}\bar{m}} - \frac{1}{\Psi_2} \cpartial' \Psi^{(0,1)}_2 + ( \text{\TH} + \rho ) \bar{a}$ & $- \frac{1}{2}  \text{\TH}' q_{\bar{m}\bar{m}} + \frac{1}{2} \cpartial' q_{n\bar{m}} - \cpartial' \bar{a}$ \\ 
		 \hline
		 $\text{\TH}\text{\TH}$ & $-\frac{1}{2}h_{nn}$ & $-\frac{1}{2}q_{nn}$ & $0$ \\ 
		 \hline
		 $\text{\TH}$ & $ ( - \text{\TH}' - 5\mu ) h_{ln} + ( \frac{1}{2} \text{\TH} + 3 \rho ) h_{nn} - \frac{1}{2} \cpartial h_{n\bar{m}} + 2 \cpartial' h_{nm} - \text{\TH}' h_{m\bar{m}} + \frac{1}{\Psi_2} ( \text{\TH}' + 3 \mu ) \Psi^{(1,0)}_2$ & $ ( - \text{\TH}' - 5\mu ) q_{ln} + ( \frac{1}{2} \text{\TH} + 3 \rho ) q_{nn} - \frac{1}{2} \cpartial q_{n\bar{m}} + 2 \cpartial' q_{nm} - \text{\TH}' q_{m\bar{m}} + \frac{1}{\Psi_2} ( \text{\TH}' + 3 \mu ) \Psi^{(0,1)}_2 - \cpartial \bar{a}$ & $( \frac{1}{2} \text{\TH}' +  \frac{1}{2} \mu ) q_{n\bar{m}} - \frac{1}{2} \cpartial' q_{nn} + ( \text{\TH}' + \mu ) \bar{a}$ \\ 
		 \hline
		 $\text{\TH}\text{\TH}'$ & $-\frac{1}{2}h_{ln}$ & $-\frac{1}{2}q_{ln}$ & $0$ \\ 
		 \hline
		 $\cpartial\cpartial'$ & $-\frac{1}{2}h_{m\bar{m}}$ & $-\frac{1}{2}q_{m\bar{m}}$ & $0$ \\ 
		 \hline
		 $1$ & $ ( - \frac{1}{2} \rho \text{\TH} - 2 \rho^2 ) h_{nn} + ( - \frac{1}{2} \cpartial' \cpartial + \text{\TH}' \text{\TH} + 5\mu \text{\TH} + \rho \text{\TH}' + 4 \rho \mu - \Psi_2) h_{ln} + ( \text{\TH}' \text{\TH}' + \frac{11}{2} \mu \text{\TH}' + 2 \mu^2 ) h_{ll} + \frac{1}{2} \rho \cpartial h_{n\bar{m}} + ( - \frac{1}{2} \cpartial' \text{\TH} - \frac{5}{2} \rho \cpartial' ) h_{nm} + ( - \frac{1}{2} \cpartial \text{\TH}' - 2 \mu \cpartial ) h_{l\bar{m}} + ( 2 \cpartial' \text{\TH}' + \frac{11}{2} \mu \cpartial' )h_{lm} - \cpartial' \cpartial' h_{mm} + ( \cpartial \cpartial' - \frac{1}{2} \text{\TH} \text{\TH}' - \frac{5}{2} \mu \text{\TH} + \rho \text{\TH}' ) h_{m\bar{m}} - \frac{1}{\Psi_2} ( \rho \text{\TH}' + 3 \rho \mu - 3 \Psi_2 ) \Psi^{(1,0)}_2$ & $ ( - \frac{1}{2} \rho \text{\TH} - 2 \rho^2 ) q_{nn} + ( - \frac{1}{2} \cpartial' \cpartial + \text{\TH}' \text{\TH} + 5\mu \text{\TH} + \rho \text{\TH}' + 4 \rho \mu - \Psi_2) q_{ln} + ( \text{\TH}' \text{\TH}' + \frac{11}{2} \mu \text{\TH}' + 2 \mu^2 ) q_{ll} + \frac{1}{2} \rho \cpartial q_{n\bar{m}} + ( - \frac{1}{2} \cpartial' \text{\TH} - \frac{5}{2} \rho \cpartial' ) q_{nm} + ( - \frac{1}{2} \cpartial \text{\TH}' - 2 \mu \cpartial ) q_{l\bar{m}} +( 2 \cpartial' \text{\TH}' + \frac{11}{2} \mu \cpartial' ) q_{lm} - \cpartial' \cpartial' q_{mm} + (  \cpartial \cpartial' - \frac{1}{2} \text{\TH} \text{\TH}' - \frac{5}{2} \mu \text{\TH} + \rho \text{\TH}' ) q_{m\bar{m}} - \frac{1}{\Psi_2} ( \rho \text{\TH}' + 3 \rho \mu - 3 \Psi_2 ) \Psi^{(0,1)}_2 + \rho \cpartial \bar{a}$ & $( \frac{1}{2} \cpartial' \text{\TH} + 2 \rho \cpartial' ) q_{nn} + ( - \frac{5}{2} \cpartial' \text{\TH}' - 6 \mu \cpartial' ) q_{ln} + \frac{1}{2} \cpartial \text{\TH}' q_{\bar{m}\bar{m}} - \frac{5}{2} \cpartial' \text{\TH}' q_{m\bar{m}} + ( - \frac{1}{2} \cpartial' \cpartial - \frac{1}{2} \text{\TH} \text{\TH}' - \frac{3}{2} \rho \text{\TH}' - \frac{1}{2} \mu \text{\TH} - 3 \rho \mu + \frac{1}{2} \Psi_2 ) q_{n\bar{m}} + \frac{5}{2} \cpartial' \cpartial' q_{nm} +( \frac{5}{2} \text{\TH}' \text{\TH}'  + 6 \mu \text{\TH}' - 6 \rho^2 ) q_{l\bar{m}} + \frac{4\mu}{\Psi_2} \cpartial' \Psi^{(1,0)}_2 + ( - 4 \cpartial \cpartial' + 4 \text{\TH}' \text{\TH} +  2\rho \text{\TH}' + 16 \mu \text{\TH} + 18 \rho \mu - 18 \Psi_2 ) \bar{a}$ \\ 
		 \hline
		 $\cpartial'\cpartial$ & $-\frac{1}{2}h_{m\bar{m}}$ & $-\frac{1}{2}q_{m\bar{m}}$ & $0$ \\ 
		 \hline
		 $\text{\TH}'\text{\TH}$ & $-\frac{1}{2}h_{ln}$ & $-\frac{1}{2}q_{ln}$ & $0$ \\ 
		 \hline
		 $\text{\TH}'$ & $ ( \text{\TH}' + \rho ) h_{ln} + ( - \frac{3}{2} \text{\TH}' - 3 \mu ) h_{ll} + \frac{3}{2} \cpartial' h_{lm} - \frac{1}{2} \cpartial h_{l\bar{m}} - \text{\TH} h_{m\bar{m}}$ & $ ( \text{\TH}' + \rho ) q_{ln} + ( - \frac{3}{2} \text{\TH}' - 3 \mu ) q_{ll} + \frac{3}{2} \cpartial' q_{lm} - \frac{1}{2} \cpartial q_{l\bar{m}} - \text{\TH} q_{m\bar{m}}$ & $ ( \frac{3}{2} \text{\TH}' - \frac{3}{2} \mu ) q_{l\bar{m}} - \frac{3}{2} \cpartial' q_{ln} + \frac{1}{\Psi_2} \cpartial' \Psi^{(1,0)}_2 + ( 3 \text{\TH} + 3 \rho ) \bar{a}$ \\ 
		 \hline
		 $\text{\TH}'\text{\TH}'$ & $-\frac{1}{2}h_{ll}$ & $-\frac{1}{2}q_{ll}$ & $0$ \\ 
		 \hline
		 $\cpartial'$ & $ - \frac{1}{2} \cpartial h_{ln} + ( - \frac{1}{2} \text{\TH} - \frac{3}{2} \rho ) h_{nm} + ( \frac{3}{2} \text{\TH}' + \frac{7}{2} \mu) h_{lm} + \cpartial h_{m\bar{m}} - \frac{3}{2} \cpartial' h_{mm}$ & $ - \frac{1}{2} \cpartial q_{ln} + ( - \frac{1}{2} \text{\TH} + \frac{3}{2} \rho ) q_{nm} + ( \frac{3}{2} \text{\TH}' + \frac{7}{2} \mu) q_{lm} + \cpartial q_{m\bar{m}} - \frac{3}{2} \cpartial' q_{mm} $ & $- \frac{3}{2}  \text{\TH}' q_{m\bar{m}} + \frac{3}{2} \cpartial' q_{nm} - \frac{1}{\Psi_2} (  \text{\TH}' + 3 \mu ) \Psi^{(1,0)}_2 - 3 \cpartial \bar{a}$ \\ 
		 \hline
		 $\cpartial'\text{\TH}$ & $h_{nm}$ & $q_{nm}$ & $0$ \\ 
		 \hline
		 $\cpartial'\text{\TH}'$ & $h_{lm}$ & $q_{lm}$ & $0$ \\ 
		 \hline
		 $\cpartial'\cpartial'$ & $-\frac{1}{2}h_{mm}$ & $-\frac{1}{2}q_{mm}$ & $0$ \\ [1ex] 
		 \hline
	\end{tabular}
\end{adjustbox}
\end{widetext}

\newpage
\bibliography{master}

\end{document}